\documentclass[twocolumn, float fix, prb, aps, showpacs, longbibliography]{revtex4-2}
\usepackage{graphicx, amssymb, color, amsmath}
\usepackage{nicefrac}
\usepackage{multirow, array, booktabs}
\usepackage{longtable}
\usepackage{romannum}
\usepackage{mathtools}
\usepackage{bbm}
\usepackage{mathrsfs}
\usepackage{IEEEtrantools}
\usepackage[titletoc, title]{appendix}
\usepackage[colorlinks, bookmarks=true, citecolor=blue, linkcolor=red, urlcolor=blue]{hyperref}
\usepackage{orcidlink}
\AtBeginDocument{\pagenumbering{arabic}}
\begin{document}

\newcommand{\addPRD}[1]{{\color{red}{#1}}}
\newcommand{\addACB}[1]{{\color{blue}{#1}}}

\title{Competition between fractional quantum Hall liquid and electron solid phases in the Landau levels of multilayer graphene}
 
\author{Rakesh K. Dora\orcidlink{0009-0009-0043-2982} and Ajit C. Balram\orcidlink{0000-0002-8087-6015}}
\affiliation{Institute of Mathematical Sciences, CIT Campus, Chennai, 600113, India}
\affiliation{Homi Bhabha National Institute, Training School Complex, Anushaktinagar, Mumbai 400094, India}
\date{\today}

\begin{abstract}
    We study the competition between the electron liquid and solid phases, such as  Wigner crystal and bubbles, in partially filled Landau levels (LLs) of multilayer graphene. Graphene systems offer a versatile platform for controlling band dispersion by varying the number of its stacked layers. The band dispersion determines the LL wave functions, and consequently, the LL-projected Coulomb interaction in graphene and its multilayers is different from that in conventional semiconductors like GaAs. As a result, the energies of the liquid and solid phases are different in the different LLs of multilayer graphene, leading to an alternative phase diagram for the stability of these phases, which we work out. The phase diagram of competing solid and liquid phases in the LLs of monolayer graphene has been studied previously. Here, we primarily consider $AB{-}$ or Bernal$-$stacked bilayer graphene (BLG) and $ABC{-}$stacked trilayer graphene (TLG) and focus on the Laughlin fractions. We determine the cohesive energy of the solid phase using the Hartree-Fock approximation, and the energy of the Laughlin liquid is computed analytically via the plasma sum rules. We find that at the Laughlin fillings, the electron liquid phase has the lowest energy among the phases considered in the $\mathcal{N}{=}0, 1, 2$ LLs of BLG, as well as in the $\mathcal{N}{=}3, 4$ LLs of TLG, while in the $\mathcal{N}{>}2$ LLs of BLG and $\mathcal{N}{>}4$ LLs of TLG, the solid phases are more favorable. We also discuss the effect of impurities on the above-mentioned phase diagram. 
\end{abstract}

\maketitle

\section{Introduction}
 Two-dimensional electron systems (2DESs) placed under a strong perpendicular magnetic field and cooled to low temperatures provide a fertile platform to study interaction-driven electronic phases. Electron-electron interactions in this regime stabilize a multitude of phases including the fractional quantum Hall effect (FQHE)~\cite{Tsui82, Laughlin83}, charge density wave (CDW) states~\cite{Anderson79} like Wigner crystal (WC)~\cite{Koulakov95, Levesque84}, bubble phases~\cite{Koulakov95, Lilly99}, stripes~\cite{ Koulakov95, Du99, JainHalperinBook, Papic22}, etc. Traditionally, these phases have been studied in 2DESs formed in semiconducting quantum wells or heterojunctions of GaAs/AlGaAs systems. With the advent of van der Waals heterostructures, graphene and a few layers of stacked graphene have provided a new platform for realizing 2DESs, which have properties that are different from those of conventional GaAs systems. These graphene-based systems are of particular interest due to their highly tunable band structure~\cite{Castro07, Avetisyan09, Avetisyan10}, presence of additional degrees of freedom such as spin, orbital, valley, etc., and the ability to control the interelectronic interactions~\cite{Abanin12} through knobs such as electric bias and magnetic field.

Owing to the linear dispersion of electrons around the Dirac points and nearly degenerate states for spin and valley degrees of freedom~\cite{Neto09}, monolayer graphene (MLG) displays anomalous behavior in both the integer quantum Hall (IQH) effect and FQHE regimes~\cite{Gusynin05, Novoselov05, Du09, Bolotin09, Dean11, Feldman12, Feldman13}. Depending on the number of layers and stacking configurations, the electronic properties of multilayer graphene systems differ from those of MLG~\cite{Guinea06, Aoki07}. In the case of $ABC{\cdots}{-}$stacked $J{-}$layer graphene ($J{-}$LG), which we will focus on, electrons near the band touching points possess a Berry phase of $J\pi$, which distinguishes them fundamentally from MLG where electrons have a Berry phase of $\pi$~\cite{Barlas12}. As a result of this Berry phase, when exposed to a perpendicular magnetic field, a zero-energy manifold consisting of $J$-fold degenerate LLs, characterized by an orbital quantum number $\mathcal{N}{=}0, 1, {\dots}, J{-}1$ is formed in $J{-}$LG. These orbital degrees of freedom are absent in MLG. Taking spin and valley degrees of freedom into account, the zero-energy manifold becomes $4J$-fold degenerate, and we will refer to this manifold as the zeroth LL (ZLL) of $J{-}$LG. The presence of this ZLL leads to an unconventional sequence of IQH states in $J{-}$LG at fillings $\nu{=}{\pm}4 (\mathfrak{n}{+}J/2)$, where $\mathfrak{n}{=}0, 1, {\dots}$~\cite{McCann06, Min08}. In the presence of the Coulomb interaction between electrons, new IQH states emerge as the degeneracy associated with spin, valley, and orbital quantum numbers gets lifted via the mechanism of quantum Hall ferromagnetism~\cite{Sondhi93, Nomura06, Goerbig06, Alicea06, Abanin07}. In our study, we will consider the first few LLs of graphene multilayers, such as $AB{-}$ or Bernal${-}$stacked bilayer graphene (BLG) where $J{=}2$, and $ABC{-}$stacked trilayer graphene (TLG)~\cite{Min08, Barlas12} where $J{=}3$.

A strong perpendicular magnetic field in conjunction with a symmetry-breaking term quenches the dynamics of electrons to a particular LL with a given spin, valley, and orbital degree of freedom. The interaction physics in the LL of interest is governed by its effective Coulomb interaction, which is described by a LL-dependent form factor. Electrons confined to the lowest LL (LLL) of GaAs exhibit a plethora of FQHE liquid phases, which are predominantly described by the theory of Laughlin~\cite{Laughlin83} and Jain~\cite{Jain89}. At very low fillings in the LLL, electrons are expected to arrange themselves in a triangular lattice called the Wigner crystal~\cite{Lam84, Levesque84, Zuo20}. The second LL (SLL) of GaAs also exhibits FQHE, which can be explained in a unified manner using the framework of the parton theory~\cite{Jain89b, Balram18, Balram19, Balram21, Bose23}. In the higher LLs of GaAs, the effective interaction between electrons becomes attractive, which favors the formation of stripes or bubble phases where two, three, or more electrons cluster at a site, and these clusters are arranged in a triangular lattice~\cite{Lilly99, Du99, Cooper99a, Fogler96, Musaelian96, Moessner96, Rezayi99, Fradkin99, Haldane00}. Competition between the FQHE liquid and electron solid phases like WC and bubble phases in the higher LLs give rise to re-entrant IQH effect (RIQHE)~\cite{Goerbig03, Goerbig04a, Deng12, Deng12a, Liu12, Wang15c, Baer15}, where the longitudinal resistance vanishes around a fractional filling between the usual sequence of FQHE states and develops a plateau in the Hall conductivity with a value quantized to the nearest integer. We note that very recently, a bubble phase not of electrons but electron-vortex composites called composite fermions (CFs)~\cite{Jain89} was observed in the LLL of GaAs~\cite{Shingla23}.

Owing to the linear dispersion in MLG, the effective interaction in its LLs differs from those in conventional semiconducting systems. In the absence of LL mixing, the physics of the $\mathcal{N}{=}0$ LL of MLG is identical to the ideal LLL of nonrelativistic electrons~\cite{Shibata09, Balram15a} (We thus refer to the $\mathcal{N}{=}0$ LL of MLG as LLL.). The effective interaction in the other LLs of MLG is distinct from those of the higher LLs of GaAs~\cite{Shibata09, Kim19}. In the approximation that we will be working in, the physics of LLs with negative energy (``hole LLs") is identical to that of the corresponding LL with positive energy (``electron LLs"), and thus it suffices to consider only the latter, i.e., LLs with orbitals $\mathcal{N}{\geq}0$. The $\mathcal{N}{=}1$ LL of MLG has been studied extensively both theoretically~\cite{Shibata09, Balram15c} and experimentally~\cite{Amet15, Kim19}, and is known to stabilize liquid phases. The competition between electron solid and FQHE liquid phases in the other LLs of MLG has also been explored~\cite{Knoester16}. Recently, RIQHE has been observed in the $\mathcal{N}{=}2$ LL of MLG~\cite{Chen19}, indicating the presence of competing CDW states as predicted theoretically~\cite{Knoester16}. Very recently, bubble phases and phase transitions among bubble crystals with different numbers of electrons per bubble have also been observed in several LLs (up to $\mathcal{N}{=}4$) of MLG~\cite{Yang23}. 

In this paper, we extend the analysis of the competition between liquid and solid phases like WC and bubble phases to the LLs of Bernal${-}$stacked BLG and $ABC{-}$stacked TLG, particularly focusing on the Laughlin fractions~\cite{Laughlin83}, i.e., at fillings $\Bar{\nu}{=}1/(2s{+}1)$, where $s$ is a positive integer. For both BLG and TLG, we consider the LLs ranging from $\mathcal{N}{=}0$ to $\mathcal{N}{=}5$. In LLs with orbital quantum number $\mathcal{N}{>}5$, electron solid phases or stripes are expected to prevail. The $\mathcal{N}{=}1$ orbital of the ZLL of BLG is special in that the interelectronic interaction in it can be continuously tuned as a function of the perpendicular magnetic field~\cite{Apalkov11, Zhu20a, Balram21b}. The interaction in this LL interpolates between that in the LLL and the SLL of GaAs as the magnetic field is tuned from large values to small ones. The effective Coulomb interaction in the $\mathcal{N}{=}1$ ZLL of BLG for low magnetic fields exhibits a short-range attractive component that can promote the formation of electronic bubble phases. Like the higher LLs of GaAs and MLG~\cite{Goerbig03a, Goerbig04a, Knoester16}, the short-range part of the effective interactions in the higher LLs of BLG for $\mathcal{N}{>}2$ and TLG for $\mathcal{N}{>}3$ become attractive, favoring the formation of bubble phases. To quantitatively compute the phase diagram of competing electron solid and liquid phases, we have calculated the cohesive energy of the solid phase using the Hartee-Fock (HF) approximation while the energy of the liquid phase at the Laughlin filings is calculated analytically using the plasma sum rules. To compare with the analytically computed Laughlin liquid energies, we have determined the exact Coulomb ground-state energies at fillings $1/3$ and $1/5$ in the LLs of BLG and TLG by exact diagonalization. To further assess the accuracy of the Laughlin states as candidate wave functions at fillings $1/3$ and $1/5$ in the various LLs of BLG and TLG, we have also computed its overlap with the exact Coulomb ground states. We find for all magnetic fields, the Laughlin liquid dominates over the WC phase in the $\mathcal{N}{=}1$ ZLL of BLG while the bubble phases may appear at non-Laughlin fillings for small magnetic fields. In the $\mathcal{N}{=}2$ LL of BLG, and in the $\mathcal{N}{=}3$ and $\mathcal{N}{=}4$ LLs of TLG, the Laughlin liquid has the lowest energy among the phases considered. In the other higher LLs of both BLG and TLG, the electron bubble phases become viable. We also discuss the effect of impurities on the stability of the electron solid and liquid phases. The main results of our work are summarized in Table~\ref{table: phase_diagram_BLG}. A caveat to keep in mind is that we have not considered phases such as stripes, nematics, or Wigner crystal or bubble phases of CFs (we have considered these phases only for electrons) or Fermi liquid of CFs~\cite{Halperin93}, which can potentially also be feasible in these LLs. Thus, if any of these phases (which are difficult to deal with) happen to have the lowest energy in the LLs that we consider (for example, extensive numerical calculations have shown that at low fillings in the LLL, the Wigner crystal of CFs has lower energy than the electron crystal and strongly competes with the FQHE liquids~\cite{Zuo20}), we will not be able to capture that. Therefore, the phase that we predict has the lowest energy, only among the phases that we have considered.

\begin{table*}[tbh]
  \caption{Lowest energy states among the Laughlin liquid and $M{-}$electron bubble phases at fillings $1/3$,  $1/5$, $1/7$, and  $1/9$ in the LLs of bilayer graphene and trilayer graphene. ``Yes" (``No") represents the presence (absence) of impurities with pinning strength $E_{\rm pin}{=}10^{-4}$.}
  \label{table: phase_diagram_BLG}
  \begin{tabular}{c*{6}{c}}
\toprule
  &  &  \multicolumn{5}{c}{LLs of bilayer graphene}\\\cmidrule (l){3-7} $\Bar{\nu}$ & Impurity & $\mathcal{N}{=}1$ & $\mathcal{N}{=}2$ & $\mathcal{N}{=}3$ & $\mathcal{N}{=4}$ & $\mathcal{N}{=}5$\\
  \midrule

\multirow{2}{*}{$1/3$} & No & Laughlin liquid & ~Laughlin liquid & ~$2$-bubble & ~$3$-bubble & ~$4$-bubble\\[1ex]
   & Yes & Laughlin liquid & ~Laughlin liquid   & ~$2$-bubble & ~$3$-bubble & ~$4$-bubble \\[2ex]
   
\multirow{2}{*}{$1/5$} & No & Laughlin liquid & ~Laughlin liquid & ~Laughlin liquid & ~$2$-bubble & ~$2$-bubble\\[1ex]
   & Yes & Laughlin liquid & ~Laughlin liquid   & ~Wigner crystal & ~$2$-bubble & ~$2$-bubble \\[2ex]   

   \multirow{2}{*}{$1/7$} & No & Laughlin liquid & ~Laughlin liquid & ~Laughlin liquid & ~Laughlin liquid & ~$2$-bubble\\[1ex]
   & Yes & Wigner crystal & ~Wigner crystal  & ~Wigner crystal & ~Wigner crystal & ~$2$-bubble \\[2ex] 

   \multirow{2}{*}{$1/9$} & No & Laughlin liquid & ~Laughlin liquid & ~Laughlin liquid & ~Laughlin liquid & ~Laughlin liquid\\[1ex]
   & Yes & Wigner crystal & ~Wigner crystal  & ~Wigner crystal & ~Wigner crystal & ~Wigner crystal \\[2ex] 
  \midrule
  
  &  & & \multicolumn{3}{c}{LLs of trilayer graphene}\\\cmidrule (l){4-6}  &  & & $\mathcal{N}{=}3$ & $\mathcal{N}{=4}$ & $\mathcal{N}{=}5$ \\
  \midrule

\multirow{2}{*}{$1/3$} & No &  & Laughlin liquid & Laughlin liquid & ~$3$-bubble\\[1ex]
   & Yes &  & Laughlin liquid   & ~Laughlin liquid & ~$3$-bubble \\[2ex]

   \multirow{2}{*}{$1/5$} & No &  & Laughlin liquid & Laughlin liquid & ~$2$-bubble\\[1ex]
   & Yes &  & Laughlin liquid   & ~Wigner crystal & ~$2$-bubble \\[2ex]

    \multirow{2}{*}{$1/7$} & No &  & Laughlin liquid & Laughlin liquid & ~Laughlin liquid\\[1ex]
   & Yes &  & Wigner crystal   & ~Wigner crystal & ~Wigner crystal \\[2ex]

    \multirow{2}{*}{$1/9$} & No &  & Laughlin liquid & Laughlin liquid & ~Laughlin liquid\\[1ex]
   & Yes &  & Wigner crystal   & ~Wigner crystal & ~Wigner crystal \\[2ex]
 \bottomrule
  \end{tabular}
  \end{table*}

The rest of the paper is organized as follows: In Sec.~\ref{ssec: non_interacting_hamiltonian} we review the four-band and two-band models of BLG that describe its low-energy properties. We also discuss the effective two-band Hamiltonian of $ABC{\cdots}{-}$stacked $J{-}$LG, which for $J{=2}$ reproduces the two-band model of BLG. In Sec.~\ref{ssec: Coulomb_interaction} we derive the LL-projected Coulomb interaction in the various LLs of BLG and TLG and discuss its functional form. In Sec.~\ref{sec: solid_phase_energy} we calculate the energy of the electron solid phase, and in Sec.~\ref{sec: energy_FQH_liquid}, the energy of the Laughlin liquid. In Sec.~\ref{sec: energy_FQH_liquid} we also present results obtained from exact diagonalization. The calculated phase diagram in the various LLs of BLG and TLG is summarized in Sec.~\ref{sec: results_phase_diagram}. The effect of disorder on the phase diagram is discussed in Sec.~\ref{sec: effect of impurities}. The paper is concluded with a  discussion of our results and their experimental ramifications in Sec.~\ref{sec: discussion}. Appendixes~\ref{sec: Dirac_eqn_J_layer}-\ref{sec: qp_qh_energies} contain technical details and additional numerical results. Appendix~\ref{sec: Dirac_eqn_J_layer} discusses the spectrum of the chiral fermion model for $J{-}$LG in detail. Appendixes~\ref{sec: disk_pps_JLG} and \ref{sec: spherical_pps_JLG} present the disk and spherical Haldane pseudopotentials~\cite{Haldane83} of the Coulomb interaction in various LLs of $J{-}$LG, respectively. In Appendix~\ref{sec: overlaps} we have tabulated overlaps between the Laughlin and exact Coulomb ground states in the different LLs of $J{-}$LG. In Appendix~\ref{sec: qp_qh_energies} we present the quasiparticle and quasihole energies above the Laughlin states in the various LLs of BLG and TLG.


\section{The Model}

\subsection{Noninteracting Hamiltonian of bilayer and $J{-}$layer graphene}
\label{ssec: non_interacting_hamiltonian}

\begin{figure*}[tbh]
\includegraphics[width=0.99\textwidth]{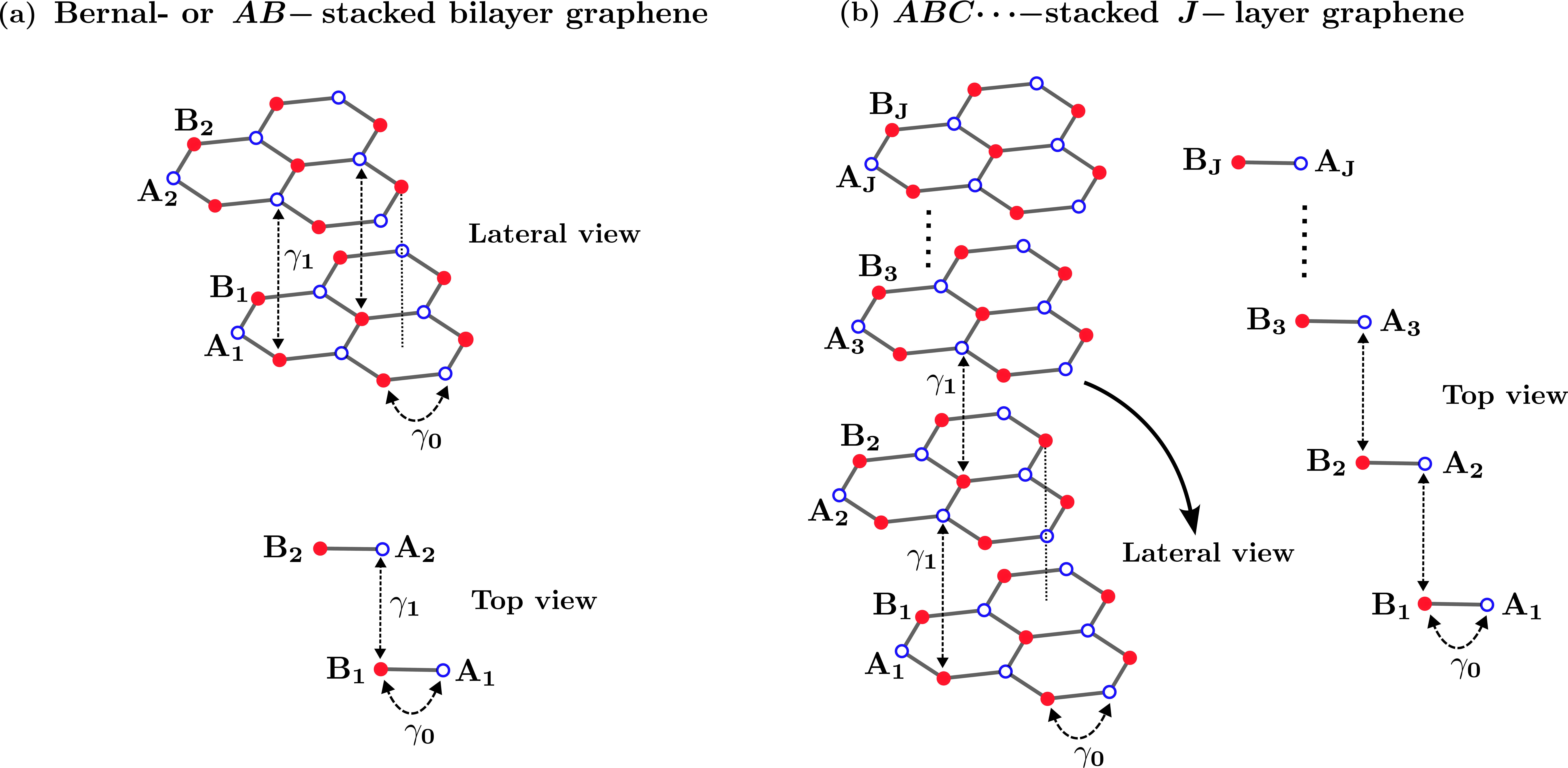}
	\caption{Lateral and top view of the stacking configuration in $(a)$ Bernal${-}$stacked bilayer graphene and $(b)$ $ABC{\cdots}{-}$stacked $J{-}$layer graphene are shown schematically. In each layer, the inequivalent sublattice sites $A$ and $B$ are shown by filled red and empty blue circles, respectively. The coupling parameters $\gamma_{0}$ and $\gamma_{1}$ denote the nearest-neighbor intralayer coupling in each layer and the nearest-neighbor interlayer coupling between adjacent layers, respectively.} 
	\label{fig: schematics_BLG_J_LG}
\end{figure*}
Bilayer graphene comprises two graphene layers as shown in Fig.~\ref{fig: schematics_BLG_J_LG}$(a)$, with the bottom layer containing sublattices $A_{1}$ and $B_{1}$, and the upper layer containing sublattices $A_{2}$ and $B_{2}$, resulting in a unit cell of four atoms. Our focus will be on Bernal$-$ or $AB{-}$stacked BLG where $A_{2}$ is positioned directly above $B_{1}$, and $B_{2}$ is directly above the center of a hexagonal plaquette of the bottom layer [see Fig.~\ref{fig: schematics_BLG_J_LG}$(a)$]. Using the Slonczwski-Weiss-McClure tight-binding parametrization~\cite{McClure57}, we consider only the nearest-neighbor intralayer coupling $\gamma_{0}$ between $A_{i}$ and $B_{i}$ sublattice sites within each layer, and the nearest-neighbor interlayer coupling $\gamma_{1}$ between the $A_{2}$ and $B_{1}$ sublattice sites. We neglect all other couplings between the sublattice sites as they have relatively minor effects on the band structure~\cite{Aoki13}. The low-energy properties of BLG are captured by a four-band model that applies in the vicinity of the two inequivalent Fermi points $\boldsymbol{K}$ and $\boldsymbol{K}^{\prime}$ (referred to as valleys) in the Brillouin zone~\cite{McCann13}. The lowest-energy conduction band and the highest-energy valence band touch each other at the $\boldsymbol{K}$ and $\boldsymbol{K}^{\prime}$ points with a quadratic dispersion, in contrast to MLG where the dispersion is linear~\cite{Neto09, McCann13}. When a uniform perpendicular magnetic field $B$ is applied, the Hamiltonian for spin-polarized electrons around the $\boldsymbol{K}$ valley can be expressed in the basis $(A_{1}, B_{2}, A_{2}, B_{1})$ as~\cite{McCann06}
\begin{equation}
\mathcal{H}_{\boldsymbol{K}}=
\begin{pmatrix}
0 & 0 & 0 & v_{0}\pi^{\dagger} \\
0 & 0&v_{0}\pi &0 \\
0 & v_{0}\pi^{\dag} &0 & \gamma_{1}\\
 v_{0}\pi &0 & \gamma_{1} & 0
\end{pmatrix}.
\label{eq: four band model}
\end{equation}
Here, the velocity $v_{0}{=}\sqrt{3}a\gamma_{0}/(2\hbar)$ where $a$ is the lattice constant. The operators $\pi^{\dag}$ and $\pi$, in the Landau gauge, coincide with the LL creation and annihilation operators, respectively~\cite{McCann06} (see Appendix~\ref{sec: Dirac_eqn_J_layer}). These operators act on the $\mathcal{N}$th LL eigenstates $|\mathcal{N}, X\rangle$ of a nonrelativistic 2DES as
\begin{equation}
\begin{aligned}
   \pi^{\dag}|\mathcal{N}, X\rangle&=i\frac{\hbar}{\ell}\sqrt{2\left (\mathcal{N}+1\right)}~|\mathcal{N}+1, X\rangle, \\
   \pi|\mathcal{N}, X\rangle&=-i\frac{\hbar}{\ell}\sqrt{2\mathcal{N}}~|\mathcal{N}-1, X\rangle~\text{for}~\mathcal{N}>0, \\
\text{and}\;\;
\pi|0, X\rangle&=0,
   \end{aligned}
   \label{eq: Action_of_LL_lowering_raising_operator}
\end{equation}
   where the magnetic length $\ell{=}\sqrt{\hbar c/ (eB)}$ and the quantum number $X$ labels the degenerate eigenstates within a LL that are associated with the guiding center coordinates. The Hamiltonian $\mathcal{H}_{\boldsymbol{K}^{\prime}}$ around the valley $\boldsymbol{K}^{\prime}$ is obtained by replacing $\pi$ by $\pi^{\dag}$ and $\pi^{\dag}$ by $\pi$ in $\mathcal{H}_{\boldsymbol{K}}$. For both the Hamiltonians $\mathcal{H}_{\boldsymbol{K}}$ and $\mathcal{H}_{\boldsymbol{K}^{\prime}}$, there exist two zero-energy LLs with orbital quantum numbers $\mathcal{N}{=}0$ and $\mathcal{N}{=}1$. The eigenstates associated with these LLs are~\cite{Apalkov11}
\begin{IEEEeqnarray}{rCrc}
\label{eq: N=0 in four band model}
\bigl|\Tilde{\Phi}_{\boldsymbol{K}, \mathcal{N}=0, X}^{ (J{=}2)}\bigr\rangle &=& &
\begin{pmatrix}
|0, X\rangle\\0\\0\\0
\end{pmatrix},
\\[1ex]
\label{eq: N=1 in four band model}
\bigl|\Tilde{\Phi}_{\boldsymbol{K}, \mathcal{N}=1, X}^{ (J{=}2)}\bigr\rangle &=& &
\begin{pmatrix*}
 \sin (\theta)|1, X\rangle\\0\\ \cos (\theta)|0, X\rangle\\0
\end{pmatrix*},
\\[1ex]
\label{eq: N=0 in k prime}
\bigl|\Tilde{\Phi}_{\boldsymbol{K}^{\prime}, \mathcal{N}=0, X}^{ (J{=}2)}\bigr\rangle &=& &
\begin{pmatrix}
 0 \\ |0, X\rangle\\0\\0
\end{pmatrix},
\\[1ex]
\label{eq: ZLL in four band model}
\bigl|\Tilde\Phi_{\boldsymbol{K}^{\prime}, \mathcal{N}=1, X}^{ (J{=}2)}\bigr\rangle &=& &
\begin{pmatrix}
0\\ \sin (\theta)|1, X\rangle\\0\\ \cos (\theta)|0, X\rangle
\end{pmatrix}.
\end{IEEEeqnarray}
The tunable parameter $\theta$ is related to the magnetic field as $\tan (\theta){=}\gamma_{1}\ell/ (\sqrt{2}\hbar v_{0}){\propto}1/\sqrt{B}$. The eigenstates and eigenvalues of the remaining LLs (those of the other spin and higher LLs) can be determined similarly. 

The low-energy effective theory of BLG can also be described by a two-band chiral fermion model~\cite{Barlas12} with chirality index $J{=}2$ (see Appendix~\ref{sec: Dirac_eqn_J_layer} for details). The single-particle chiral Hamiltonian for the spin-polarized electrons under a perpendicular magnetic field at the valley $\boldsymbol{K}$ is given by~\cite{McCann06, Min08, Barlas12}:
\begin{align}
\mathcal{H}_{\boldsymbol{K}}&= -\frac{1}{2m^{*}}
\begin{pmatrix}
0 & (\pi^{\dag})^{2} \\
 \pi^{2} &0 
\end{pmatrix},
\label{eq: two band model}
\end{align}
where the effective mass $m^{*}{=}\gamma_{1}/ (2v_{0}^{2})$. The above Hamiltonian is written on the $(B_{1}, A_{2})$ basis. The Hamiltonian $\mathcal{H}_{\boldsymbol{K}^{\prime}}$ at the valley $\boldsymbol{K}^{\prime}$ is obtained by replacing $\pi$ by $\pi^{\dag}$ and vice-versa. Due to the presence of inversion symmetry in unbiased (absence of any perpendicular electric field between the layers) BLG, the LL spectrum of the Hamiltonians $\mathcal{H}_{\boldsymbol{K}}$ and $\mathcal{H}_{\boldsymbol{K}^{\prime}}$ are identical. The single-particle spectrum at the valley $\boldsymbol{K}$ is given by~\cite{McCann06, Min08, Barlas12}:
\begin{equation*}
E_{\boldsymbol{K}, \mathcal{N}}= \pm\hbar\omega_{c}\sqrt{ \mathcal{N} (\mathcal{N}-1)},
\end{equation*}
where $\mathcal{N}$ is a non-negative integer that labels the orbital quantum number of a LL, and the cyclotron frequency $\omega_{c}{=}eB/(m^{*} c)$, where $m^{*}$ is the effective mass. The $\pm$ sign denotes the electron and hole states, respectively. The LLs with orbital quantum number $\mathcal{N}{=}0$ and $\mathcal{N}{=}1$ are degenerate at zero energy:
\begin{align*}
E_{\boldsymbol{K}, \mathcal{N}=0}=E_{\boldsymbol{K}, \mathcal{N}=1}=0.
\end{align*} 
Including the spin and valley degrees of freedom, the ZLL of BLG becomes eight fold degenerate. The eigenstates corresponding to the above LLs are given by~\cite{McCann06, Min08, Barlas12}:
\begin{IEEEeqnarray}{rCrc}
\label{eq: ZLL of BGL}
\bigl|\Phi_{\boldsymbol{K}, \mathcal{N}, X}^{ (J{=}2)}\bigr\rangle &=& &
\begin{pmatrix}
\bigl|\mathcal{N}, X\bigr\rangle\\[1ex] 0
\end{pmatrix}, \;\;\mathcal{N}=0, 1,
\\[1ex]
\label{eq: higher LL of BGL}
\bigl|\Phi_{\boldsymbol{K}, \mathcal{N}, X}^{ (J{=}2)}\bigr\rangle &=&\frac{1}{\sqrt{2}}&
\begin{pmatrix}
\mp\bigl|\mathcal{N}, X\bigr\rangle \\[1ex] \bigl|\mathcal{N}{-}2, X\bigr\rangle
\end{pmatrix}, \;\;\mathcal{N}\geq 2.
\end{IEEEeqnarray}

Similar to the four-band Hamiltonian described in Eq.~\eqref{eq: four band model}, the two-band chiral fermion model of Eq.~\eqref{eq: two band model} reproduces the $\mathcal{N}{=}0$ and $\mathcal{N}{=}1$ LLs with zero energy. Nonetheless, the structure of the eigenstates corresponding to these LLs in the four-band Hamiltonian differs from that of the two-band model. The eigenstates in the ZLL of the chiral fermion model [as seen in Eq.~\eqref{eq: ZLL of BGL}] have complete weight on one or the other layer depending upon the valley index. However, the correspondence between valley and layer within the ZLL of the four-band Hamiltonian is not exact: eigenstates of the $\mathcal{N}{=}0$ orbital [see Eqs.~\eqref{eq: N=0 in four band model} and \eqref{eq: N=0 in k prime}] reside entirely on one or the other layer depending on the valley index while wave functions of the $\mathcal{N}{=}1$ orbital [see Eqs.~\eqref{eq: N=1 in four band model} and \eqref{eq: ZLL in four band model}] have support on both the layers. The experimentally observed quantum Hall states within the ZLL of BLG can be accounted for by considering the ZLL eigenstates of the four-band Hamiltonian~\cite{Hunt17, Zibrov17, Huang22, Balram21b} but not by those of the chiral fermion model. Experimental findings in the higher LLs, particularly in the $\mathcal{N}{=}2$ LL, can be effectively described by the chiral fermion model~\cite{Shibata09}. In the $\mathcal{N}{=}2$ LL, FQHE states with odd denominators are observed while even denominator FQHE states are absent~\cite{Diankov16}. This can be understood from the structure of the LL wave functions in the chiral fermion model (although the argument we provide next does not necessarily rule out the presence of an even-denominator state in the $\mathcal{N}{=}2$ LL since one has to test using numerics if the interaction in this LL can stabilize an even-denominator state). In nonrelativistic LLs, even-denominator states are hosted by the $|1, X\rangle$ LL. The wave function of the $\mathcal{N}{=}2$ LL of BLG in the chiral fermion model comprises an equal superposition of nonrelativistic LLs $|0, X\rangle$ and $|2, X\rangle$ [as seen in Eq.~\eqref{eq: higher LL of BGL}], and neither of these orbitals supports even denominator FQHE states. In our calculations, we shall consider the eigenstates corresponding to the ZLL manifold of BLG from the four-band Hamiltonian and those corresponding to the $\mathcal{N}{\geq} 2$ LL from the two-band chiral fermion model. At the moment, we do not know of any experimental observation in the $\mathcal{N}{\geq} 2$ LLs of BLG whose explanation requires going beyond the two-band model.

Next, we consider $ABC{\cdots}{-}$stacked (also referred to as rhombohedral-stacked) $J{-}$LG where coupling between adjacent layers is $\gamma_{1}$ and the coupling within a layer is $\gamma_{0}$. Interlayer couplings beyond the adjacent layers have been neglected. In this stacking, each pair of adjacent layers forms a Bernal${-}$stacked BLG with the upper $A$ sublattice site directly above the lower $B$ sublattice, and the upper $B$ sublattice sits directly above the center of a hexagonal plaquette of the layer beneath it [see Fig.~\ref{fig: schematics_BLG_J_LG}(b)]. Like BLG, the chiral fermion model with chirality index $J$ can effectively describe the low-energy theory of $J{-}$LG for a small number of stacking layers. In a perpendicular magnetic field, the single-particle chiral Hamiltonian for $J{-}$LG, which acts on the sublattices of the outermost layers, i.e., $(B_{1}, A_{J})$ has the form~\cite{Min08, Barlas12, Sakurai12}
\begin{align}
\mathcal{H}_{\boldsymbol{K}}^{\left (J\right)} &= -\gamma_{1}\big (v_{0}/\gamma_{1}\big)^{J}
\begin{pmatrix}
0 & (\pi^{\dag})^{J} \\
 \pi^{J} &0 
\end{pmatrix}.
\label{eq: two_band_model_of_$J{-}$LG}
\end{align}
The above Hamiltonian pertains to the spin-polarized electrons at the valley $\boldsymbol{K}$. As before, the Hamiltonian at the valley $\boldsymbol{K}^{\prime}$ is obtained by replacing $\pi$ by $\pi^{\dag}$ and vice versa. The single-particle spectrum of $\mathcal{H}_{\boldsymbol{K}}^{\left (J\right)}$ is given by~\cite{Min08, Barlas12, Sakurai12}
\begin{align*}
E_{\boldsymbol{K}, \mathcal{N}}&= \pm\hbar\omega_{J}\sqrt{ \mathcal{N} (\mathcal{N}-1)\dots (\mathcal{N}-J+1)} \;, 
\end{align*}
where $\hbar\omega_{J}{=}\gamma_{1}\big (2\hbar v_{0}/ (\gamma_{1}\ell)\big)^{J}{\propto}{B^{J/2}}$. The $\pm$ sign stands for the electron and hole states, respectively. Similar to BLG, the zero energy manifold of $J{-}$LG consists of $4J$ degenerate LLs (including spin and valley degrees of freedom) since the orbitals $\mathcal{N}{=}0,1,{\dots}, J{-}1$ all have zero energy:
\begin{equation*}
E_{\boldsymbol{K}, \mathcal{N}=0}=E_{\boldsymbol{K}, \mathcal{N}=1}= \dots=E_{\boldsymbol{K}, \mathcal{N}=J-1}=0.
\end{equation*}
The corresponding eigenstates have the form~\cite{Min08, Barlas12, Sakurai12}:
\begin{IEEEeqnarray}{lrCc}
\label{eq: ZLL of M-GL}
\bigl|\Phi_{\boldsymbol{K}, \mathcal{N}, X}^{\left (J\right)}\bigr\rangle &=& & 
\begin{pmatrix}
\bigl|\mathcal{N}, X\bigr\rangle\\[1ex] 0
\end{pmatrix}, \;\;\mathcal{N}=0, 1,\dots,J-1,
\\[1ex]
\label{eq: higher LL of M-GL}
\bigl|\Phi_{\boldsymbol{K}, \mathcal{N}, X}^{\left (J\right)}\bigr\rangle &=& \frac{1}{\sqrt{2}} &
\begin{pmatrix}
\mp\bigl|\mathcal{N}, X\bigr\rangle \\[1ex] \bigl|\mathcal{N}-J, X\bigr\rangle
\end{pmatrix}, \;\mathcal{N}\geq J.
\end{IEEEeqnarray}
A derivation of the above energy spectrum can be found in Appendix~\ref{sec: Dirac_eqn_J_layer}. 

In our study, we focus on the spin and valley polarized electrons and confine them to the $\mathcal{N}$th LL of multilayer graphene. Without loss of generality, we consider spin-polarized electrons at the valley $\mathbf{K}$ and omit the explicit mention of the spin and valley indices. Within the ZLL manifold of BLG, we consider only the $\mathcal{N}{=}1$ orbital. By setting the parameter $\theta{=}0$ in the $\mathcal{N}{=}1$ ZLL eigenstates of Eq.~\eqref{eq: N=1 in four band model}, we recover the physics of the $\mathcal{N}{=}0$ ZLL. For $J{-}$LG $(J{>}2)$, we restrict ourselves to the LLs with $\mathcal{N}{\geq}J$. This is because, similar to the two-band model of BLG, the above two-band Hamiltonian for $J{-}$LG may not correctly capture the structure of its ZLL eigenstates. A four- or higher-band model may be required to accurately describe the interaction physics in the ZLL of $J{-}$LG for $J{>}2$, which is beyond the scope of the current work. We denote the partial filling of the topmost $\mathcal{N}$th LL as $\Bar{\nu}{=}\nu{-}[\nu]{=}N/N_{\phi}$, where $[\nu]$ denotes the integer part of $\nu$, $N$ is the number of electrons in this LL and $N_{\phi}{=}A/ (2\pi\ell^{2})$ is the number of flux quanta passing through the sample of area $A$. 

\subsection{The Coulomb interaction}
\label{ssec: Coulomb_interaction}
As we are interested in the phases of interacting electrons restricted to a LL, in this section, we discuss the Coulomb interaction projected to a LL that can be written in terms of the LL-projected density operators. The electron density operator projected to the $\mathcal{N}$th LL of $J{-}$LG $(J{\geq}2)$ is given by $\Bar{\rho}^{ (J)}_{\mathcal{N}} (\boldsymbol{r}){=}\Psi_{J}^{\dag} (\boldsymbol{r})\Psi_{J} (\boldsymbol{r})$, where $\Psi_{J} (\boldsymbol{r})$ is the real space electron annihilation field operator. The operator $\Psi_{J} (\boldsymbol{r})$ can be reexpressed in terms of annihilation operators $\hat{c}_{\mathcal{N}, X}$ associated with the LL basis states as
\begin{align*}
\Psi_{J} (\boldsymbol{r}) &=\begin{cases}
	     \sum_{X}\langle\mathbf{r}|\Tilde{\Phi}_{\mathcal{N}, X}^{ (J{=}2)}\rangle~\hat{c}_{\mathcal{N}, X} &\text{for}~\mathcal{N}=1\;\text{ZLL of BLG} \\\\
\sum_{X}\langle\mathbf{r}|\Phi_{\mathcal{N}, X}^{ (J)}\rangle~\hat{c}_{\mathcal{N}, X} & \text{for}~\mathcal{N} \geq J.
         \end{cases}
\end{align*}
The operator $\hat{c}_{\mathcal{N}, X}^{\dag}$ ($\hat{c}_{\mathcal{N}, X}$) creates (annihilates) a state $|\mathcal{N}, X\rangle$ in the $\mathcal{N}$th LL [see Eq.~\eqref{eq: N=1 in four band model} for the $\mathcal{N}{=}1$ ZLL of BLG and Eq.~\eqref{eq: higher LL of M-GL} for $\mathcal{N} \geq J$ LLs]. In reciprocal space, the density operator can be written as~\cite{Goerbig06, Knoester16}
\begin{align}
\Bar{\rho}^{ (J)}_{\mathcal{N}}\left (\boldsymbol{q}\right)&= F_{\mathcal{N}}^{\left (J\right)}\left (q\right)\sum_{X, X^{\prime}}\left\langle X\left|e^ {-i\mathbf{q}\cdot\mathbf{R}}\right| X^{\prime}\right\rangle c_{\mathcal{N}, X}^{\dagger} c_{\mathcal{N}, X^{\prime}}\nonumber\\
&\equiv F_{\mathcal{N}}^{\left (J\right)}\left (q\right)\Bar{\rho} (\boldsymbol{q}),
    \label{eq: projected density operator}
\end{align}
where $q$ denotes the magnitude of the wave vector $\boldsymbol{q}$, and $\mathbf{R}$ is the guiding center operator~\cite{Prange87}. The form factor $F_{\mathcal{N}}^{\left (J\right)}\left (q\right)$, which captures the form of the LL wave functions in the $\mathcal{N}$th LL, is written in terms of the $y$th order Laguerre polynomials $L_{y} (x)$ as follows:
\begin{align}
 \intertext{In the $\mathcal{N}{=}1$ ZLL of BLG $(J{=}2)$}
 \label{eq: ZLL form factor}
    F_{\mathcal{N}=1}^{\left (J{=}2\right)} (q)&= \left[\sin^{2} (\theta) L_{1}\bigg (\frac{q^{2}\ell^{2}}{2}\bigg)\right. \nonumber\\
  & \left. + \cos^{2} (\theta) L_{0}\bigg (\frac{q^{2}\ell^{2}}{2}\bigg)\right]e^{-q^{2}\ell^{2}/4},
   \intertext{and in all other LLs with $ \mathcal{N}{\geq} J$}
  \label{eq: higher LL form factor}
    F_{\mathcal{N}}^{\left (J\right)}\left (q\right)&= \frac{1}{2}\left[ L_{\mathcal{N}}\left (\frac{q^{2}\ell^{2}}{2}\right)\right.\nonumber\\
    &\left.+  L_{\mathcal{N}-J}\left (\frac{q^{2}\ell^{2}}{2}\right)\right]e^{-q^{2}\ell^{2}/4}.
  \end{align}
 For some special values of the parameter $\theta$, the form factor of different physical systems can be obtained from the form factor of the $\mathcal{N}{=}1$ ZLL of BLG [see Eq.~\eqref{eq: ZLL form factor}]. At $\theta{=}0$, the form factor $F_{\mathcal{N}{=}1}^{\left (J{=}2\right)}\left (q\right)$ coincides with the form factor of the $\mathcal{N}{=}0$ orbital of the ZLL of $J{-}$LG and also the $n{=}0$ LL (LLL) of nonrelativistic 2DESs~\cite{Goerbig04a}. For $\theta{=}\pi/4$ and $\pi/2$, the form factor of the $\mathcal{N}{=}1$ ZLL of BLG coincides with that of the $\mathcal{N}{=}1$ LL of MLG~\cite{Balram15c, Knoester16} and the $n{=}1$ LL (second LL) of GaAs, respectively~\cite{Goerbig04a}. In $J{-}$LG, the form factor of the LLs with $\mathcal{N}{\geq} J$ [see Eq.~\eqref{eq: higher LL form factor}] is an equal weighted admixture of the form factors associated with the $\mathcal{N}$th and $\left (\mathcal{N}{-}J\right)$th LLs of nonrelativistic 2DESs. We note here that an even more accurate description of the ZLL of BLG (or $ABA{-}$stacked TLG) can be obtained by introducing some admixture of the $|2, X\rangle$ state (a more sophisticated version of the four-band model that we presented can result in such a term), i.e., to consider the form factor~\cite{Aoki13} 
\begin{align}
\label{eq: ZLL_form_factor_accurate}
    F_{\mathcal{N}=1}^{\left (J{=}2\right)} (q)&= \left[\sin^{2} (\phi)\sin^{2} (\theta) L_{2}\bigg (\frac{q^{2}\ell^{2}}{2}\bigg)\right. \nonumber\\
  & \left. +\cos^{2} (\phi)\sin^{2} (\theta) L_{1}\bigg (\frac{q^{2}\ell^{2}}{2}\bigg)\right.\nonumber\\
  & \left. + \cos^{2} (\theta) L_{0}\bigg (\frac{q^{2}\ell^{2}}{2}\bigg) \right]e^{-q^{2}\ell^{2}/4},
    \end{align} 
where it suffices to consider $0{\leq}\theta,\phi{\leq}\pi/2$ since the form factor is only dependent on $\sin^{2}(\phi)$ and $\sin^{2}(\theta)$. (Similar ideas can be applied to higher LLs and for other values of $J$.) However, it is not precisely clear how exactly the weight of the $|2, X\rangle$ state is dependent on external parameters, such as the magnetic field. Moreover, it remains to be seen if this admixture of the $|2, X\rangle$ state is significant and necessary to account for any experimental observation. Therefore, we shall not be considering the model given in Eq.~\eqref{eq: ZLL_form_factor_accurate} in the current work.

Since the electrons are restricted to a particular LL, their kinetic energies are quenched, and the Hamiltonian consists of only the Coulomb interaction $v(\boldsymbol{r})$ [which is spherically symmetric so $v(\boldsymbol{r}){\equiv}v(r)$, where $r{=}|\boldsymbol{r}|$] between them, which we write as
\begin{align*}
  H &= \frac{1}{2}\int{d^{2}\boldsymbol{r}d^{2}\boldsymbol{r^{\prime}}\Bar{\rho}^{\left (J\right)}_{\mathcal{N}}\left (\boldsymbol{r}\right) v (|\boldsymbol{r}-\boldsymbol{r}^{\prime}|)\Bar{\rho}^{\left (J\right)}_{\mathcal{N}} (\boldsymbol{r}^{\prime})} . 
\end{align*}
Moving to the momentum space and using the expression of the projected density operator defined in Eq.~\eqref{eq: projected density operator}, the above Hamiltonian can be rewritten as
\begin{align}
    H &= \frac{1}{2}\sum_{\boldsymbol{q}} v_{\mathcal{N}}^{\left (J\right)}\left (q\right) \Bar{\rho}\left (\boldsymbol{-q}\right) \Bar{\rho}\left (\boldsymbol{q}\right),
    \label{eq: Coulomb interaction}
\end{align}
where $\sum_{\boldsymbol{q}}{=}A\int{d^{2}\boldsymbol{q}/ (2\pi)^{2}}$ and $v_{\mathcal{N}}^{\left (J\right)} (q)$ is the effective interaction in the $\mathcal{N}$th LL of $J{-}$LG, which is related to the Fourier transform of the usual Coulomb interaction $v (q){=}2\pi e^{2}/\epsilon q$ as follows:
\begin{align}
   v_{\mathcal{N}}^{\left (J\right)}\left (q\right) = v\left (q\right)\left[F_{\mathcal{N}}^{\left (J\right)}\left (q\right)\right]^{2}, 
\end{align}
where $\epsilon$ denotes the dielectric constant of the background host material. Throughout this study, unless otherwise specified, we shall consider the bare Coulomb interaction and neglect the effects of screening by gates, LL mixing, and disorder. Before proceeding to the detailed ground-state energy calculations, we present some qualitative analysis on the possibility of the CDW states, in particular, the electronic bubble phases, in the different LLs of $J{-}$LG. The analysis is based on the structure of the real space form of the effective potential $v_{N}^{\left (J\right)} (q)$. Previously, it has been shown that the real space form of the effective potential $v_{n} (q){=}v (q)[L_{n} (q^{2}\ell^{2}/2)]^{2}e^{-q^{2}\ell^{2}/2}$ corresponding to the $n{>}0$ LLs of GaAs satisfies the scaling law $v_{n} (r){=}\Tilde{v} (r/R_{C})/ (R_{C}/\ell)$~\cite{Goerbig03a}, where the cyclotron radius $R_{C}{=}\sqrt{2n{+}1}\ell$. For all $n{>}0$, the scaled interaction $\Tilde{v} (r)$ exhibits a plateau of width $2R_{C}$. Due to the presence of this plateau, two electrons separated by $r{<}2R_{C}$ can come close together with a low-energy penalty, thus allowing for the formation of a cluster of electrons that can arrange themselves to form electronic bubble phases~\cite{Fogler96}.

\begin{figure}[tbh]
	\includegraphics[width=0.99\columnwidth]{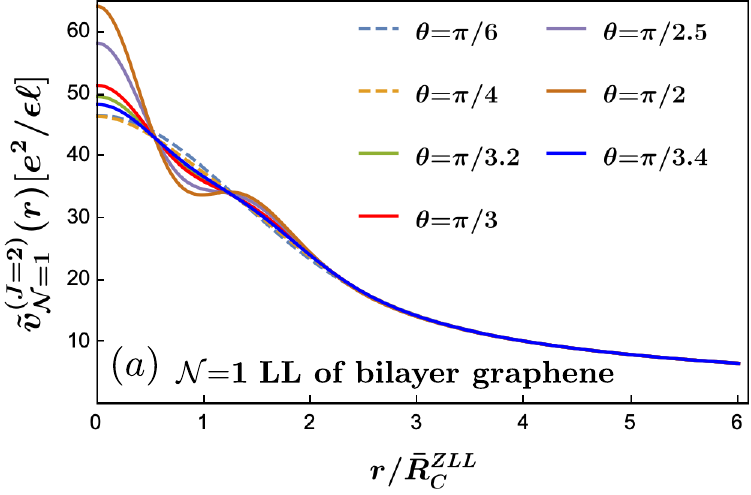}
        \includegraphics[width=0.99\columnwidth]{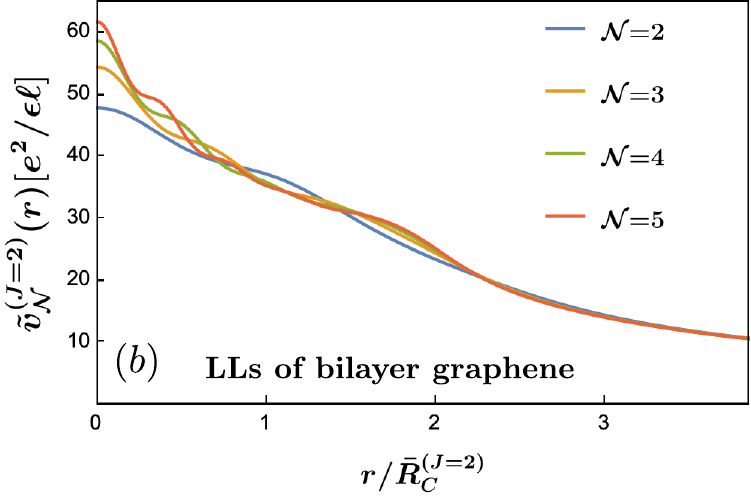}
        \includegraphics[width=0.99\columnwidth]{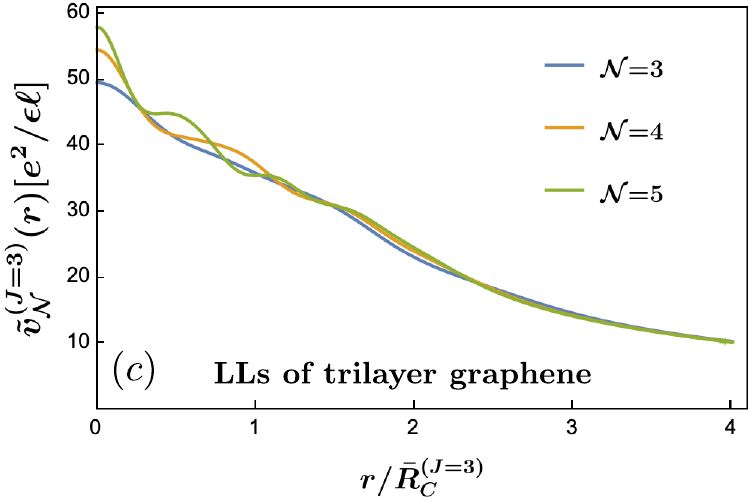}
	\caption{Effective scaled real-space Coulomb potentials in $\left (a\right)$ $\mathcal{N}{=}1$ zeroth LL of bilayer graphene for several values of the parameter $\theta$ [defined in Eq.~\eqref{eq: ZLL in four band model}], $\left (b\right)$ nonzero energy LLs of bilayer graphene, $\left (c\right)$ nonzero energy LLs of trilayer graphene.} 
	\label{fig: scaled_potentials}
\end{figure}

In general, the form factor of $J{-}$LG is built from the form factors of two different LLs of GaAs. To construct a scaled real-space potential for $J{-}$LG, we define an appropriately averaged cyclotron radius as follows: $\Bar{R}_{C}^{\rm ZLL}{=}\ell[\sqrt{3}\sin^{2} (\theta)+\cos^{2}{\theta} ]$ for the $\mathcal{N}{=}1$ ZLL of BLG $(J{=}2)$ and $\Bar{R}_{C}^{\left (J\right)}{=} (\ell/2)\bigl[\sqrt{2\mathcal{N}{+}1}+\sqrt{2 (\mathcal{N}{-}J){+}1}~\bigr]$ for the $\mathcal{N}{\geq} J$ LL of $J{-}$LG $(J{\geq} 2)$. The corresponding real-space scaled potentials $\Tilde{v}^{\left (J\right)} (r)$ are defined as
\begin{align*}
    v^{\left (J{=}2\right)}_{\mathcal{N}=1} (r) &= \frac{\Tilde{v}^{\left (J{=}2\right)}_{\mathcal{N}=1} (r/\Bar{R}_{C}^{\rm ZLL})}{\Bar{R}_{C}^{\rm ZLL}/ \ell},
\\[1ex]
    v^{\left (J\right)}_{\mathcal{N}} (r)&= \frac{\Tilde{v}^{\left (J\right)}_{\mathcal{N}} (r/\Bar{R}_{C}^{\left (J\right)})}{\Bar{R}_{C}^{\left (J\right)}/ \ell}.
\end{align*}
In Fig.~\ref{fig: scaled_potentials}$(a)$, the scaled potentials in the $\mathcal{N}{=}1$ ZLL of BLG for different values of the parameter $\theta$ are shown. A plateau-like structure starts to emerge around $\theta{=}\pi/3.4$, at length scales $\Bar{R}_{C}^{\rm ZLL}{<} r {<}2\Bar{R}_{C}^{\rm ZLL}$ and becomes fully developed at $\theta{=}\pi/2$. This can be understood by noting that the form factor in the ZLL of BLG is a weighted superposition of the form factors of the $n{=}0$ and $n{=}1$ LLs of GaAs, where the weights are controlled by $\theta$ as seen in Eq.~\eqref{eq: ZLL form factor}. The $n{=}0$ LL interaction potential develops no plateau, so it remains scale-free while the $n{=}1$ LL interaction potential develops a plateau around $2R_{C}$~\cite{Goerbig04a, Goerbig03a}. As the parameter $\theta$ increases, the amplitude of the $n{=}0$ LL decreases. Beyond a critical value of $\theta$, a length scale of $2\Bar{R}_{C}^{\rm ZLL}$ starts to emerge. Consequently, we anticipate that bubble phases in the ZLL of BLG can occur only for $\theta{>}\pi/3.4$. In the higher LLs of BLG as shown in Fig.~\ref{fig: scaled_potentials}$(b)$, the scaled potentials fall on top of each other and develop a plateau at length scales $\Bar{R}_{C}^{\left (J{=}2\right)}{<}r{<}2\Bar{R}_{C}^{\left (J{=}2\right)}$, except in the $\mathcal{N}{=}2$ LL. As in BLG, the scaled potentials in TLG [see Fig.~\ref{fig: scaled_potentials}$(c)$], for LLs with $\mathcal{N}{>}3$ fall on top of each other at length scales $\Bar{R}_{C}^{\left (J{=}3\right)} {<}r {<}2\Bar{R}_{C}^{\left (J{=}3\right)}$, except in the $\mathcal{N}{=}3$ LL. As shown in Fig.~\ref{fig: scaled_potentials}$(c)$, the scaled potentials in the $\mathcal{N}{>}3$ LLs of TLG also exhibit plateaus in the range $r{\leq} R_{C}^{\left (J{=}3\right)}$. The absence of a plateau in the scaled interaction in the $\mathcal{N}{=}2$ LL of BLG and the $\mathcal{N}{=}3$ LL of TLG can be attributed to the fact that these LLs have a form factor that has 50\% weight in the LLL, which does not exhibit any plateau. 


\section{Energy of electron solid phases}
\label{sec: solid_phase_energy}

\begin{figure}[tbh]
\includegraphics[width=0.99\columnwidth]{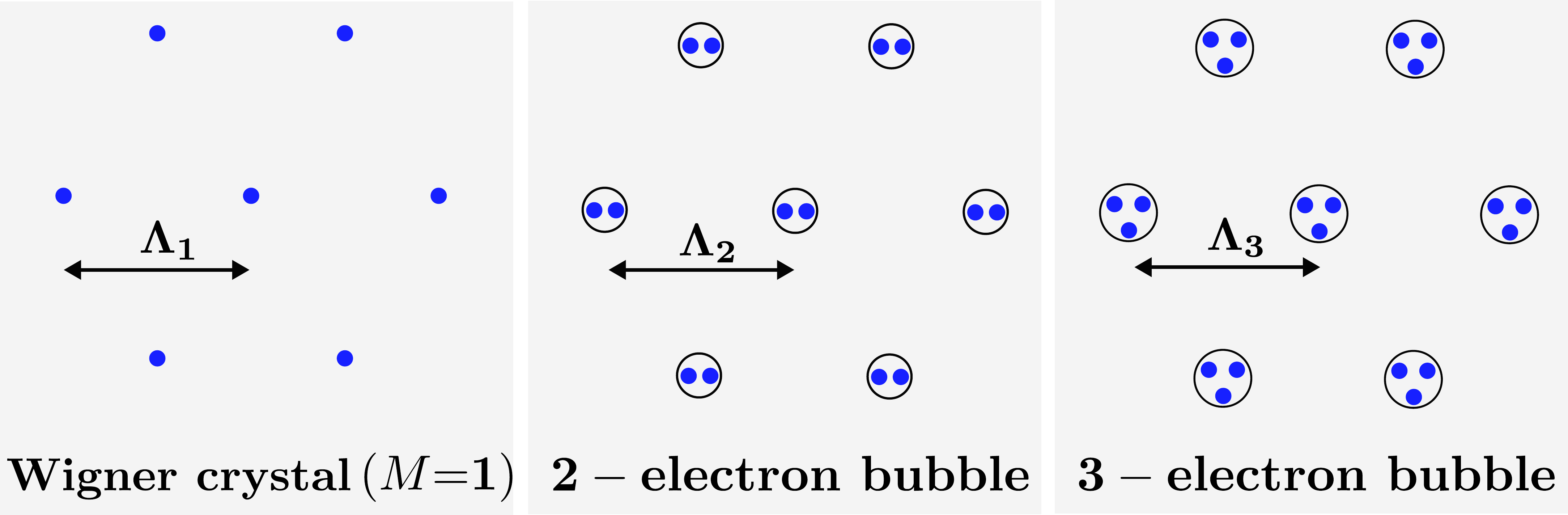}
	\caption{Sketch of $M{-}$electron bubble phases for $M{=}1,2$, and $3$, with lattice constants $\Lambda_{M}$. Solid blue circles represent electrons, and black circles represent bubbles.} 
	\label{fig: schematics_bubbles}
\end{figure}

We will use the Hartree-Fock (HF) mean-field approximation, which is known to give an accurate description of the electronic solid phases like WC and bubbles and stripes~\cite{Fogler96, Goerbig04a}, to find the energy of the solids under consideration. The HF mean-field analysis provides a quantitative understanding of the RIQHE observed in higher LLs of GaAs that arises due to the presence of the bubble phases~\cite{Goerbig04a}. Very recently, phase transitions between bubble crystals in the LLs of MLG have been observed and shown to be in quantitative agreement with the HF results~\cite{Yang23}. This suggests that the same HF method can give a reasonable description of the bubble crystals in few-layer graphene. Following Ref.~\cite{Goerbig04a}, the interaction Hamiltonian $H$ [see Eq.~\eqref{eq: Coulomb interaction}] in the HF approximation is given by:
\begin{align*}
H_{\rm HF} &=\frac{1}{2}\sum_{\boldsymbol{q}}u_{\mathcal{N}}^{{\rm HF}, \left (J\right)} (\boldsymbol{q})\langle  \Bar{\rho} (\boldsymbol{-q})\rangle \Bar{\rho} (\boldsymbol{q}),
\intertext{where}
 u_{\mathcal{N}}^{{\rm HF}, \left (J\right)} (\boldsymbol{q})&= u_{\mathcal{N}}^{H, \left (J\right)} (\boldsymbol{q}) - u_{\mathcal{N}}^{F, \left (J\right)} (\boldsymbol{q}).
\end{align*}
Here, the Hartree term $u_{\mathcal{N}}^{H, \left (J\right)}$ is equivalent to the effective interaction, i.e., $u_{\mathcal{N}}^{H, \left (J\right)}{=}v_{\mathcal{N}}^{\left (J\right)} (q)$, and the Fock exchange term is related to the effective interaction $v_{\mathcal{N}}^{\left (J\right)} (q)$ as~\cite{Goerbig04a}
\begin{align*}
    u_{\mathcal{N}}^{F, \left (J\right)} (\boldsymbol{q})&= \frac{1}{N_{\phi}}\sum_{\boldsymbol{p}}v_{\mathcal{N}}^{\left (J\right)} (p)e^{-i \ell^{2} (\boldsymbol{p}\times \boldsymbol{q})_{z}}.
\end{align*}
 The cohesive energy $E_{\rm coh}{=}\langle H_{{\rm HF}}\rangle/N$ corresponding to a CDW state with the order parameter $\Delta (\boldsymbol{q}){=}\langle \Bar{\rho} (\boldsymbol{q})\rangle/N_{\phi}$ is given by~\cite{Goerbig04a}:
 \begin{align*}
     E_{\rm coh}^{{\rm CDW}, \left (J\right)}= \frac{N_{\phi}}{2A\Bar{\nu}}\sum_{\boldsymbol{q}}u_{\mathcal{N}}^{{\rm HF}, \left (J\right)} (\boldsymbol{q})|\Delta (\boldsymbol{q})|^{2}.
 \end{align*}
 We consider the order parameter corresponding to a $M$-electron bubble phase, where clusters of $M$ ($M$ is a positive integer) electrons are arranged in a triangular lattice of lattice constant $\Lambda_{M}{=}\sqrt{4\pi M/ (\sqrt{3}\Bar{\nu})}\ell$. Note that the electron bubble phase with $M{=}1$ is equivalent to the WC. A schematic of electron bubble phases is shown in Fig.~\ref{fig: schematics_bubbles}. The cohesive energy of the $M{-}$electron bubble is~\cite{Goerbig04a}
 \begin{align}
 E_{\rm coh}^{B, \left (J\right)}\left (\mathcal{N};M, \Bar{\nu}\right)&= \frac{N_{\phi}\Bar{\nu}}{AM}\sum_{l}u_{\mathcal{N}}^{{\rm HF}, \left (J\right)}\left (\boldsymbol{G}_{l}\right)\nonumber\\
 & \times\frac{J_{1}^{2}\left (\sqrt{2M}\ell|\boldsymbol{G}_{l}|\right)}{\ell^{2}\boldsymbol{G}_{l}^{2}},
     \label{eq: solid phase energy}
 \end{align}
where $\boldsymbol{G}_{l}$ are the reciprocal lattice vectors of the triangular lattice formed by the clusters of electrons and $J_{1} (x)$ is the first-order Bessel function. The summation in the cohesive energy converges rapidly, and thus, it suffices to obtain it by summing over a few dozen reciprocal lattice vectors $\boldsymbol{G}_{l}$ of different magnitudes. 

\begin{figure}[tb]
	\includegraphics[width=0.99\columnwidth]{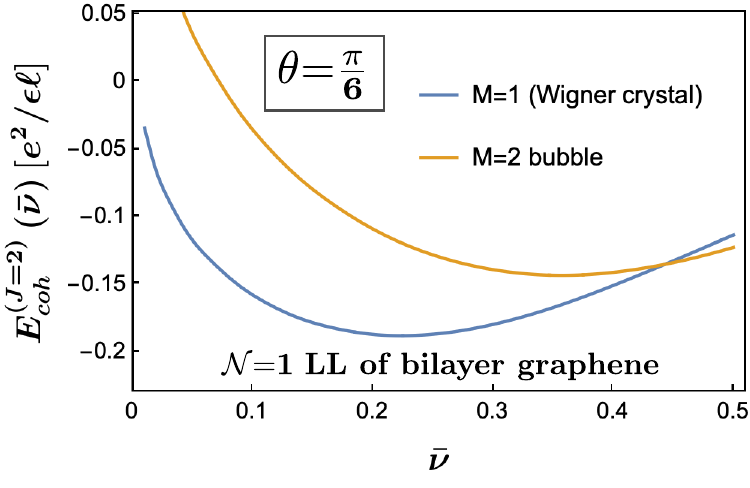}
	\caption{Cohesive energies of the Wigner crystal $(M{=}1)$ and two-electron bubble phase in the $\mathcal{N}{=}1$ zero-energy LL of bilayer graphene at $\theta{=}\pi/6$ [see Eq.~\eqref{eq: ZLL in four band model} for definition of the parameter $\theta$].} 
	\label{fig: Wc_vs_bubble}
\end{figure}

Figure~\ref{fig: Wc_vs_bubble}, shows the computed energy of the WC and two-electron bubble phase in the $\mathcal{N}{=}1$ zero-energy LL of BLG at $\theta{=}\pi/6$. Other higher electron bubble phases have larger energies and are not shown in the figure. This particular value of $\theta$ serves as a representative of the observed trends in the energy of solid phases for the entire range of the parameter $\theta$ from $0$ to $\pi/2$. We observe that for fillings $\Bar{\nu}{>}0.44$, the two-electron bubble phase has lower energy than the WC. Similarly, across all other values of $\theta$, the two-electron bubble exhibits the lowest energy near half-fillings. However, for $0{<}\theta{<}\pi/3.4$, we anticipate the WC phase to have the lowest energy at all fillings as discussed in Sec.~\ref{ssec: Coulomb_interaction}. This slight deviation from the expectation may be due to the first-order approximation employed in the computation of the solid phase energy. 

 \section{FQH liquid}
 \label{sec: energy_FQH_liquid}
 \subsection{Energy of the Laughlin liquid and its excitations}
 
Unlike in the solid phases, the HF approximation does not capture the electronic correlations present in the FQHE liquids. Thus, to evaluate the energy of the liquid state, we rely on trial wave functions and exact diagonalization. We restrict ourselves to the states at filling fractions $\Bar{\nu}{=}1/ (2s{+}1)$, where $s$ is a positive integer. In the lowest LL $ (\mathcal{N}{=}0)$, the FQH states at these fillings are well-described by Laughlin's wave functions~\cite{Laughlin83}. In higher LLs, the true FQH ground state (if FQH is stabilized in the first place) may not necessarily have a good overlap with the Laughlin ansatz~\cite{Ambrumenil88, Balram13b, Kusmierz18}, and other candidate FQHE states could compete with it~\cite{Peterson15, Jeong16, Kusmierz18, Balram20b, Faugno21}. Nevertheless, in this work, we consider only the possibility that a Laughlin state is stabilized since unlike other competing states, its energy can be computed semi-analytically. 

Following Ref.~\cite{Goerbig04a}, the total energy of the Laughlin state in the $\mathcal{N}$th LL of $J{-}$LG can be written as
\begin{align}
U &=  E_{\rm coh}^{L, \left (J\right)} (\mathcal{N}, s) - \frac{\Bar{\nu}}{2 A} \sum_{\boldsymbol{q}} v_{\mathcal{N}}^{\left (J\right)} (q),
\end{align}
where $E_{\rm coh}^{L, \left (J\right)} (\mathcal{N}, s)$ is the cohesive energy of the Laughlin liquid. The cohesive energy can be written in terms of the Haldane pseudopotentials~\cite{Haldane83}
\begin{align}
    V_{m}^{\left (J, ~\mathcal{N}\right)}= \frac{1}{N_{\phi}}\sum_{\boldsymbol{q}}v_{\mathcal{N}}^{\left (J\right)} (q)L_{m} (q^{2}\ell^{2})e^{-q^{2}\ell^{2}/2},
    \label{eq: Pseudopotentials_J_LG}
\end{align}
which is the energy of two electrons in a state of relative angular momentum $m$, where $m$ is an integer. Since we consider electrons confined to a single LL, only the odd Haldane pseudopotentials are relevant to determine the electron-electron interaction energy. The pseudopotentials for the ZLL of BLG and the higher LLs of BLG and TLG are given in Appendixes~\ref{sec: disk_pps_JLG} and~\ref{sec: spherical_pps_JLG}. The cohesive energy of the Laughlin state at filling $1/ (2s{+}1)$ is given by~\cite{Goerbig04a}
\begin{align}
    E_{\rm coh}^{L, \left (J\right)} (\mathcal{N}, s) = \frac{\Bar{\nu}N_{\phi}}{\pi A}\sum_{m=0}^{\infty} c_{2m+1}^{s} V_{2m+1}^{\left (J, ~\mathcal{N}\right)},
    \label{liquid phase energy}
\end{align}
where $c_{2m+1}^{s}$ are dimensionless coefficients. By exploiting the plasma analogy of the Laughlin wave function~\cite{Laughlin83}, these coefficients are found to satisfy three sum rules that are related to the charge neutrality, perfect screening, and compressibility of the two-dimensional plasma~\cite{Girvin86}. The coefficients are obtained by using the sum rules as constraints and fitting the pair-distribution function of the Laughlin state to large-scale Monte Carlo calculations~\cite{Girvin86}. Alternatively, they can also be computed analytically by assuming $c_{2m+1}^{s}{=}0$ for $m{\geq} s{+}3$, along with the aforementioned sum rules and using the condition that electrons at short distance repel each other, which yields $c_{2m+1}^{s}{=}{-}1$ for $m{<}s$~\cite{Goerbig02}. The coefficients obtained by this procedure can be found in Table~\Romannum{2} of Ref.~\cite{Knoester16} and will be utilized here for the analytic computation of energy of the Laughlin state. The total energies, which include the contribution of the positively charged background, $U$ of the Laughlin liquid at $\Bar{\nu}{=}1/3$ and $1/5$ in the LLs of BLG and TLG are given in Table~\ref{table: comparison_of_exact_analytic_results}.

\begin{table}[tbh!]
  \caption{Comparison of the analytically obtained energy $U$ of the Laughlin liquid in the thermodynamic limit with the numerically obtained energies (extrapolated to the thermodynamic limit from finite-size systems) of the exact Coulomb ground state and the Laughlin state at $\Bar{\nu}{=}1/3$ and $1/5$ in the spherical geometry using the spherical pseudopotentials in the following LLs of bilayer graphene (BLG) and trilayer graphene (TLG): $(a)$ In the $\mathcal{N}{=}1$ zero-energy LL of BLG for $\theta{=}0$, $\theta{=}\pi/4$, and $\theta{=}\pi/2$, $(b)$ In the $\mathcal{N}{=}2$ LL of BLG, $(c)$ In the $\mathcal{N}{=}3$ LL of TLG. All energies are quoted in units of $e^{2}/ (\epsilon\ell)$.}
  \label{table: comparison_of_exact_analytic_results}
  \begin{tabular}{*{5}{c}}
\toprule
\multirow{4}{*}{LL} & \multirow{4}{*}{$\Bar{\nu}$} & \multicolumn{1}{c}{Analytical}  & \multicolumn{2}{c}{Numerical}\\\cmidrule (l){3-3}\cmidrule (l){4-5} & & Laughlin &  Laughlin & Exact\\
  & & $\left (U\right)$ & ansatz & ground state\\ \midrule
\multicolumn{1}{c}{$\mathcal{N}{=}1$~\text{BLG}} &  &   & \\\cmidrule (l){1-1}  & & & & \\ 
 \multirow{2}{*}{$\theta{=}0$} & $1/3$ & $-0.409$ & $-0.40982(1)$ & $-0.41015(5)$~\cite{Balram20b}\\\\
    (LLL) & 1/5 & $-0.3265$ & $-0.3275(1)$ & $-0.328(1)$\\[1ex]\hline\\
   \multirow{2}{*}{$\theta{=}\pi/4$} & $1/3$ & $-0.37$ & $-0.37130(1)$ & $-0.3718(1)$\\\\
    ($\mathcal{N}{=}1$~MLG) & 1/5 & $-0.311$ & $-0.3119(2)$ & $-0.3128(4)$\\[1ex] \hline\\
   \multirow{2}{*}{$\theta{=}\pi/2$} & $1/3$ & $-0.3245$ & $-0.32644(2)$ & $-0.328(2)$~\cite{Balram20b}\\\\
    (SLL) & 1/5 & $-0.294$ & $-0.2952(2)$ & $-0.296(1)$\\[1ex]\midrule \midrule\\
 \multirow{2}{*}{$\mathcal{N}{=}2$~BLG} & $1/3$ & $-0.334$ & $-0.3356 (7)$ & $-0.336 (1)$\\[1ex]
   & 1/5 & $-0.292$ & $-0.294 (1)$ & $-0.296 (2)$\\[1ex] \midrule\midrule\\
\multirow{2}{*}{$\mathcal{N}{=}3$~TLG} & $1/3$ & $-0.297$ & $-0.2993 (2)$ & $-0.3001 (4)$\\[1ex]
 & 1/5 & $-0.272$ & $-0.275 (1)$ & $-0.276 (2)$\\ 
 \bottomrule
  \end{tabular}
  \end{table}

    \begin{figure*}[tbh]
        \includegraphics[width=0.99\columnwidth]{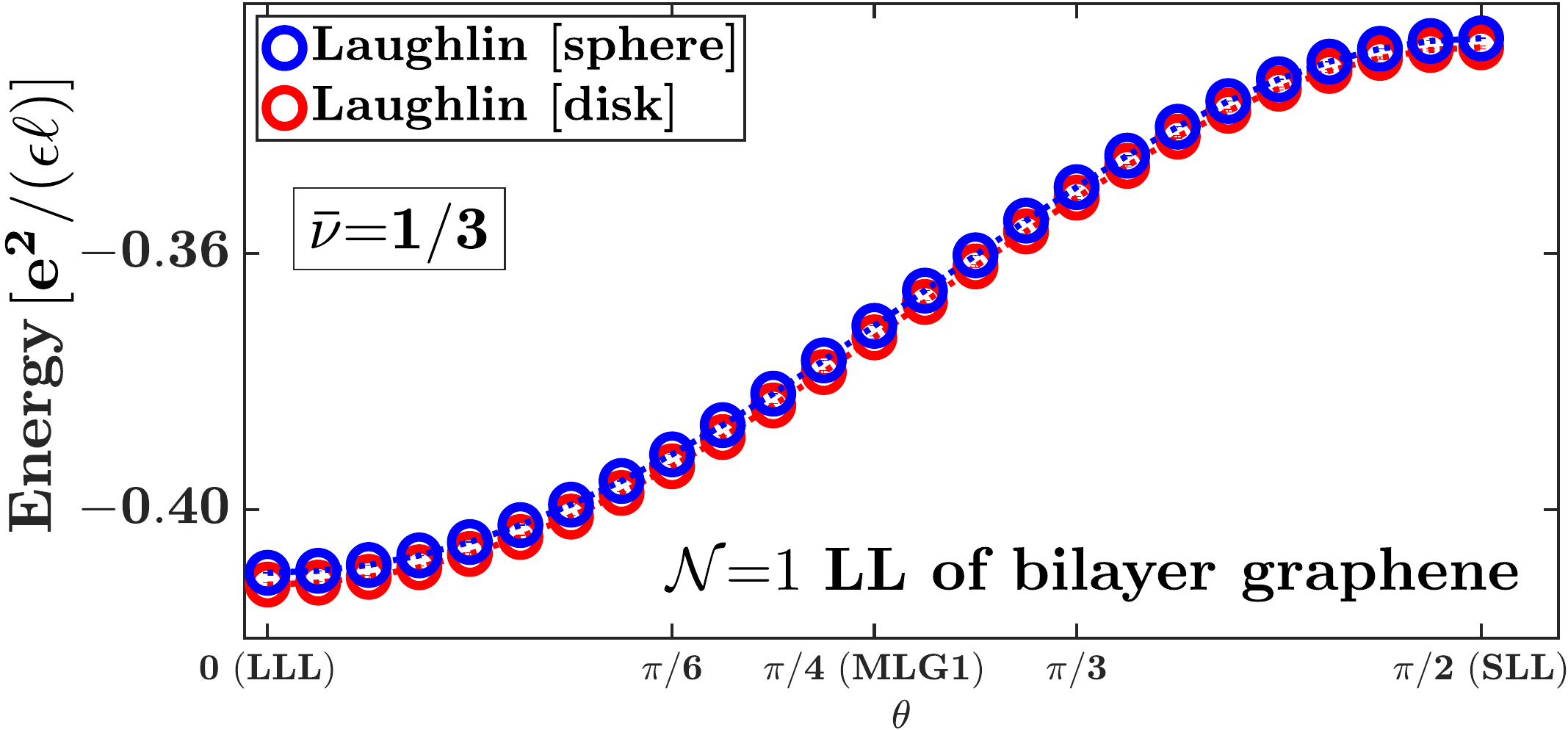}
        \includegraphics[width=0.99\columnwidth]{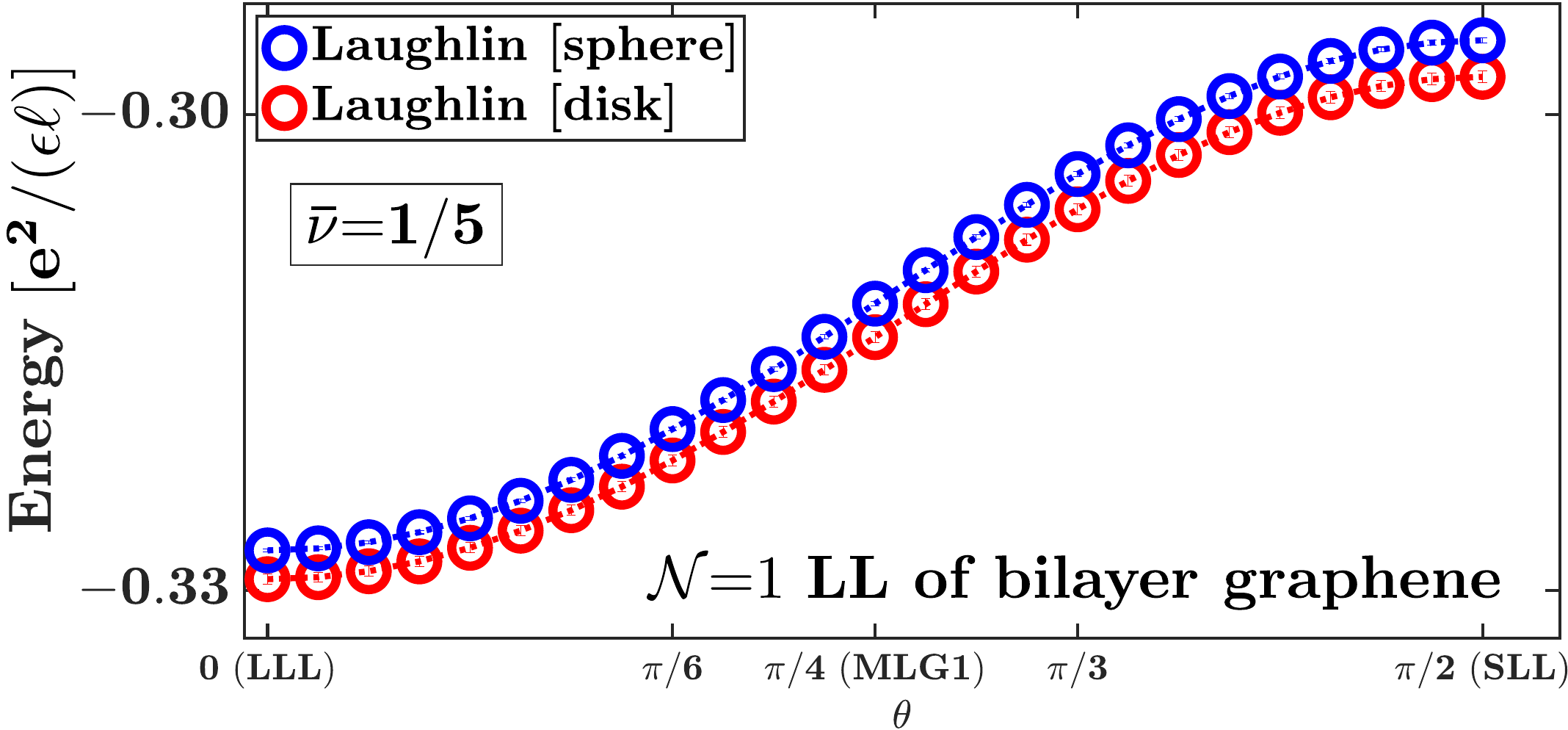}
	\caption{Thermodynamic per-particle density-corrected Coulomb energies of the $\Bar{\nu}{=}1/3$ and $1/5$ Laughlin states in the $\mathcal{N}{=}1$ zero-energy LL of BLG as a function of the parameter $\theta$ obtained using the spherical and disk pseudopotentials in the spherical geometry. The error bars are obtained from the uncertainty in the linear fit as a function of $1/N$. ``MLG1'' represents the $\mathcal{N}{=}1$ LL of MLG, which is equivalent to the $\mathcal{N}{=}1$ LL of BLG for $\theta{=}\pi/4$ [see Eq.~\eqref{eq: ZLL in four band model}]. Similarly, for $\theta{=0}$ and $\theta{=}\pi/2$, the $\mathcal{N}{=}1$ LL of BLG, is equivalent to the LLL ($n{=}0$) and SLL ($n{=}1$) of GaAs, which are indicated adjacent to these $\theta$ values.}
	\label{fig: Laughlin_Coulomb_energies_ZLL_BLG}
\end{figure*}


Around the immediate vicinity of the Laughlin filling $\Bar{\nu}{=}{1/ (2s{+}1)}$, the system can be viewed as consisting of a dilute density of quasiparticles (for $\Bar{\nu}_{+}{>}\Bar{\nu}$) or quasiholes (for $\Bar{\nu}_{-}{<}\Bar{\nu}$). The cohesive energy at fillings $\Bar{\nu}_{+}$ or $\Bar{\nu}_{-}$ can be obtained by considering the energy cost for creating quasiparticles or quasiholes above the Laughlin liquid, respectively. The energy cost associated with the formation of quasiparticles ($\Delta_{+}$) and quasiholes ($\Delta_{-}$) can be obtained analytically by using Murthy and Shankar's Hamiltonian theory~\cite{Shankar01, Murthy03}. For the $n$th LL of GaAs, they are given by:
\begin{align}
\label{eq: quasiparticle energy}
\Delta_{+}^n (s, p) &=  \frac{1}{2} \int_{\mathbf{q}} v_n (q)\langle p|\Bar{\rho}^p (-\mathbf{q}) \Bar{\rho}^{p} (\mathbf{q})| p\rangle\nonumber \\
& -\int_{\mathbf{q}} v_n (q) \sum_{j^{\prime}=0}^{p-1} | \langle p\left|\Bar{\rho}^p (\mathbf{q})\right | j^{\prime}\rangle|^ 2
\intertext{and}
\label{quasihole energy}
\Delta_{-}^n (s, p) &=  -\frac{1}{2} \int_{\mathbf{q}} v_n (q)\langle p-1|\Bar{\rho}^p (-\mathbf{q}) \Bar{\rho}^{p} (\mathbf{q})| p-1\rangle \nonumber\\
& +\int_{\mathbf{q}} v_n (q) \sum_{j^{\prime}=0}^{p-1} | \langle p-1\left|\Bar{\rho}^p (\mathbf{q})\right | j^{\prime}\rangle|^ 2.
\intertext{The matrix elements in the above equations are given explicitly as}
\langle j|\Bar{\rho}^p (\boldsymbol{q})| j^{\prime}\rangle &=  \sqrt{\frac{j^{\prime} !}{j !}}\left (\frac{-i\left (q_x-i q_y\right) \ell^* \mathcal{C}}{\sqrt{2}}\right)^{j-j^{\prime}}\nonumber \\
& \times e^{-|q|^2 \ell^{* 2} \mathcal{C}^2 / 4}\left[L_{j^{\prime}}^{j-j^{\prime}}\left (\frac{|q|^2 \ell^{* 2} \mathcal{C}^2}{2}\right)\right. \nonumber\\
& \left.-\mathcal{C}^{2\left (1-j+j^{\prime}\right)} e^{-|q|^2  / 2 \mathcal{C}^2} L_{j^{\prime}}^{j-j^{\prime}}\left (\frac{|q|^2 \ell^{* 2}}{2 \mathcal{C}^2}\right)\right]\nonumber,
\end{align}
where $\ell^*{=}\ell/\sqrt{1{-}\mathcal{C}^{2}}$, $\mathcal{C}^{2}{=}2ps/ (2p{+}1)$ with $s$ and $p$ being integers, and $L_{y}^{j}(x)$ is the $y$th order associated Laguerre polynomials for some integer $j$. For the Laughlin states, $p{=}1$ since the Hamiltonian theory uses the CF states, and in the CF language, the Laughlin state of electrons at filling $1/ (2s{+}1)$ can be viewed as CFs carrying $2s$ vortices that fill the lowest CF-LL $ (p{=}1)$ completely~\cite{Jain89}. 

The quasiparticle and quasihole energies in the $\mathcal{N}$th LL of $J{-}$LG are calculated by substituting $v_{\mathcal{N}}^{\left (J\right)} (q)$ for $v_n (q)$ in Eqs.~\eqref{eq: quasiparticle energy} and \eqref{quasihole energy}, and are denoted as $\Delta_{+}^{\left (J\right)}\left (\mathcal{N}, s\right)$ and $\Delta_{-}^{\left (J\right)}\left (\mathcal{N}, s\right)$, respectively. The computed quasiparticle and quasihole energies of the Laughlin states in the several LLs of BLG and TLG are given in Appendix~\ref{sec: qp_qh_energies}. The cohesive energy of the liquid phase around $\Bar{\nu}{=}1/ (2s{+}1)$ formed by quasiholes at $\Bar{\nu}_{-}$ and quasiparticles at $\Bar{\nu}_{+}$ can be approximated as
\begin{align}
 \label{eq: liquid_energy_away_from_Laughlin_filling}
E_{\rm coh}^{L, \left (J\right)}\left (\Bar{\nu}_{\pm}, \mathcal{N}, s\right) & = E_{\rm coh}^{L, \left (J\right)}\left (\mathcal{N}, s\right)
     +\left[\Bar{\nu}_{\pm} (2s{+}1)\right.\nonumber\\
&\left.-1\right]\Delta_{\pm}^{\left (J\right)}\left (\mathcal{N}, s\right).   
\end{align}
The energy of the liquid phase between two Laughlin fillings is computed by taking the minimum of the liquid phase energy formed by quasiparticles around the lower Laughlin filling and quasiholes around the higher Laughlin filling. We reemphasize that the energy of the liquid phase away from the Laughlin fillings obtained from Eq.~\eqref{eq: liquid_energy_away_from_Laughlin_filling} is an approximate one and used to just compare with the energy of the solid phase. Note that the cohesive energy of the liquid phase, shown by black dotted lines in Figs.~\ref{fig: ZLL_phase_diagram}$(b)$,~\ref{fig: ZLL_phase_diagram}$(c)$, and~\ref{fig: Higher_LL_BLG_phase_diagram}$(a)$, does not form a continuous curve. This is due to the very small energy of quasiholes, which prevents the energy of the liquid phase formed by quasiholes of the higher Laughlin filling and that of quasiparticles of the lower Laughlin filling from intersecting with each other. 

\subsection{Exact diagonalization results}
\label{ssec: exact_diagonalization}

  \begin{figure*}[tbh]
        \includegraphics[width=0.99\columnwidth]{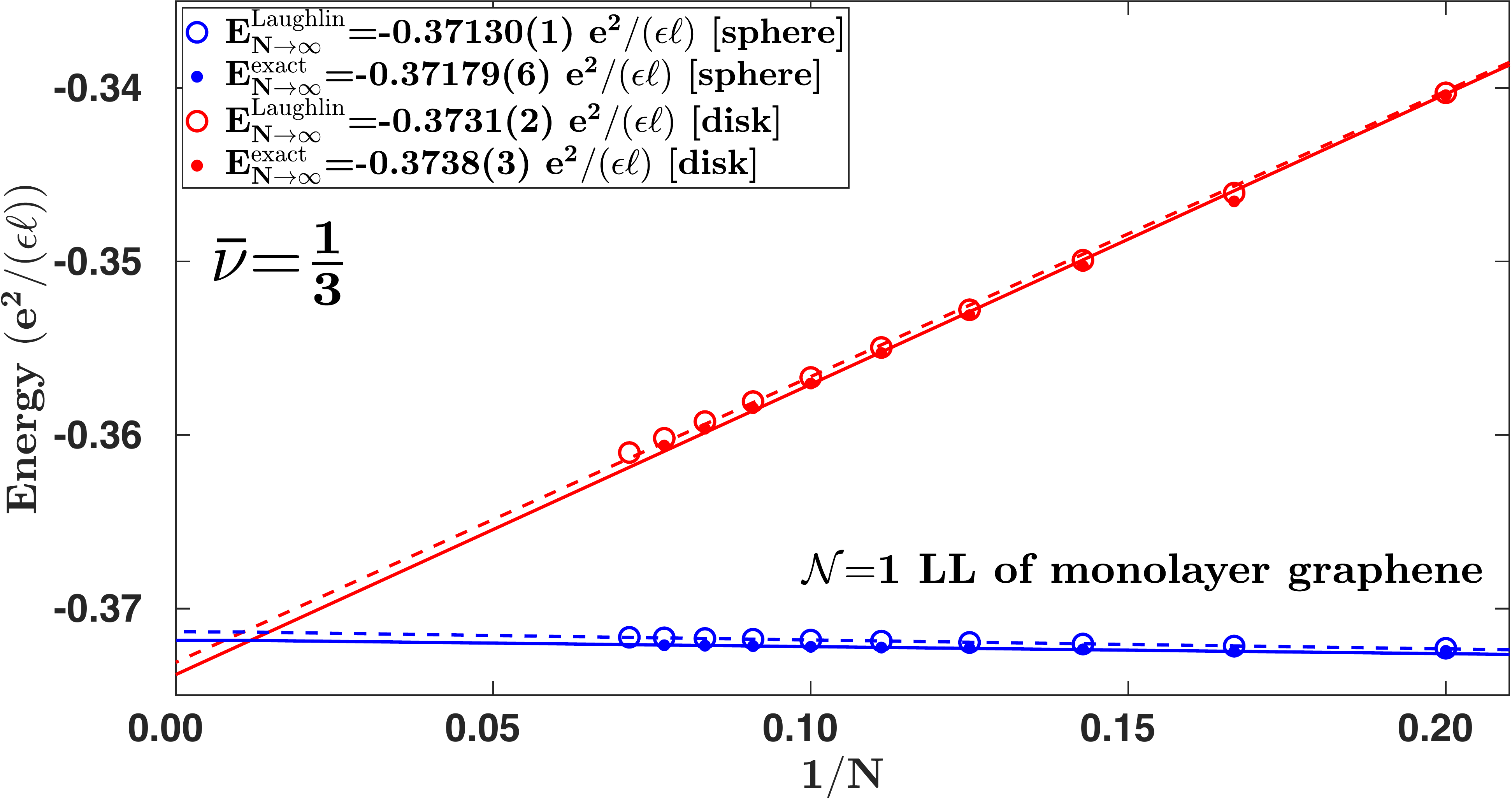}
        \includegraphics[width=0.99\columnwidth]{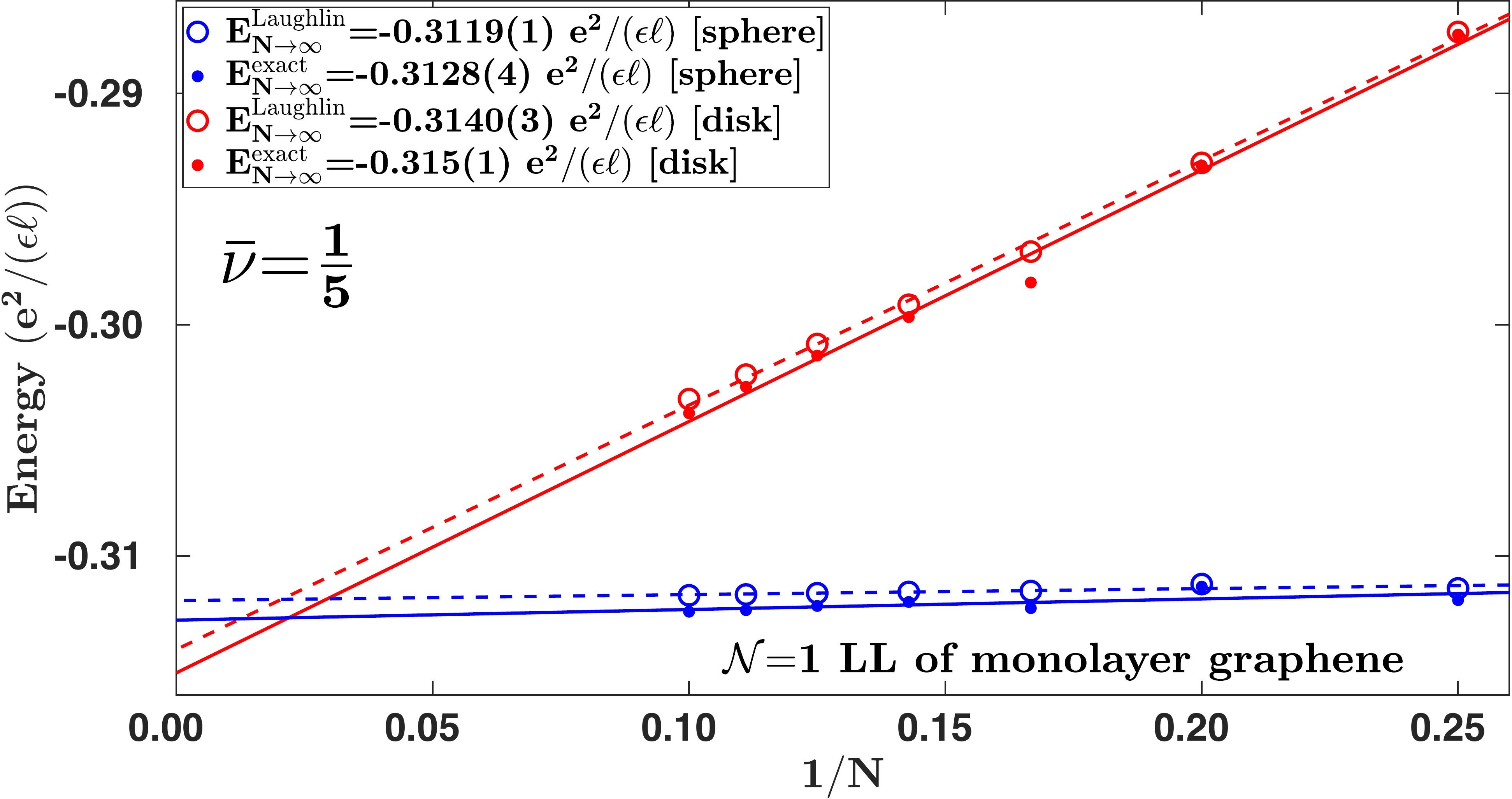} \\
        \includegraphics[width=0.99\columnwidth]{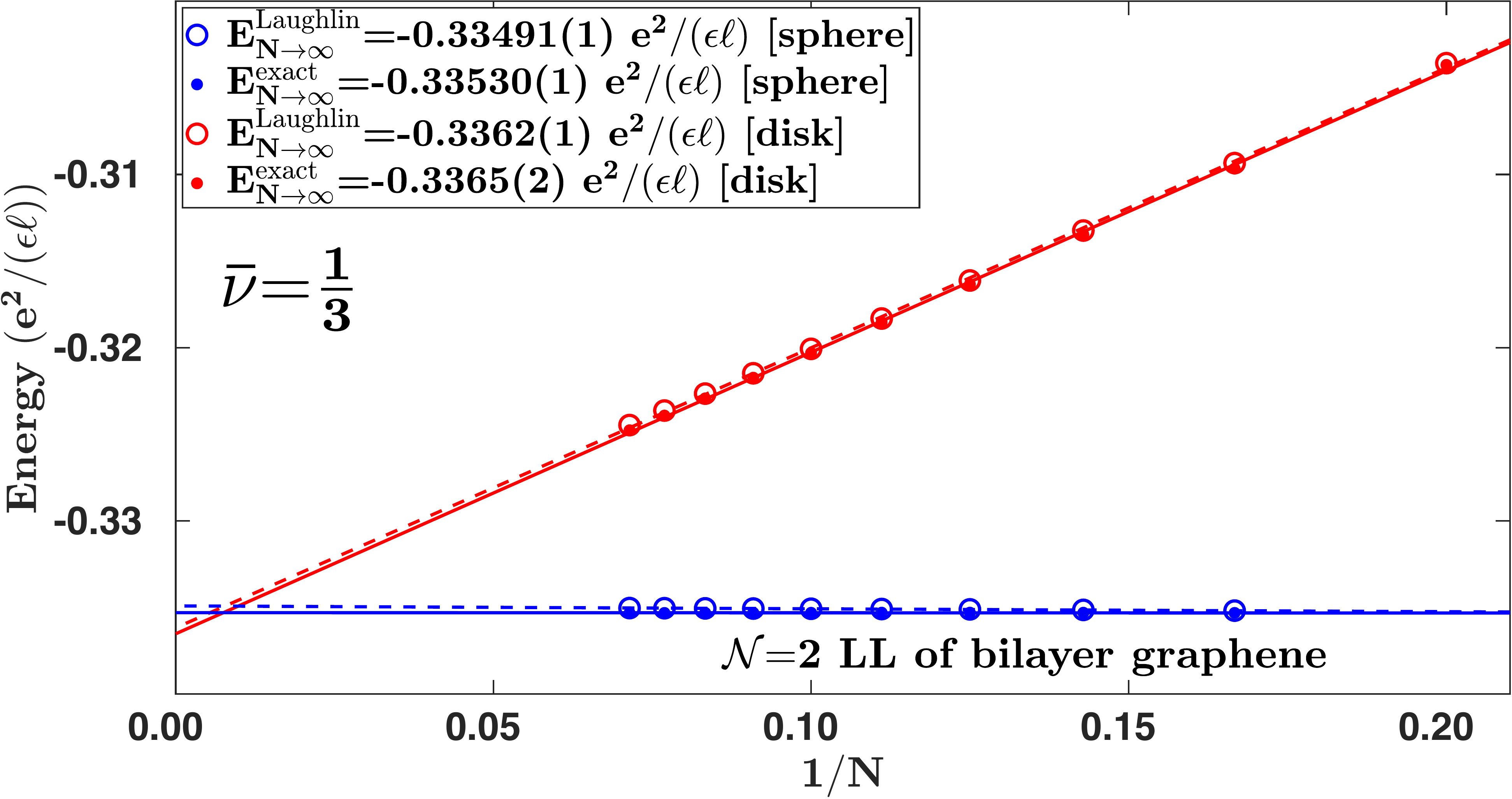}
        \includegraphics[width=0.99\columnwidth]{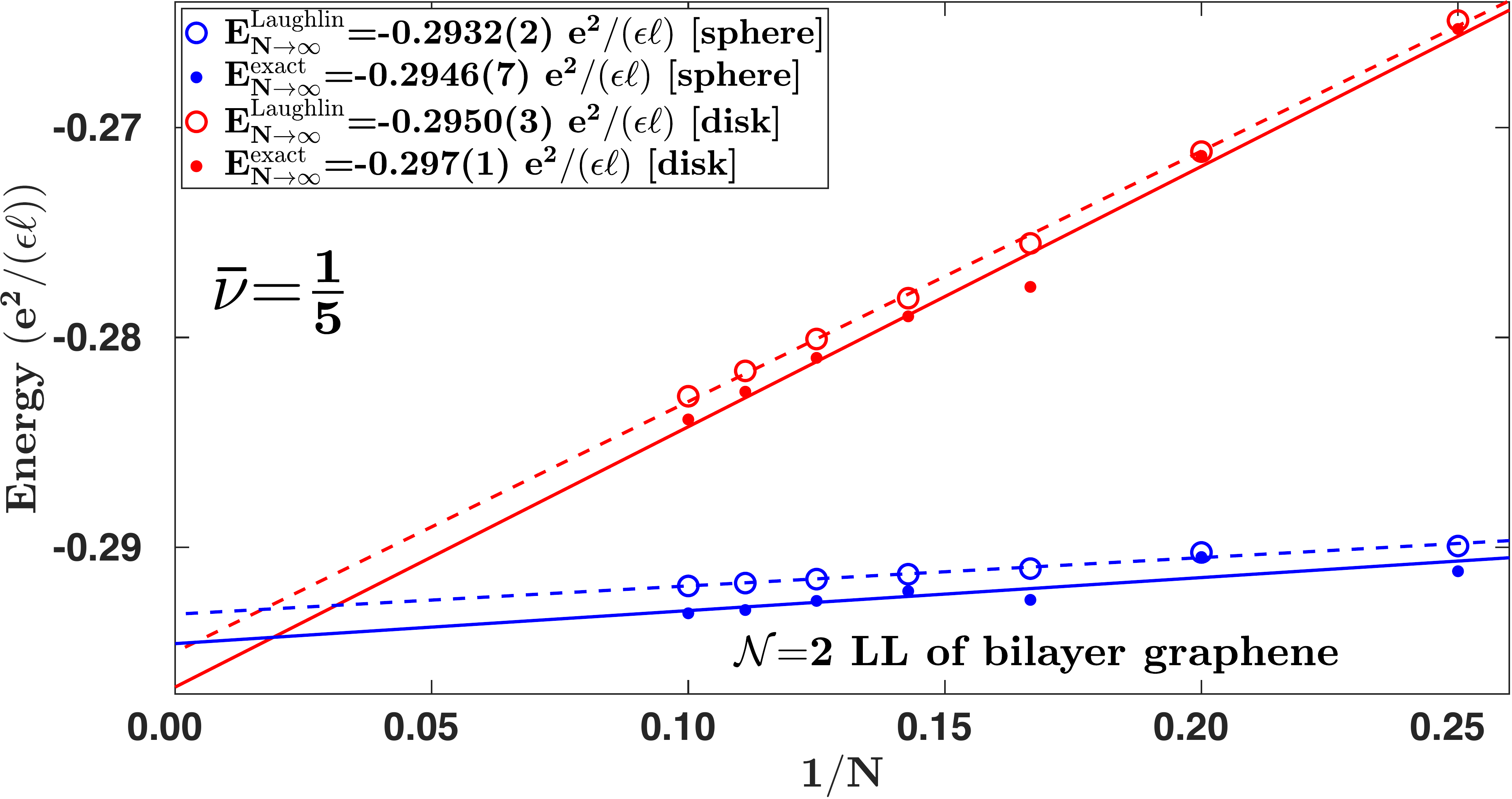} \\
        \includegraphics[width=0.99\columnwidth]{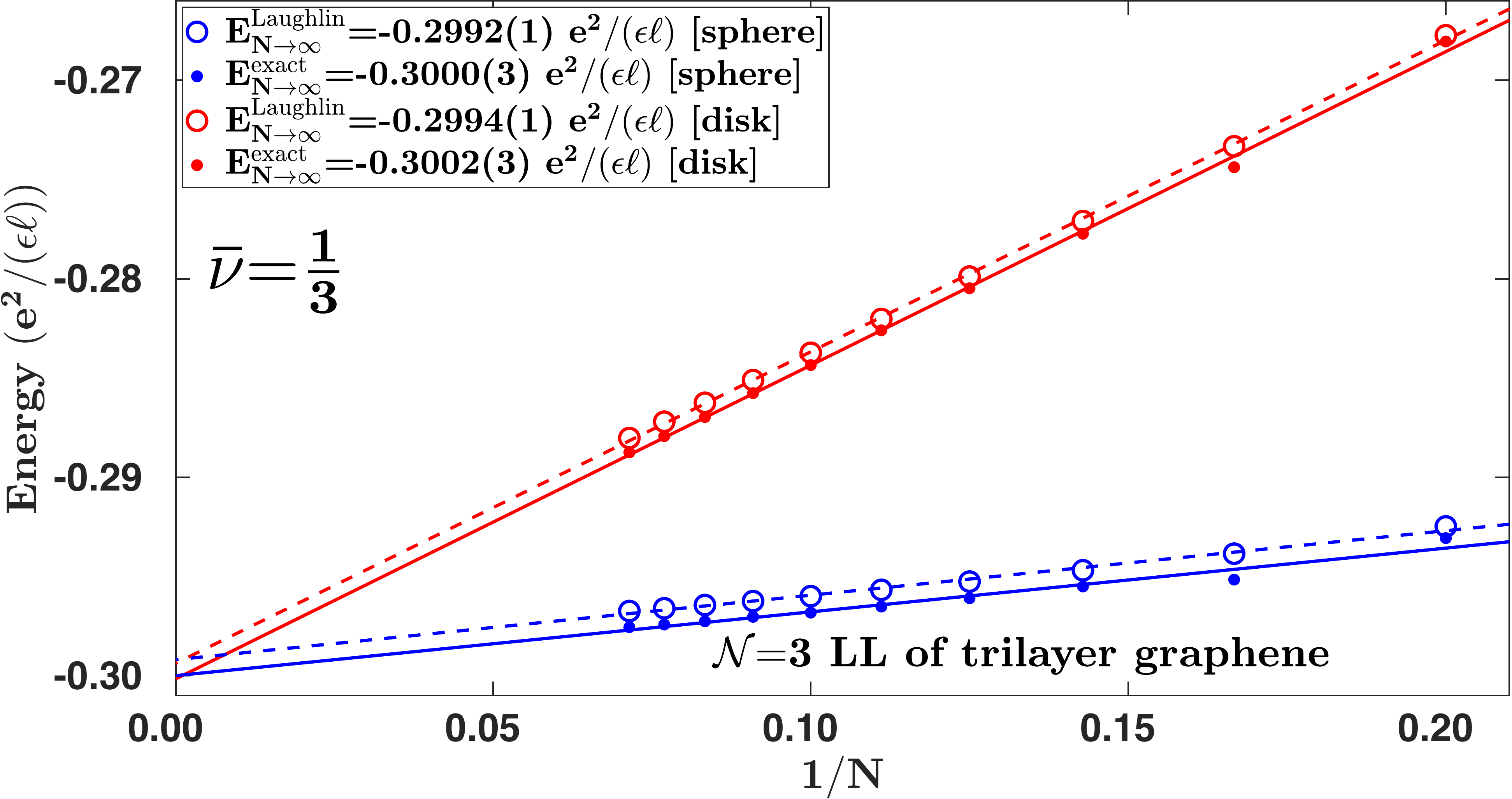}
        \includegraphics[width=0.99\columnwidth]{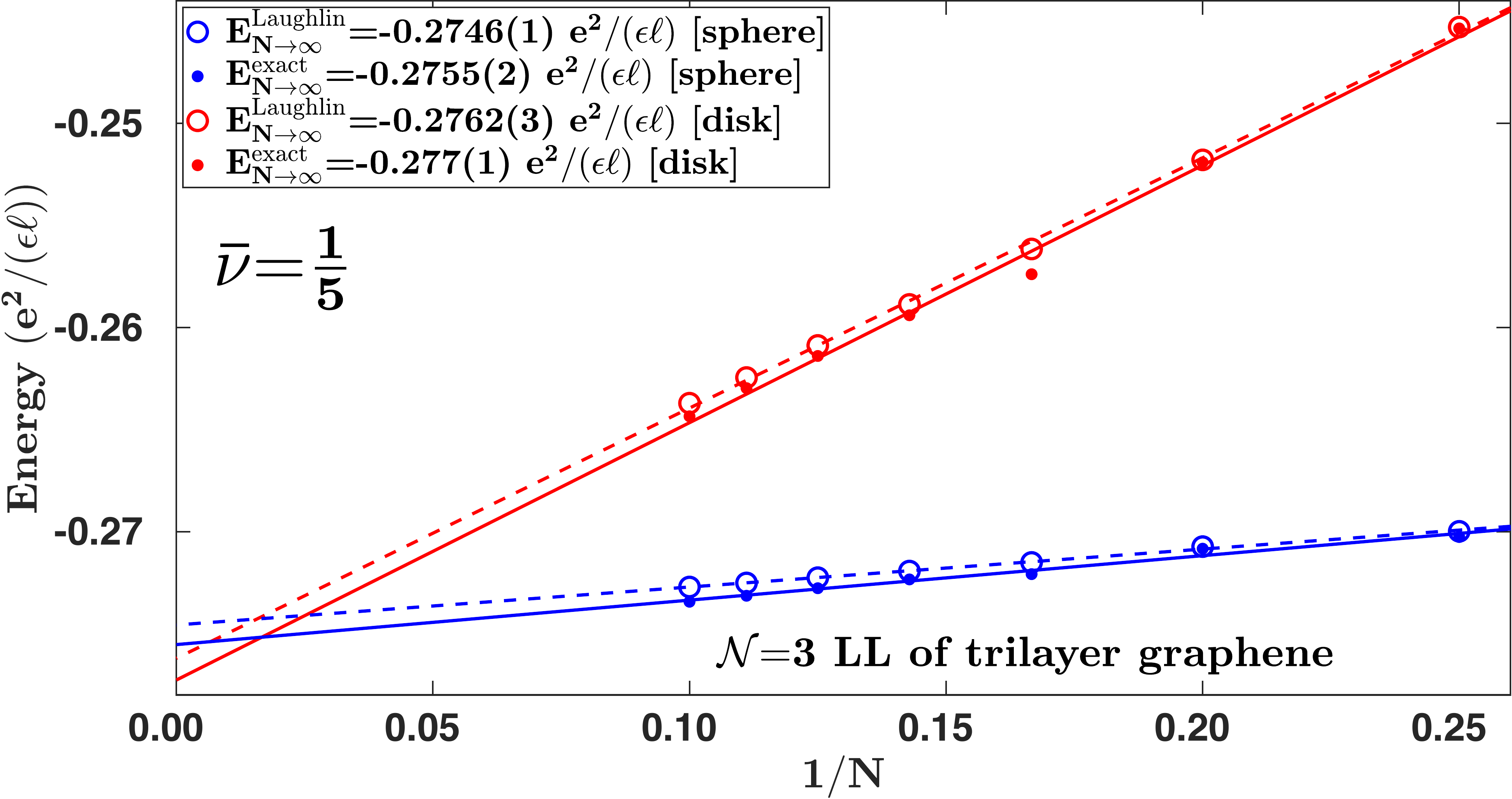} \\
	\caption{Thermodynamic extrapolation of the per-particle density-corrected energies of the $\Bar{\nu}{=}1/3$ and $1/5$ Laughlin and exact Coulomb ground states in the $\mathcal{N}{=}1$ Landau level of MLG, $\mathcal{N}{=}2$ Landau level of BLG, and $\mathcal{N}{=}3$ Landau level of TLG obtained using the spherical and disk pseudopotentials in the spherical geometry. The numbers shown in parentheses are the uncertainty in the linear fit as a function of $1/N$.}
	\label{fig: exact_Laughlin_Coulomb_energies_N_1_MLG_and_N_2_BLG_and_N_3_TLG}
\end{figure*}

In this section, we will discuss results obtained from numerical exact diagonalization, which allows us to check the accuracy of the Laughlin wave function in the LLs of $J{-}$LG. All our numerical calculations are carried out in the spherical geometry~\cite{Haldane83} where $N$ electrons move on the surface of a sphere that is pierced by a radial magnetic flux of strength $2Qhc/e$ that emanates from a magnetic monopole that sits at the center of the sphere. For the LL indexed by $\mathcal{N}$, the shell-angular momentum $l$ is related to the flux as $l{=}Q{+}\mathcal{N}$. On the sphere, the $\Bar{\nu}{=}1/ (2s{+}1)$ Laughlin state occurs when $2l{=}(N{-}1)/\Bar{\nu}$. We shall use the planar disk and spherical pseudopotentials (see Appendixes~\ref{sec: disk_pps_JLG} and \ref{sec: spherical_pps_JLG}) to carry out exact diagonalization at $\Bar{\nu}{=}1/3$ and $1/5$ in the different Landau levels of multilayer graphene. We have also obtained the energy of the Laughlin state (which serves as a variational upper bound) in these Landau levels and overlaps of the Laughlin state with the exact Coulomb ground states. The contribution of the positively charged background, which is required to obtain the per-particle energies, is accounted for following the procedure outlined in the supplemental material of Ref.~\cite{Balram20b}. Before an extrapolation to the thermodynamic limit, the per-particle energies are density-corrected~\cite{Morf86}, i.e., the energies are multiplied by a factor of $\sqrt{2l \Bar{\nu}/N}$, which mitigates the $N$ dependence of the energies. 

In Figs.~\ref{fig: Laughlin_Coulomb_energies_ZLL_BLG} and~\ref{fig: exact_Laughlin_Coulomb_energies_N_1_MLG_and_N_2_BLG_and_N_3_TLG}, numerically computed energies along with their thermodynamic extrapolation of the exact Coulomb ground state and the Laughlin state at $\Bar{\nu}{=}1/3$ and $1/5$ are shown for the $\mathcal{N}{=}1$ ZLL of BLG and the $\mathcal{N}{=}2$ LL of BLG and $\mathcal{N}{=}3$ LL of TLG, respectively. Additionally, in Fig.~\ref{fig: exact_Laughlin_Coulomb_energies_N_1_MLG_and_N_2_BLG_and_N_3_TLG}, we have also presented the exact Coulomb ground-state energies in the $\mathcal{N}{=}1$ LL of MLG (equivalent to the $\mathcal{N}{=}1$ LL of BLG for $\theta{=}\pi/4$) obtained using spherical geometry, and find that these energies are consistent with those obtained using torus geometry in Ref.~\cite{Knoester16}. In Table~\ref{table: comparison_of_exact_analytic_results}, we have compared the energies of the Laughlin liquid with the exact Coulomb ground-state energy in these LLs and find a reasonable agreement between them, which suggests that the Laughlin liquid phase could be stabilized in these LLs. We find an excellent agreement between the analytically and numerically computed energies of the Laughlin liquid, which validates the analytic procedure described in the previous section to determine the energy of the Laughlin state. The overlaps between the Laughlin and exact Coulomb ground state in the several  LLs of BLG and TLG are given in Appendix~\ref{sec: overlaps}. Below we briefly summarize our results.


\begin{figure*}[tbh]
	\includegraphics[width=0.99\columnwidth]{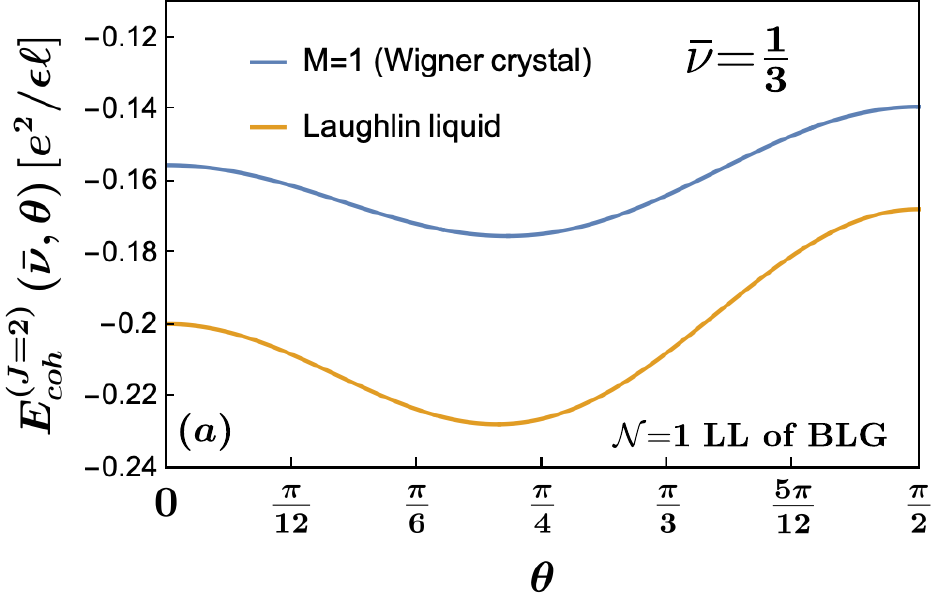}
 \includegraphics[width=0.99\columnwidth]{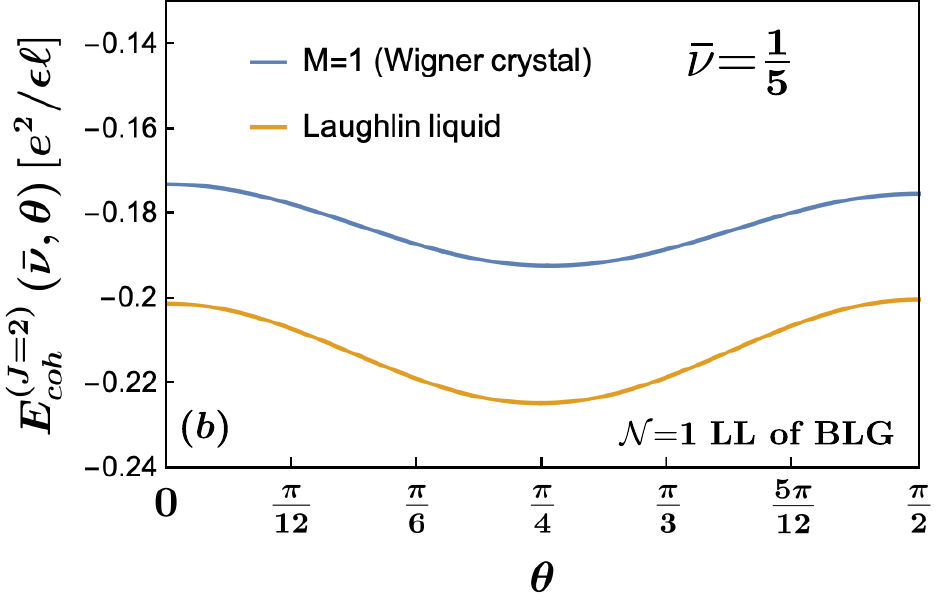}
        \includegraphics[width=0.99\columnwidth]{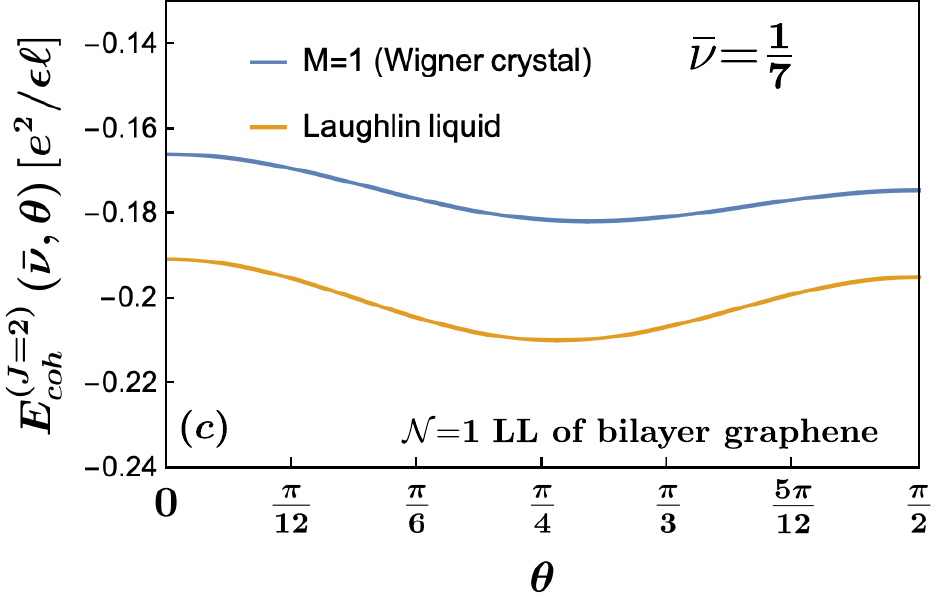}
           \includegraphics[width=0.99\columnwidth]{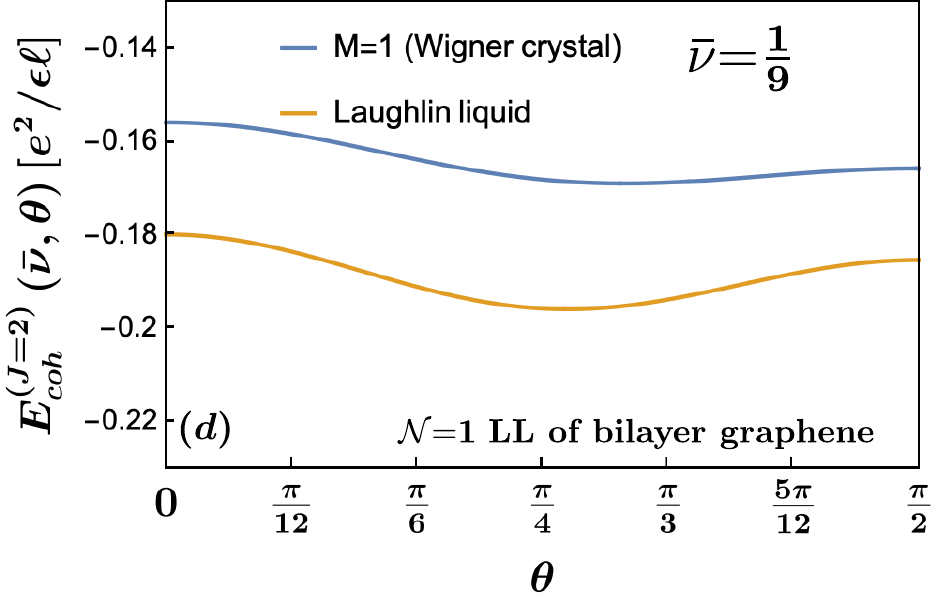} 
	\caption{Cohesive energies of the WC ($M{=}1$) and electron liquid at the Laughlin fillings $(a)~\Bar{\nu}{=}1/3$, $(b)~\Bar{\nu}{=}1/5$, $(c)~\Bar{\nu}{=}1/7$, $(d)~\Bar{\nu}{=}1/9$ as a function of the parameter $\theta$ [see Eq.~\eqref{eq: ZLL in four band model} for definition of the parameter $\theta$] in the $\mathcal{N}{=}1$ zero-energy LL of BLG. Note that the Laughlin state energies in panels $(a)$ and $(b)$ are different from those shown in Fig.~\ref{fig: Laughlin_Coulomb_energies_ZLL_BLG} since in Fig.~\ref{fig: Laughlin_Coulomb_energies_ZLL_BLG} the total energy of the Laughlin states is shown while in this figure only the cohesive part of the total energy is shown.}
	\label{fig: ZLL_phase_Laughlin_filling}
\end{figure*}


The overlaps of the Laughlin state with the exact Coulomb ground state in the LLs of BLG and TLG follow certain rules (determined empirically and tabulated in Table~\ref{table: LLs_with_high_overlaps}) that depend on the weight of the $n{=}0$ or $n{=}1$ LLs of nonrelativistic 2DESs in its form factor [see Eqs.~\eqref{eq: N=0 in four band model} to~\eqref{eq: ZLL in four band model} for zero-energy LLs and Eq.~\eqref{eq: higher LL of BGL} for nonzero energy LLs of BLG; for TLG, see Eqs.~\eqref{eq: ZLL of M-GL} and~\eqref{eq: higher LL of M-GL}]. In the LLs that have at least half of the $n{=}0$ LL wave function, the Laughlin state at $1/3$ filling shows a very high overlap with the exact Coulomb ground state. This is because the Laughlin states build good correlations to efficiently minimize the short-range repulsion that is strongest in the $n{=}0$ LL. Specifically, in the $\mathcal{N}{=}0$ ZLL of BLG, in the $\mathcal{N}{=}1$ ZLL of BLG for $0{\leq}\theta{\leq}\pi/4$, in the $\mathcal{N}{=}2$ LL of BLG and in the $\mathcal{N}{=}0, 3$ LLs of TLG, the Laughlin state at $1/3$ filling has high overlap with the exact Coulomb ground state. In the LLs of BLG and TLG that contain at least half of the $n{=}0$ or $n{=}1$ LLs, the $\Bar{\nu}{=}1/5$ Laughlin state has a good overlap with the exact Coulomb ground state. To be specific, we observe a very high overlap between the Laughlin state and exact Coulomb ground state at $\Bar{\nu}{=}1/5$, in the $\mathcal{N}{=}1$ ZLL of BLG for all $\theta$, in the $\mathcal{N}{=}2, 3$ LLs of BLG, and in the $\mathcal{N}{=}0, 1, 3$ LLs of TLG. As we shall see later, the above rules are also consistent with our studies of the energy of the liquid and solid phases that find the Laughlin liquid state to prevail in the above-mentioned LLs. In the $\mathcal{N}{=}$3 LL of BLG at $\Bar{\nu}{=}1/3$, the solid phase dominates [see Sec.~\ref{ssec:higher_LLs_phase_diagram_BLG}] and exact diagonalization shows the absence of a uniform ground state. Similarly, in the $\mathcal{N}{=}4$ LL of BLG at $\Bar{\nu}{=}1/3$ and $1/5$, in exact diagonalization, we find no uniform ground state or negligibly small overlap of the Laughlin state with the exact Coulomb ground state. Consistently, our studies of the energy of the liquid and solid phases given in Sec.~\ref{ssec:higher_LLs_phase_diagram_BLG} indicate the dominance of the solid phase at these fillings. 

We note one exception to the rules mentioned above. In the $\mathcal{N}{=}4$ LL of TLG, we find reasonably high overlap between the Laughlin and exact Coulomb ground state at $\Bar{\nu}{=}1/3$ but a low number for the corresponding overlap at $\Bar{\nu}{=}1/5$. This is surprising since the $\mathcal{N}{=}4$ LL of TLG contains the $n{=}1$ LL as a component (and no $n{=}0$ LL), which would suggest a high overlap of the $\Bar{\nu}{=}1/5$ Laughlin state with the exact Coulomb ground state. In our study of competing phases in the $\mathcal{N}{=}4$ LL of TLG [see Sec.~\ref{ssec:higher_LLs_phase_diagram_TLG}], too, we find the $1/5$ Laughlin liquid to have the lowest energy among the phases considered. It is plausible that phases like stripes or nematics that we have not considered may have even lower energy than the Laughlin liquid at these fillings in the $\mathcal{N}{=}4$ LL of TLG, which can explain the anomalous overlaps we find.

For completeness, in Table~\ref{table: LLs_with_high_overlaps}, we have also mentioned the LLs of MLG where the Laughlin states have high overlap with the exact Coulomb ground states~\cite{Kusmierz18} [see also Appendix.~\ref{sec: overlaps}]. Surprisingly, even though the Laughlin state at $\Bar{\nu}{=}1/5$ in the $\mathcal{N}{=}1$ LL of MLG has high overlap with the exact Coulomb ground state (see Table~\Romannum{4} of Ref.~\cite{Kusmierz18}), to the best of our knowledge, an FQHE state at this filling has not yet been observed in an experiment~\cite{Lin14, Kim19}.

\begin{table}[tbh!]
  \caption{Landau levels in MLG, BLG, and TLG where the Laughlin states at $\Bar{\nu}{=}1/3$ and $1/5$ have high overlap with the exact Coulomb ground states.}
  \label{table: LLs_with_high_overlaps}
  \begin{tabular}{c*{3}{c}}
\toprule
$\Bar{\nu}$ & MLG & BLG & TLG\\ \midrule
$1/3$ & $\mathcal{N}{=}0, 1$ & ~~$\mathcal{N}{=}0, 1~\big(\text{for}~0\leq\theta\leq \pi/4\big)$,~2 & ~~~$\mathcal{N}{=}0, 3, 4$\\ [2ex]
1/5 & ~$\mathcal{N}{=}0, 1, 2$ & $\mathcal{N}{=}0, 1~\big(\text{for all}~\theta\big),~2, 3$ & ~~$\mathcal{N}{=}0, 1, 3$\\ \bottomrule
\end{tabular}
   \end{table}


\section{Competition between electron solid and electron liquid phases}
\label{sec: results_phase_diagram}
In this section, we discuss the phase diagram in the various LLs of BLG and TLG by comparing the energy of the electron solid phase computed using Eq.~\eqref{eq: solid phase energy} with the energy of the liquid phase that is computed using Eqs.~\eqref{liquid phase energy} and \eqref{eq: liquid_energy_away_from_Laughlin_filling}. Due to the $\Bar{\nu}{\leftrightarrow}1{-}\Bar{\nu}$ particle-hole symmetry in a given LL~\cite{Girvin84}, we limit our discussion to filling fractions in the $\Bar{\nu}{\leq} 1/2$ range. Next, we will discuss the possible phases in the different LLs individually. For ease of access, the calculated phase diagram at the Laughlin filling fractions is summarized in Table~\ref{table: phase_diagram_BLG}. 

\subsection{Phases in the $\mathcal{N}{=}1$ ZLL of BLG}
\label{ssec:ZLL_phase_diagram_results}

\begin{figure}[tbh]
	\includegraphics[width=0.99\columnwidth]{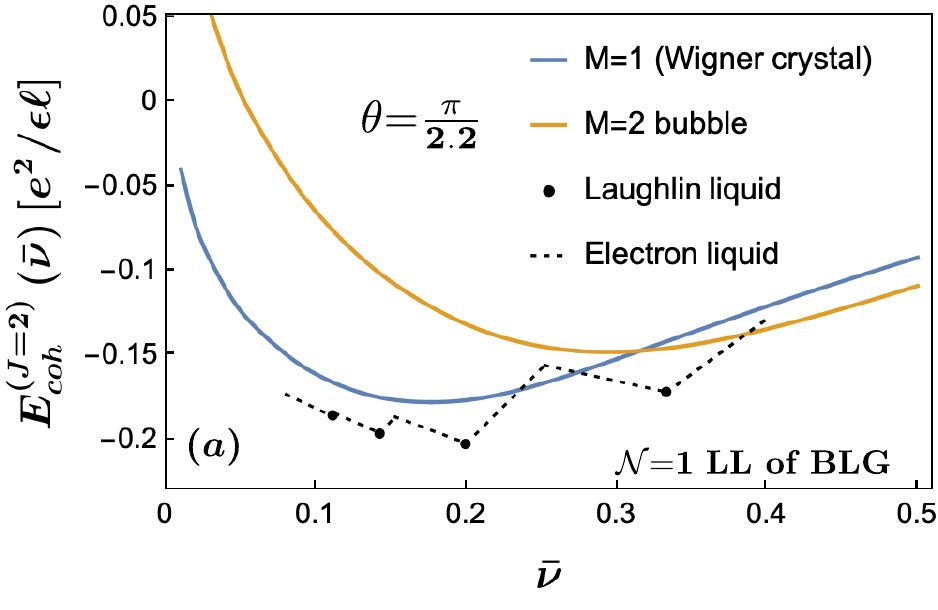}
 \includegraphics[width=0.99\columnwidth]{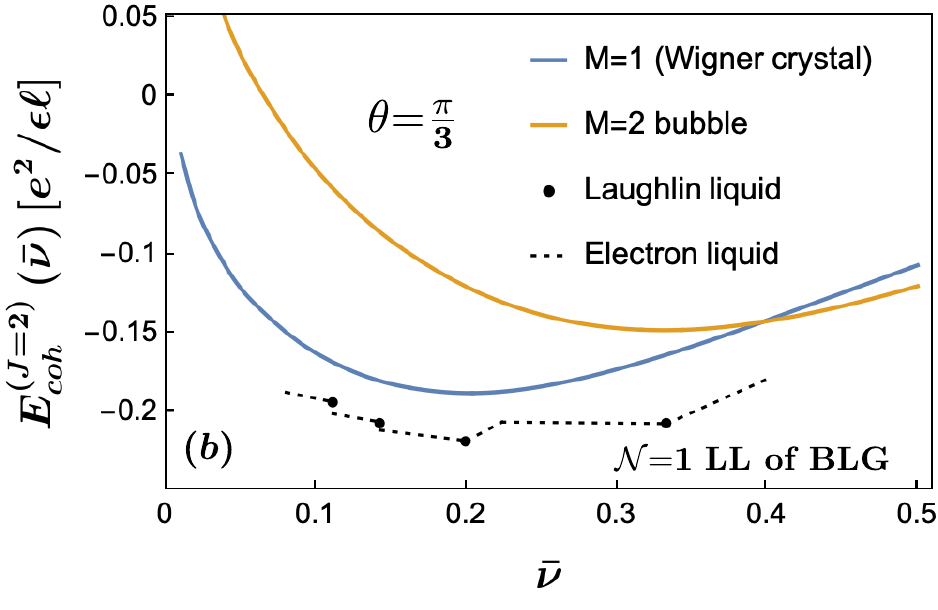}
        \includegraphics[width=0.99\columnwidth]{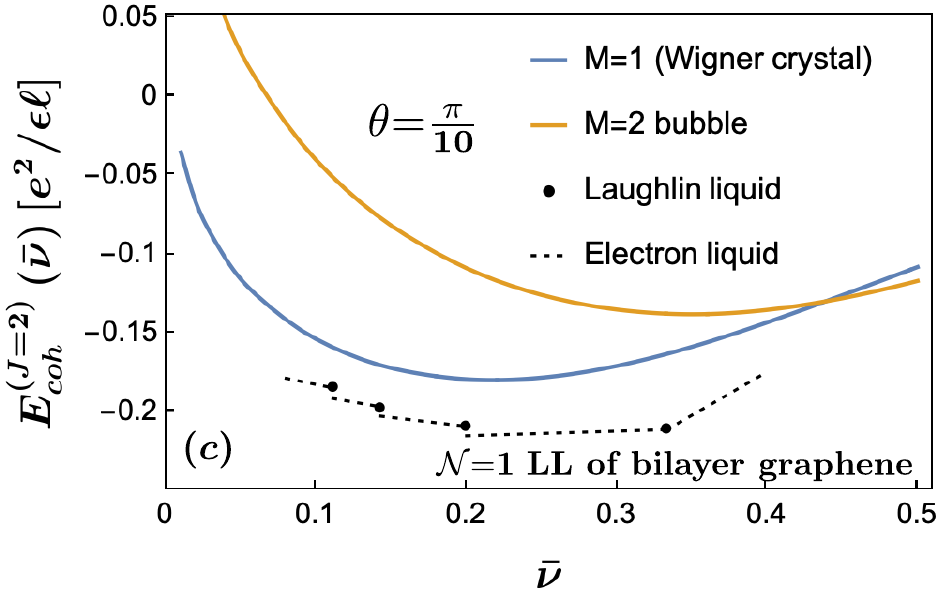}
	\caption{Cohesive energies of the $M{-}$electron bubble and liquid phases in the $\mathcal{N}{=}1$ zero-energy LL of BLG at $\left (a\right)$ $\theta{=}\pi/2.2$, $\left (b\right)$ $\theta{=}\pi/3$, $\left (c\right)$ $\theta{=}\pi/10$ [the parameter $\theta$ is defined in Eq.~\eqref{eq: ZLL in four band model}] as a function of the filling fraction $\Bar{\nu}$.}
	\label{fig: ZLL_phase_diagram}
\end{figure}

 From Fig.~\ref{fig: ZLL_phase_Laughlin_filling} we observe that at $\Bar{\nu}{=}1/ (2s{+}1)$, where $s{=}1,2,3,4$, among the phases we consider, the Laughlin states have the lowest energy in the $\mathcal{N}{=}1$ ZLL of BLG, for all values of the parameter $\theta$. At $\Bar{\nu}{=}1/7$ and $\Bar{\nu}{=}1/9$, for $\theta{=}0$ (LLL), it is known that a Wigner crystal of CFs is in close competition with the liquid~\cite{Zuo20}. At fillings $1/3$ and $1/5$, FQHE has been observed in the $n{=}0$ ($\theta{=}0$) and $n{=}1$ ($\theta{=}\pi/2$) LLs of GaAs~\cite{Tsui82, Willett87}. In the $\mathcal{N}{=}1$ LL of MLG ($\theta{=}\pi/4$), FQHE has been observed only at 1/3 but not at 1/5 yet~\cite{Lin14, Amet15, Kim19}. As one deviates from the Laughlin fillings, the energy of the liquid phase [computed using Eq.~\eqref{eq: liquid_energy_away_from_Laughlin_filling}] increases due to the finite energy of the quasiparticle or quasihole excitations and shows nonmonotonic behavior. Due to this nonmonotonicity, the ground state alternates between the solid and liquid phases for $\pi/2.4{<}\theta{\leq}\pi/2$. In Fig.~\ref{fig: ZLL_phase_diagram}$(a)$, we show the energies of bubble and liquid phases at $\theta{=}\pi/2$ that represent the observed trend in their energies in the region $\pi/2.4{<}\theta{\leq}\pi/2$. At $\Bar{\nu}{<}0.23$, the liquid phases have the lowest energy, while for $0.23{<}\Bar{\nu}{<}0.28$, the WC phase becomes the ground state. The solids quantum melt to the liquid phase around $0.28{<}\Bar{\nu}{<}0.4$. For $\Bar{\nu}{>}0.4$, the two-electron bubble phase appears as the ground state. For all other values of the parameter, i.e., $0{\leq}\theta{<}\pi/2.4$, the liquid state has the lowest energy [a couple of representative phase diagrams are shown in Figs.~\ref{fig: ZLL_phase_diagram}$(b)$ and~\ref{fig: ZLL_phase_diagram}$(c)$]. The reason for the liquid state to prevail in this parameter range is that its energy has a small slope that keeps its energy below that of the solid phases. The phase diagram of the $\mathcal{N}{=}1$ ZLL of BLG in this parameter regime resembles that of the LLL and the $\mathcal{N}{=}1$ LL of MLG~\cite{Knoester16, Goerbig04a}. 

We note that LL mixing (that we have neglected) between the $\mathcal{N}{=}0$ and $\mathcal{N}{=}1$ orbitals in the ZLL of BLG, which can be controlled by the applied electric bias, can stabilize the WC phase at $\Bar{\nu}{=}1/3$ instead of the Laughlin liquid~\cite{Le23}. We have not taken into account the potential presence of other phases, such as stripe phases, a CF Fermi liquid state, or incompressible FQH states that might arise near half-fillings. At $\Bar{\nu}{=}1/2$, in the LLL, the CF Fermi liquid is the ground state while an FQH state occurs in the SLL~\cite{Willett87}. Recent numerical calculations show that at $\Bar{\nu}{=}1/2$, the system transitions from the CF Fermi liquid to a paired FQH state as the parameter $\theta$ is increased from $0$ to $\pi/2$~\cite{Zhu20a, Balram21b}. 


\subsection{Phases in the higher LLs of BLG}
\label{ssec:higher_LLs_phase_diagram_BLG}

\begin{figure}[tbh]
	\includegraphics[width=0.83\columnwidth]{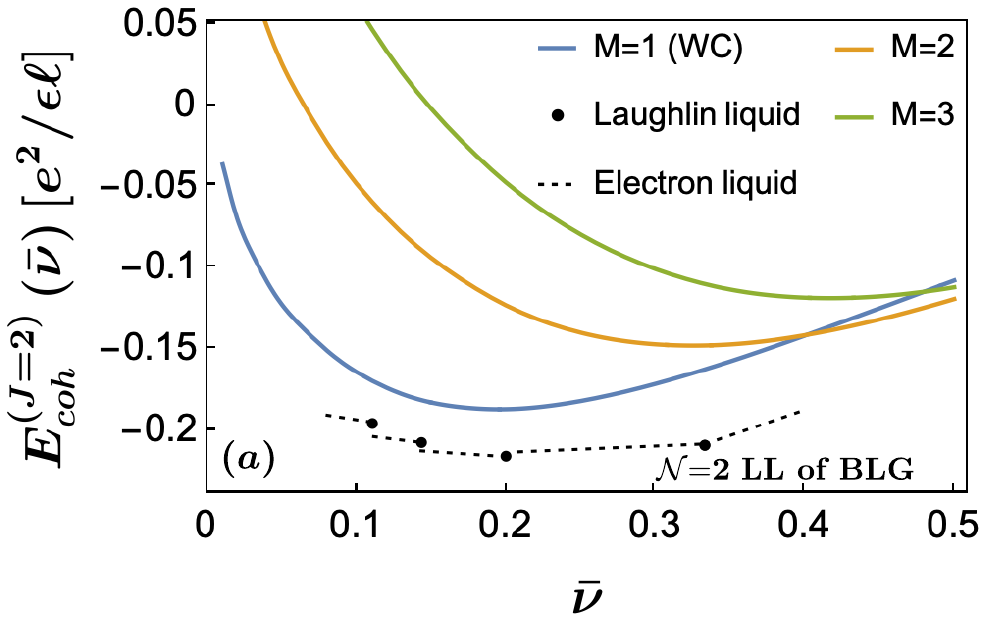}
        \includegraphics[width=0.83\columnwidth]{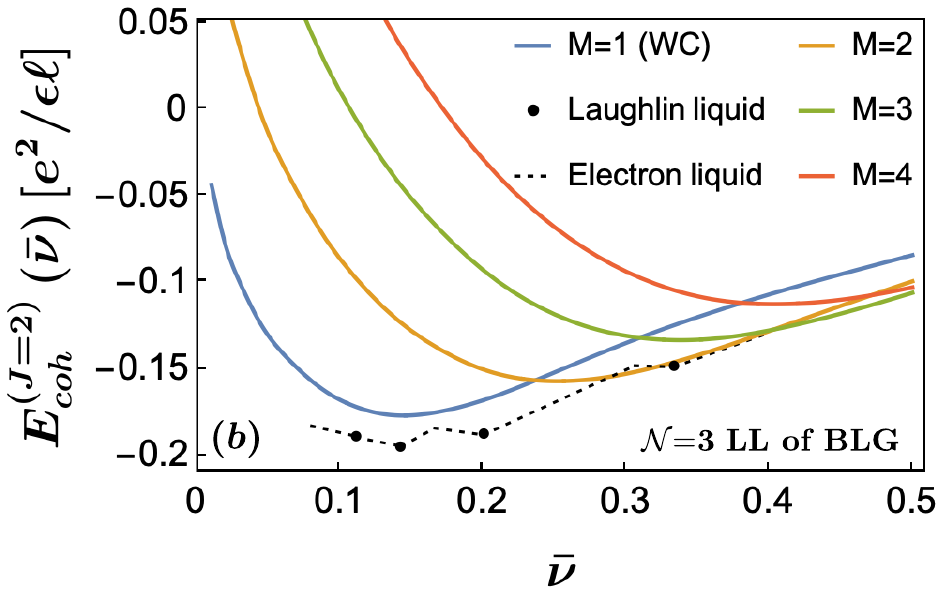} 
        \includegraphics[width=0.83\columnwidth]{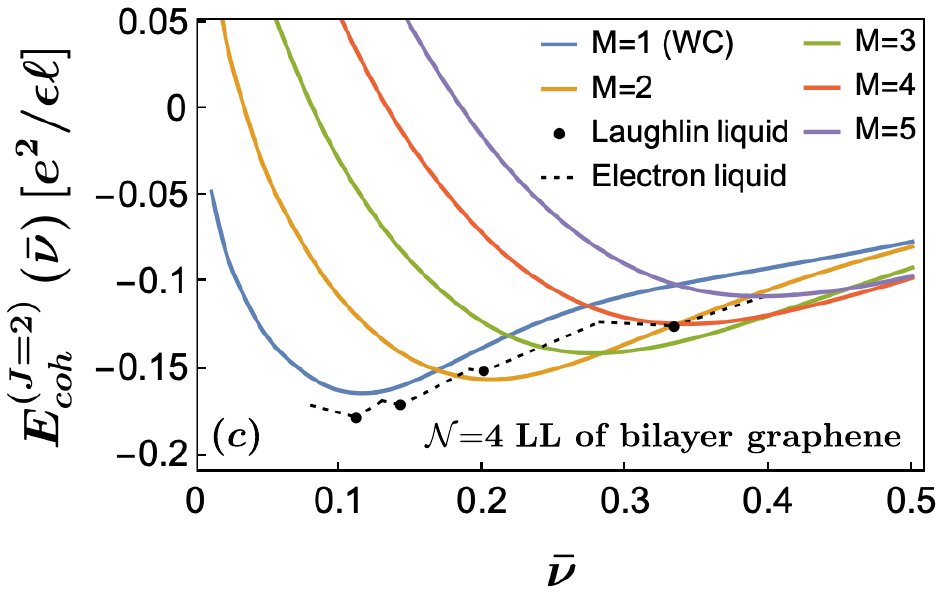}
         \includegraphics[width=0.83\columnwidth]{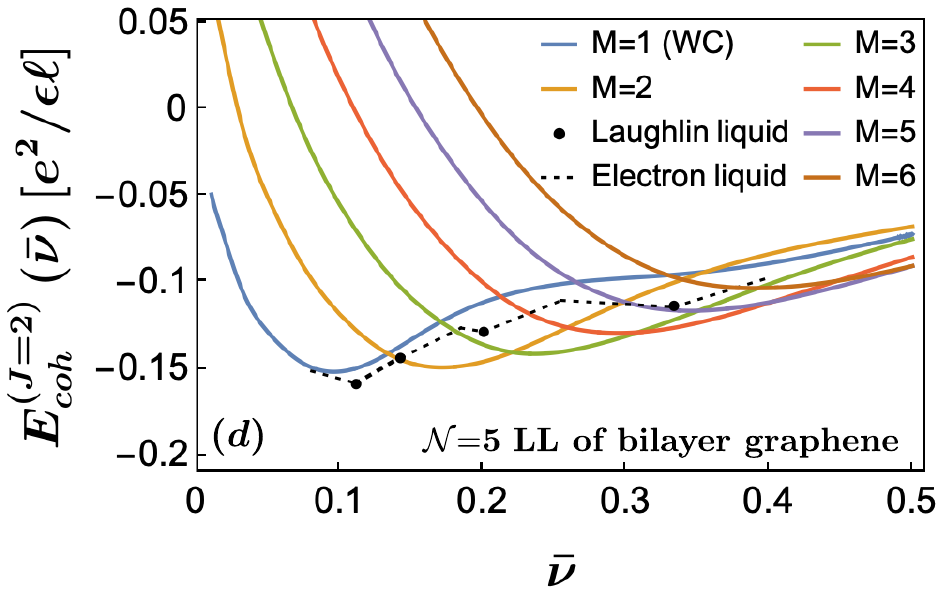}
	\caption{Cohesive energies of the $M{-}$electron bubble and liquid phases in the higher LLs of BLG: $\left (a\right)$ $\mathcal{N}{=}2$ LL, $\left (b\right)$ $\mathcal{N}{=}3$ LL, $\left (c\right)$ $\mathcal{N}{=}4$ LL, $\left (d\right)$ $\mathcal{N}{=}5$ LL.}
	\label{fig: Higher_LL_BLG_phase_diagram}
\end{figure}

In the $\mathcal{N}{=}2$ LL of BLG, we find that the Laughlin states are lower in energy than the solid phases [see Fig.~\ref{fig: Higher_LL_BLG_phase_diagram}$(a)$]. The $1/3$ Laughlin state is more stable than other Laughlin states, as evidenced by the fact that it has the largest energy difference from its closest competitor. Consistent with our findings, in the $\mathcal{N}{=}2$ LL of BLG, FQHE has been observed experimentally at $\Bar{\nu}{=}1/3$, as well as at certain other fillings that correspond to the Jain sequence $\Bar{\nu}{=}p/ (2p{\pm} 1)$~\cite{Diankov16, Hu23}. FQHE has not been observed at $\Bar{\nu}{=}1/5$ and lower fillings. As we discuss in the next section (see Sec.~\ref{sec: effect of impurities}), this may be due to the presence of impurities that can effectively decrease the energy of the solid phase at low fillings, thereby favoring it over the liquid phase. 

\begin{figure}[tbh]
        \includegraphics[width=0.99\columnwidth]{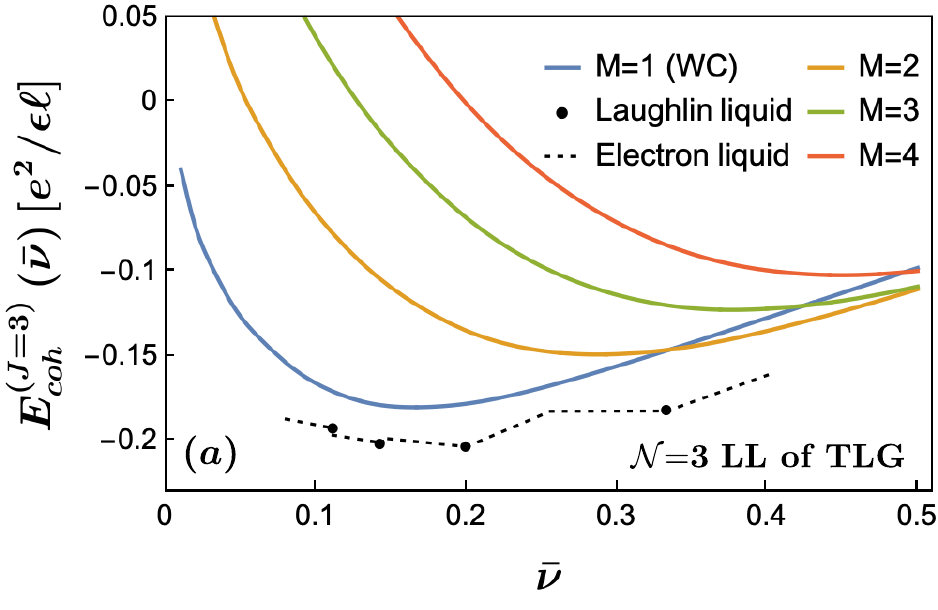} 
        \includegraphics[width=0.99\columnwidth]{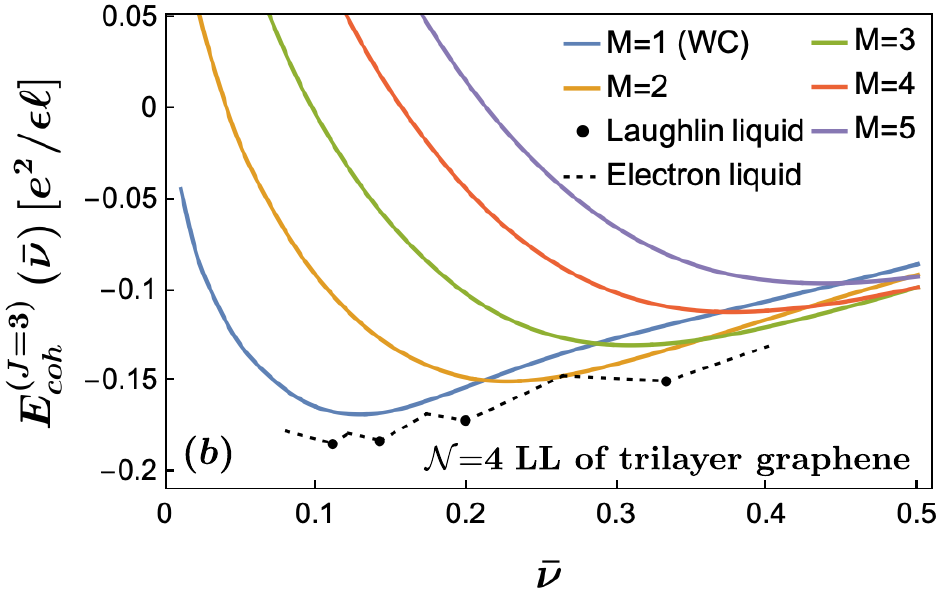}
         \includegraphics[width=0.99\columnwidth]{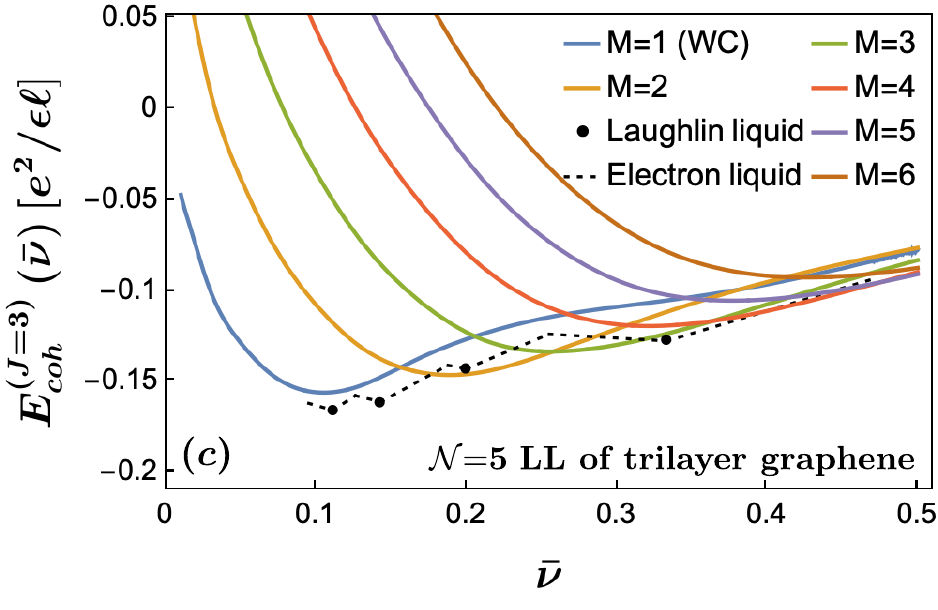}
	\caption{Cohesive energies of the $M{-}$electron bubble and liquid phases in the higher LLs of TLG: $\left (a\right)$ $\mathcal{N}{=}3$ LL, $\left (b\right)$ $\mathcal{N}{=}4$ LL, $\left (c\right)$ $\mathcal{N}{=}5$ LL.}
	\label{fig: Higher_LL_TLG_phase_diagram}
\end{figure}

\begin{figure*}[tb]
	\includegraphics[width=0.99\columnwidth]{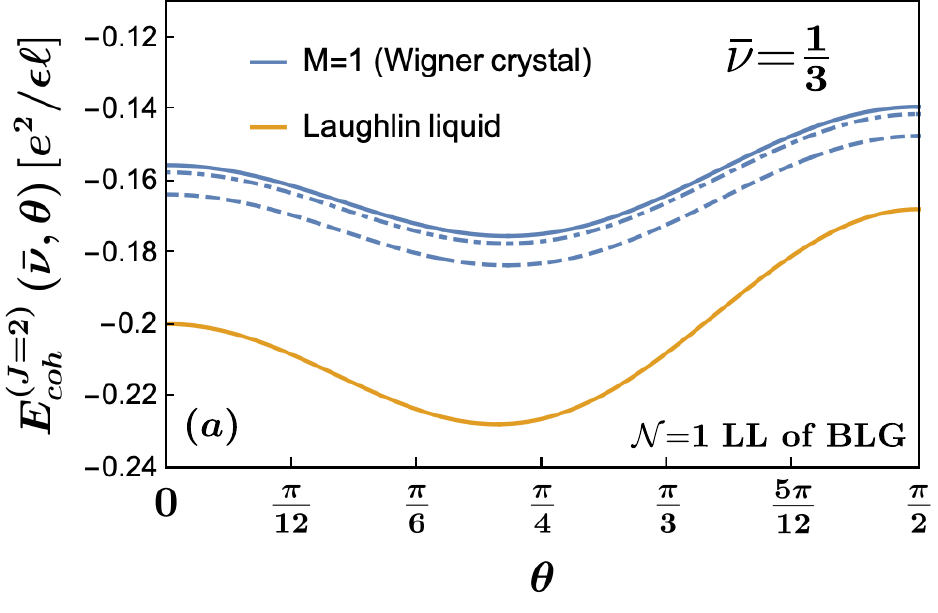}
        \includegraphics[width=0.99\columnwidth]{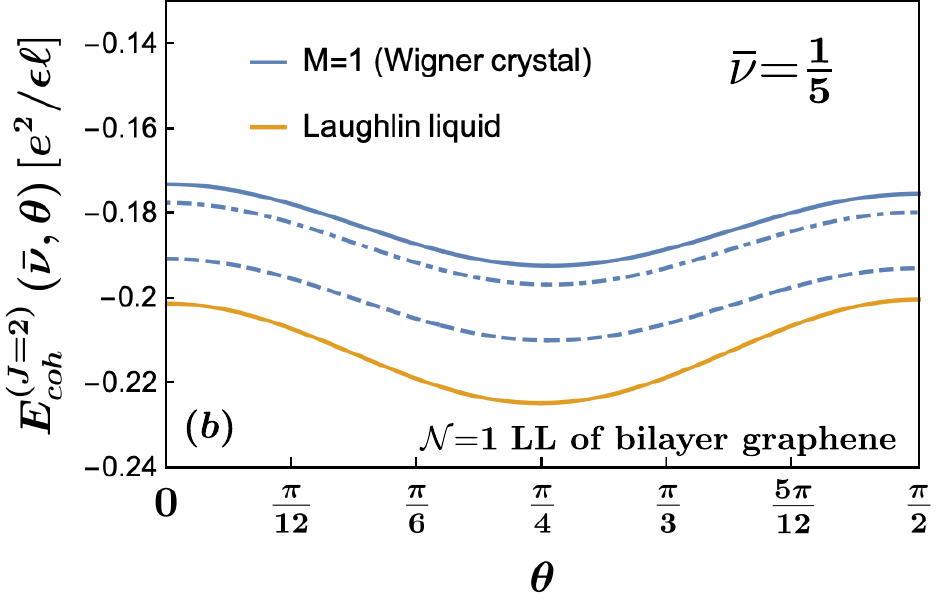}
        \includegraphics[width=0.99\columnwidth]{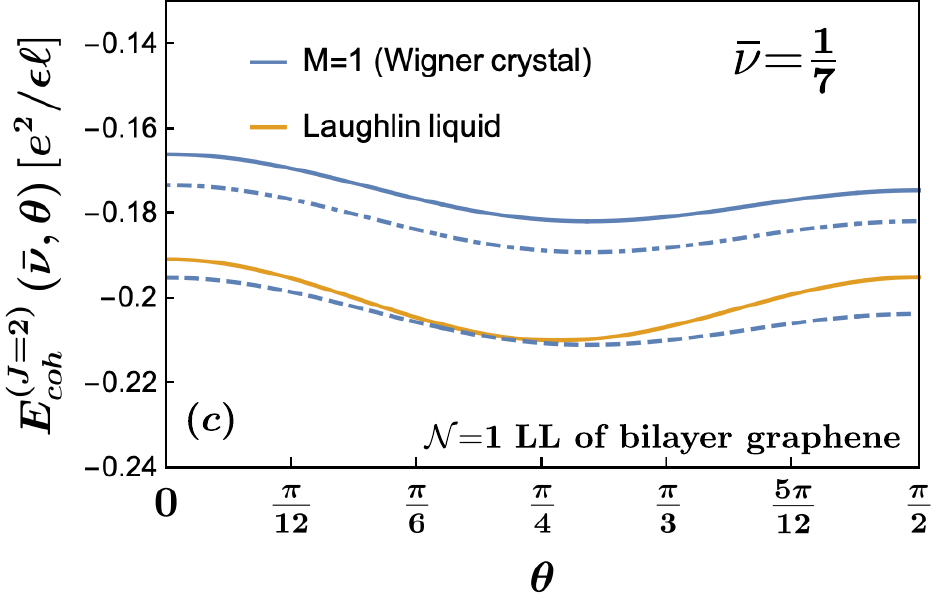}
        \includegraphics[width=0.99\columnwidth]{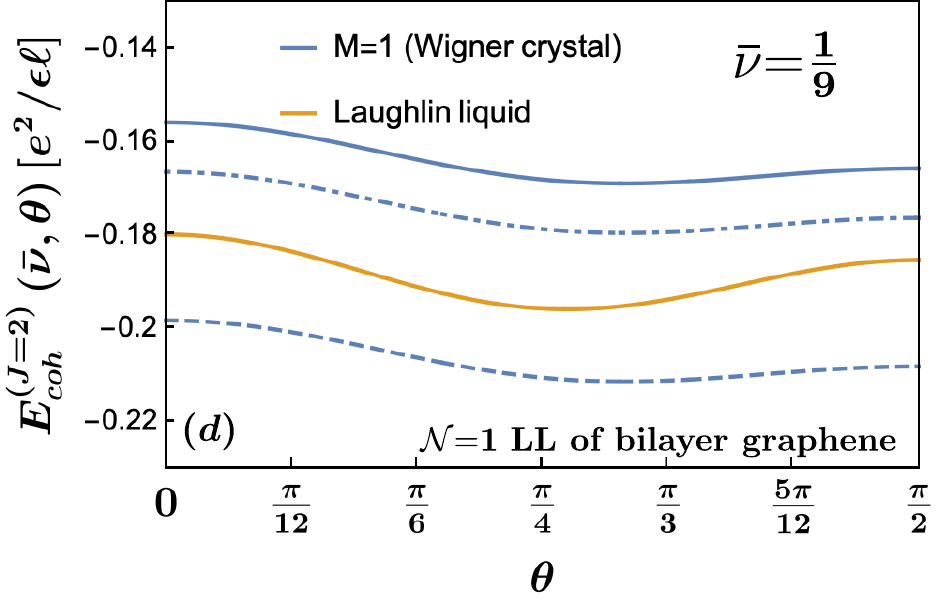}
	\caption{Cohesive energies of the WC ($M{=}1$) and electron liquid at the Laughlin fillings as a function of parameter $\theta$ [defined in Eq.~\eqref{eq: ZLL in four band model}], in the presence of impurities in the $\mathcal{N}{=}1$ zero energy LL of BLG. The dashed (dash-dotted) lines indicate the energies of the WC phase in an impurity potential with $E_{\rm pin}{=}10^{-4} (E_{\rm pin}{=}2.5\times10^{-5})$. Cohesive energies of the WC at the Laughlin fillings in the absence of impurities are also shown by solid lines for comparison.}
	\label{fig: Impurity_ZLL_phase_Laughlin_filling}
\end{figure*}

In the $\mathcal{N}{=}3$ LL of BLG, as depicted in Fig.~\ref{fig: Higher_LL_BLG_phase_diagram}$(b)$, the energy of the Laughlin state at $\Bar{\nu}{=}1/3$ is comparable to that of the two-electron bubble phase, suggesting that the ground state at this filling is a solid phase. The liquid phase has the lowest energy at other Laughlin fillings. In the range $0.28{<}\Bar{\nu}{<}0.4$, the two-electron bubble phase is the ground state while for $\Bar{\nu}{>}0.4$, the three-electron bubble phase has the lowest energy. In the vicinity of the half-filled LL, the energy difference between the two-,three-, and four-electron bubble phases is very small, which prevents us from definitively determining which among these is the true ground state.

In the $\mathcal{N}{=}4$ LL of BLG, as shown in Fig.~\ref{fig: Higher_LL_BLG_phase_diagram}$(c)$, at $\Bar{\nu}{>}0.19$, multiple bubble phases are energetically more favorable than the liquid phase. This suggests that FQHE is absent at $\Bar{\nu}{=}1/3$ and $1/5$. In the range of fillings $0.19{<}\Bar{\nu}{<}0.28$, the two-electron bubble phase has the lowest energy while the three-electron bubble phase is the ground state for $0.28{<}\Bar{\nu}{<}0.4$. The four-electron bubble phase becomes the ground state for $\Bar{\nu}{>}0.4$. The energy differences between the various bubble phases are quite small for $\Bar{\nu}{>}0.4$. In particular, around the half-fillings, the energy of the five-electron bubble phase is in close competition with that of the four-electron bubble phase. For $\Bar{\nu}{<}0.19$, the solid phases may undergo quantum melting to the liquid phase, leading to the appearance of the Laughlin states at $\Bar{\nu}{=}1/7$ and $1/9$. 

In the $\mathcal{N}{=}5$ LL of BLG, as shown in Fig.~\ref{fig: Higher_LL_BLG_phase_diagram}$(d)$, the solid phases have the lowest energy at the fillings $1/3$, $1/5$, and $1/7$ while the liquid state has the lowest energy at $1/9$. Therefore, predominantly solid phases are stabilized in this LL. In the range $0.13{<}\Bar{\nu}{<}0.22$ and $0.22{<}\Bar{\nu}{<}0.32$, the two- and three-electron bubble phases have the lowest energy, respectively. At higher fillings, the four- and five-electron bubble phases successively become the ground state, but their energy differences remain small.


 \subsection{Phases in the higher LLs of TLG}
 \label{ssec:higher_LLs_phase_diagram_TLG}
 
For the $\mathcal{N}{=}3,4$ LLs of TLG, as shown in Figs.~\ref{fig: Higher_LL_TLG_phase_diagram}$(a)$ and ~\ref{fig: Higher_LL_TLG_phase_diagram}$(b)$, the liquid states have the lowest energy at all fillings. In the $\mathcal{N}{=}4$ LL, the two-electron bubble phase can potentially be stabilized between fillings $\Bar{\nu}{=}1/3$ and $1/5$. In the $\mathcal{N}{=}5$ LL of TLG, as shown in Fig.~\ref{fig: Higher_LL_TLG_phase_diagram}$(c)$, the solid phases dominate at $\Bar{\nu}{=}1/3$ and $1/5$, suggesting the absence of FQHE at these fractions. At $\Bar{\nu}{=}1/7$ and $1/9$, the Laughlin states have the lowest energy. In the range of fillings $0.18{<}\Bar{\nu}{<}0.24$, the two-electron bubble phase has the lowest energy. Similar to the $\mathcal{N}{=}4$ and $\mathcal{N}{=}5$ LL of BLG, in this LL as well, with the increase in fillings, the higher $M{-}$electron bubble phases $ (2{<}M{\leq} 5)$ successively become the lowest energy states, although their relative energy separations remain small. 


\begin{figure*}[tb]
	\includegraphics[width=0.99\columnwidth]{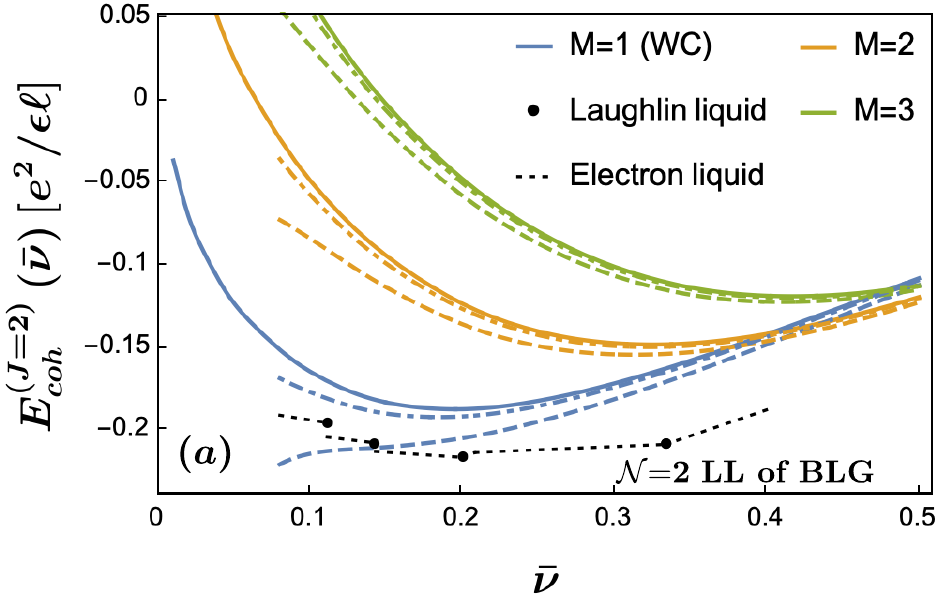}
        \includegraphics[width=0.99\columnwidth]{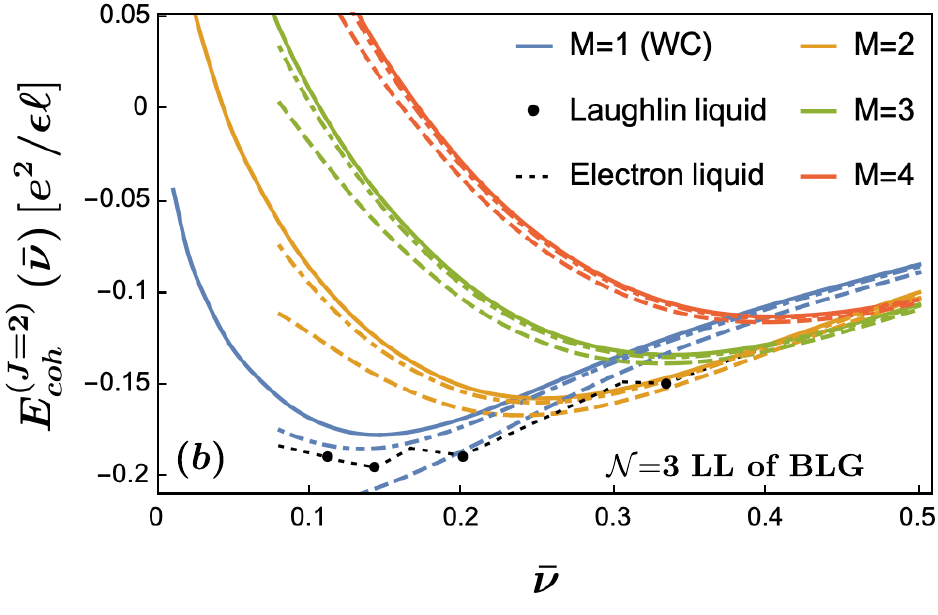} 
        \includegraphics[width=0.99\columnwidth]{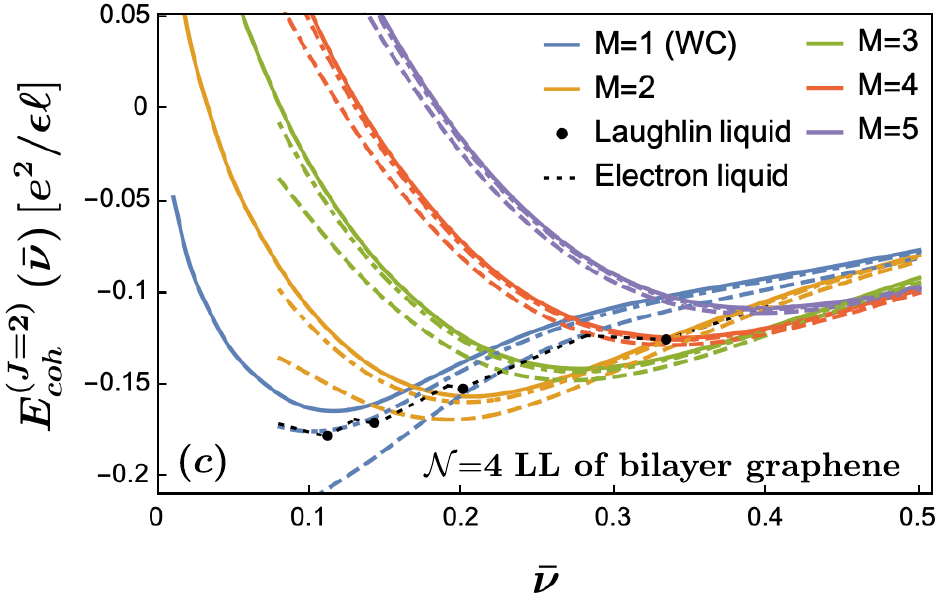}
         \includegraphics[width=0.99\columnwidth]{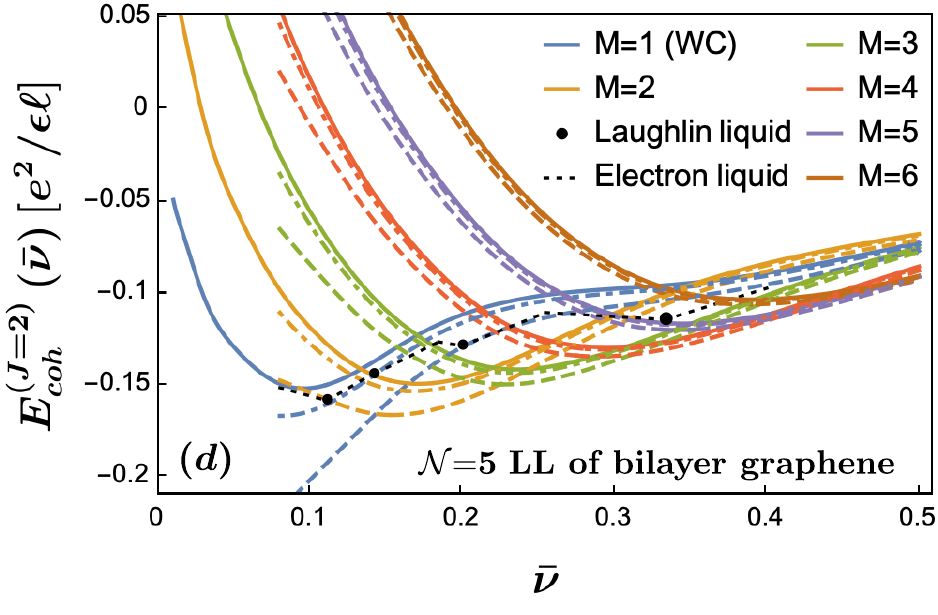}    
          \includegraphics[width=0.99\columnwidth]{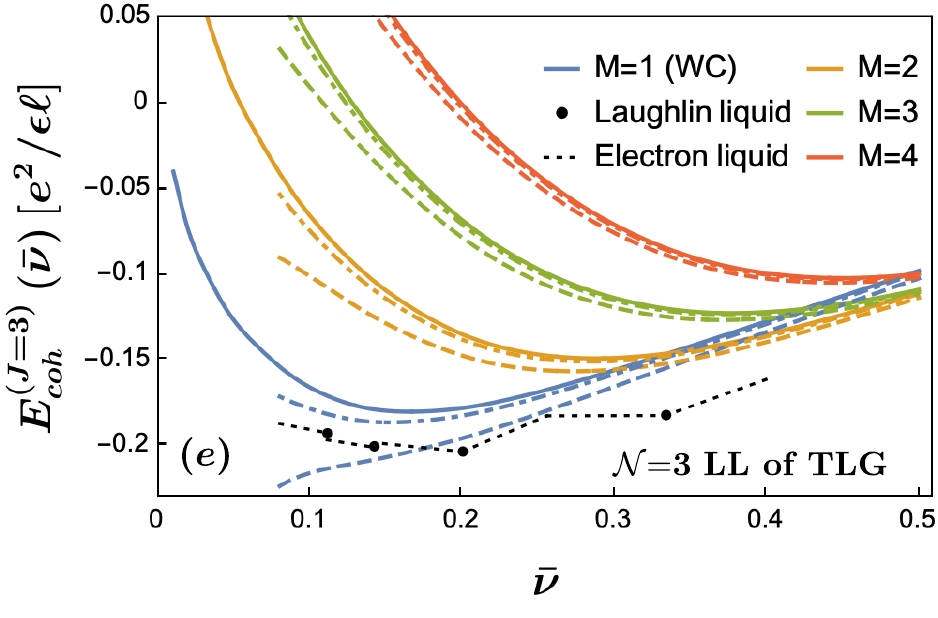} 
        \includegraphics[width=0.99\columnwidth]{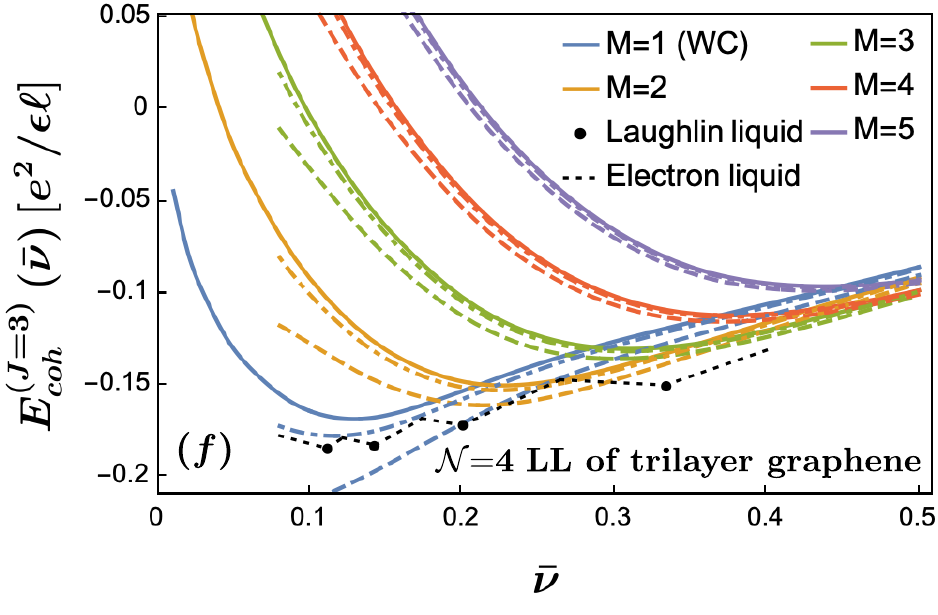}
         \includegraphics[width=0.99\columnwidth]{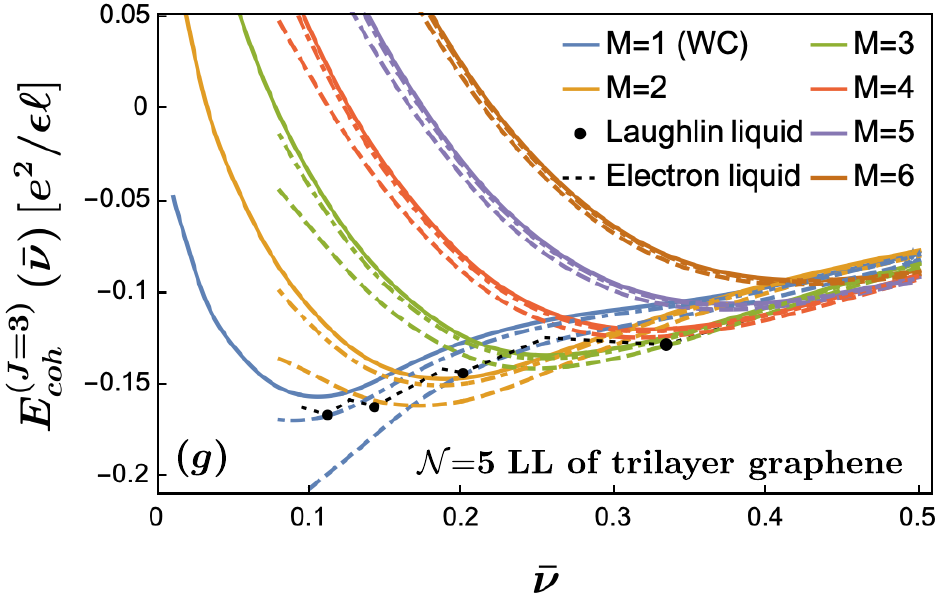}   
	\caption{Cohesive energies of the $M$-electron bubble and liquid phases in many higher LLs of BLG and TLG in the presence of impurities. The dashed (dash-dotted) lines indicate the energies of the $M{-}$electron bubble phases in an impurity potential with pinning strength $E_{\rm pin}{=}10^{-4} (E_{\rm pin}{=}2.5\times10^{-5})$. For comparison, the cohesive energies of the electron solid phases in the absence of impurities are also shown by solid lines.}
	\label{fig: impurity_higher_LL_BLG_TLG_phase_diagram}
\end{figure*}

\section{The effect of impurities on the competition between phases}
\label{sec: effect of impurities}
To compare against experiments, it is important to take into account the effects of impurities on the phase diagram. For weak disorder, the cohesive energy of the incompressible liquid phases changes by a small amount that we can neglect. However, the solid phases can lower their energies by following the landscape of the impurity potential. In the weak-pinning limit, wherein the impurities individually can not deform the electron solid, allowing it to maintain its local ordering, we model the impurity by a short-range Gaussian potential with strength $V_{0}$ and correlation length $\xi$. A collection of these impurities can deform the lattice at a pinning length $L_{0}{\gg}\Lambda$. The energy density associated with the competition between the elastic energy cost for the deformation of the solid and the energy gain resulting from the collective effect of impurities is given by~\cite{Chitra98, Fogler00}:
\begin{equation}
\epsilon (L_{0})= \frac{\mu\xi^{2}}{L_{0}^{2}}- V_{0}\frac{\sqrt{n_{\rm el}}}{L_{0}},
\label{eq: energy_density_impurity}
\end{equation}
where $n_{\rm el}{=}N/A$ is the density of electrons in the uppermost LL; the elasticity constant $\mu$ of the $M{-}$electron bubble is $0.25M^{2}e^{2}n_{M}^{3/2}/\epsilon$, where $n_{M}{=}\Bar{\nu}/ (2\pi M\ell^{2})$ is the density of the corresponding bubble phase~\cite{Goerbig04a}. The reduction in the cohesive energy of the solid phase can be determined by minimizing the energy density given in Eq.~\eqref{eq: energy_density_impurity} with respect to $L_{0}$, and is given by~\cite{Goerbig04a}:
\begin{align*}
\delta E_{\rm coh}^{B} (M, \Bar{\nu})= -\frac{e^{2}}{\epsilon\ell}\frac{ (2\pi)^{3/2}}{\Bar{\nu}^{3/2}\sqrt{M}}E^{2}_{\rm pin},
\end{align*}
where the dimensionless pinning energy $E_{\rm pin}{=} (V_{0}/\xi)/ (e^{2}/\epsilon\ell^{2})$. The energy of the $M{-}$electron bubble phase in $J{-}$LG [see Eq.~\eqref{eq: solid phase energy}] in the presence of impurities is modified as
\begin{align*}
   E_{\rm coh}^{B, \left (J\right)} (M, \Bar{\nu}) \rightarrow E_{\rm coh}^{B, \left (J\right)} (M, \Bar{\nu}) +\delta E_{\rm coh}^{B} (M, \Bar{\nu}).
\end{align*}
We have reevaluated the cohesive energy of the solid phases in the presence of impurities (the strength of the impurity is in the range given in Ref.~\cite{Goerbig04a}), and the results are presented in Figs.~\ref{fig: Impurity_ZLL_phase_Laughlin_filling} and~\ref{fig: impurity_higher_LL_BLG_TLG_phase_diagram}, and the modified phase diagram in the presence of impurities is summarized in Table~\ref{table: phase_diagram_BLG}. 

In the $\mathcal{N}{=}1$ ZLL of BLG, the FQH states at $\Bar{\nu}{=}1/3$ and $1/5$ survive even in the presence of impurities since their energies are well separated from that of the WC phase [see Figs.~\ref{fig: Impurity_ZLL_phase_Laughlin_filling}$(a)$ and~\ref{fig: Impurity_ZLL_phase_Laughlin_filling}$(b)$], across the entire range of the parameter $\theta$. In the presence of strong impurity potentials, the WC phase dominates over the FQH states at $\Bar{\nu}{=}1/7$ and $1/9$, for all values of $\theta$ as indicated by the dashed lines in Figs.~\ref{fig: Impurity_ZLL_phase_Laughlin_filling}$(c)$ and~\ref{fig: Impurity_ZLL_phase_Laughlin_filling}$(d)$. Thus, our results suggest that the FQHE state at these fillings likely arises only in pristine samples. 

In the presence of impurities, the Laughlin state at $\Bar{\nu}{=}1/3$ in the $\mathcal{N}{=}2$ LL of BLG shown in Fig.~\ref{fig: impurity_higher_LL_BLG_TLG_phase_diagram}$(a)$ has significantly lower energy than those of the solid phases, indicating that it is the ground state consistent with its experimental observation~\cite{Diankov16}. Strong impurity potentials [see dashed lines in Fig.~\ref{fig: impurity_higher_LL_BLG_TLG_phase_diagram}$(a)$] may wipe out the FQH states at $\Bar{\nu}{=}1/7$ and $1/9$. However, the FQH state at $\Bar{\nu}{=}1/5$ possibly remains stable. At $\Bar{\nu}{<}0.2$, the WC phase becomes the ground state. In the $\mathcal{N}{=}3$ LL of BLG, as depicted in Fig.~\ref{fig: impurity_higher_LL_BLG_TLG_phase_diagram}$(b)$, strong impurity potentials lead to the dominance of the WC phase over the FQH states at $\Bar{\nu}{<}0.22$. Consequently, only solid phases, ranging from the WC phase at low fillings to the three-electron bubble phase at high fillings, prevail in this LL. In the $\mathcal{N}{=}4$ LL of BLG, as shown by the dash-dotted lines in Fig.~\ref{fig: impurity_higher_LL_BLG_TLG_phase_diagram}$(c)$, the presence of very weak impurity potentials can destabilize FQH states at $\Bar{\nu}{=}1/7$ and $1/9$. At $\Bar{\nu}{<}0.18$, the WC phase becomes energetically favorable. As shown in Fig.~\ref{fig: impurity_higher_LL_BLG_TLG_phase_diagram}$(d)$, in the $\mathcal{N}{=}5$ LL of BLG, the impurity potentials lower the energy of the WC phase, which then becomes the ground state at low fillings. For strong impurities [see dashed lines in Fig.~\ref{fig: impurity_higher_LL_BLG_TLG_phase_diagram}$(d)$], the WC phase starts to appear around $\Bar{\nu}{<}0.14$.

In the $\mathcal{N}{=}3$ LL of TLG, we expect to observe the FQH state at $\Bar{\nu}{=}1/3$ even in the presence of strong impurities, as its energy is well separated from that of electron solids [see Fig.~\ref{fig: impurity_higher_LL_BLG_TLG_phase_diagram}$(e)$]. At lower fillings, in the presence of strong impurities, the WC phase dominates over the liquid phase. In particular, strong impurities may suppress the FQH state at $\Bar{\nu}{=}1/5$, as its energy is in close proximity to that of the WC. In the $\mathcal{N}{=}4$ LL of TLG  [see dashed lines in Fig.~\ref{fig: impurity_higher_LL_BLG_TLG_phase_diagram}$(f)$], at $\Bar{\nu}{\leq}0.2$ the presence of a large number of impurities stabilizes the WC. At $\Bar{\nu}{=}1/3$, the Laughlin state remains the ground state since the cohesive energy of the solid phases changes by only a very small amount in the presence of impurities. In the $\mathcal{N}{=}5$ LL of TLG [see dash-dotted lines in Fig.~\ref{fig: impurity_higher_LL_BLG_TLG_phase_diagram}$(g)$], even weak impurities can wipe out the possible FQH states at low fillings, resulting in complete dominance of the solid phases in this LL.


\section{Discussion}
\label{sec: discussion}
We have investigated the competition between the electron solid and liquid phases in the various LLs of bilayer and trilayer graphene. We focused at filling fractions $\Bar{\nu}{=}1/(2s{+}1)$ and studied the competition between the $M{-}$electron bubble phases $ (M{=}1{-}6)$ and the Laughlin liquid. 

In the $\mathcal{N}{=}1$ zero-energy LL of BLG, the electron-electron interaction can be continuously varied from LLL-like (for large magnetic fields or small $\theta$) to SLL-like (for small magnetic fields or $\theta$ around $\pi/2$). In this LL, at the Laughlin fillings $1/3$ and $1/5$, for all magnetic fields, we find that the liquid phase dominates over the solid phases. This behavior is similar to the lowest two LLs of GaAs, where the liquid phase dominates at $1/3$ and $1/5$. In the LLs of BLG and TLG, with increasing LL index, as expected from the form factor, we observe the range of fillings in the vicinity of the Laughlin fractions where the liquid phase is stabilized diminishes while the range of filling fractions where solid phases have the lowest energy expands. Considering only the solid phases, we observe that in a given LL, as the filling fraction increases, the number of electrons per site in the lowest energy bubble crystal increases to minimize the Coulomb interaction energy.  

Due to the presence of the WC phase between the FQH states at $\Bar{\nu}{=}1/3$ and $1/5$ in the $\mathcal{N}{=}1$ ZLL of BLG for $\pi/2.4 {<}\theta{\leq} \pi/2$, one expects RIQHE to be observed in this parameter regime. In the presence of impurity potentials, the WC phase becomes the ground state for $\Bar{\nu}{\lesssim}1/7$ across all values of the magnetic field, thereby broadening the Hall plateau at the integer fillings~\cite{Goerbig04a}. In the $\mathcal{N}{=}2$ LL of BLG and the $\mathcal{N}{=}3$ LL of TLG, the Laughlin state at $\Bar{\nu}{=}1/3$ survives even in the presence of strong impurities while the WC phase becomes the ground state at low fillings. In the $\mathcal{N}{=}3$ LL of BLG, for weak impurities, together with the FQH state at $\Bar{\nu}{=}1/5$, RIQHE is anticipated due to the dominance of two-electron bubble phase between $0.28{<}\Bar{\nu}{<}0.4$, suppressing the $\Bar{\nu}{=}1/3$ FQH state. Similarly, in the $\mathcal{N}{=}4$ LL of TLG, RIQHE is expected between the FQH states at $\Bar{\nu}{=}1/3$ and $1/5$ due to the presence of energetically favored two-electron bubble phase. The $\mathcal{N}{=}4$ and $5$ LLs of BLG, as well as the $\mathcal{N}{=}5$ LL of TLG, are entirely dominated by the bubble phases. This suggests the plateau of the IQH effect extends all the way up to the half-fillings in these LLs. However, this simplified perspective may be incomplete, as there could be other phases, such as the stripes, that we have not considered here, which may have even lower energies than the states we considered. 

Apart from the observations of RIQHE, the bubble phases can also be inferred from longitudinal conductance measurements upon microwave irradiation~\cite{Lewis02, Lewis04, Lewis05, Chen04, Chen06} or from the radio frequency absorption measurements~\cite{Andrei88}. In the low-energy effective description of graphene multilayers, the electron bubble phases reside on the outer layers; thus, the local density of states (LDOS) can be measured in the scanning tunneling microscopy to tell them apart from FQH liquids~\cite{Poplavskyy09, Papic18, Pu22, Coissard22, Farahi23, Hu23}. 

In our calculations, we have focused on the spin-valley polarized LLs, and for the ZLL of BLG, complete orbital polarization is assumed. The experimental filling fraction range of the ZLL of BLG corresponds to $-4{\leq} \nu{\leq}4$. The effective single-particle LL diagram given in~\cite{Li18} suggests that in the presence of a large electric bias, there exists fully spin-valley polarized LL with $\mathcal{N}{=}1$ orbital character in the filling fraction range ${-}2{<}\nu{<}{-}3$. This would be the ideal setting to test our computed phase diagram for the $\mathcal{N}{=}1$ ZLL of BLG. In general, the $\mathcal{N}$th LL of BLG corresponds to the experimental fillings $4\left(\mathcal{N}{-}1\right){<}\nu{\leq} 4\mathcal{N}$ (for $\mathcal{N}{\geq} 2$) and that of TLG corresponds to $2\left (2\mathcal{N}{-}3\right){<}\nu{\leq} 2 \left (2\mathcal{N}{-}1\right)$ (for $\mathcal{N}{\geq} 3$). For instance, in experiments, the electron solid phases in the $\mathcal{N}{=}3$ LL of BLG are expected to occur around $\nu_{0}{+}1/3$, where $\nu_{0}{\in} \left (8, 9, 10, 11\right)$. Methods used in the recent experimental observation of a series of bubble phases in multiple higher LLs of MLG~\cite{Yang23} can be adapted to test our computed phase diagram in LLs of BLG and TLG. Experimentally~\cite{Yang23}, the physics of bubble phases in MLG is entirely governed by the orbital quantum number of a LL and is independent of its spin and valley quantum numbers, which suggests that our assumption of a spin-valley polarized LL could serve as a valid starting point.      

Although we have neglected the effect of LL mixing throughout this work, for the sake of completeness, we will estimate its strength and discuss its effects on the phase diagram. We shall discuss the effect of LL mixing on the two-band model of multilayer graphene. The two-band model remains valid when the energy of the LLs is much smaller than the interlayer coupling $\gamma_1$, i.e., the energy of the LLs is smaller than the energy of higher energy bands. For BLG and TLG for up to $\mathcal{N}{=}$5 LLs, this regime is accessed for $B{<}28$ T and $B{<}13$ T, respectively. The LL mixing $\kappa$ is parametrized by the ratio of the Coulomb energy to the energy difference between the LLs. The Coulomb energy $E_c$ scales as $E_c{=}e^2/(\epsilon\ell){\sim}(55/\epsilon_{r})\sqrt{B}$ meV, where $\epsilon$ is the relative dielectric constant of the surrounding environment, which we take to be $6$~\cite{Ohba01}. In MLG, the LL gap around the charge neutrality point is $\Delta^{(J{=}1)}{=}3.6\sqrt{B}$ meV, which leads to $\kappa{=}0.21$. (Here, we set $\gamma_{0}{=}3$ eV and $\gamma_{1}{=}0.39$ eV~\cite{Zhang08a}.) In contrast to nonrelativistic 2DEGs, the LL mixing parameter in MLG is independent of the magnetic field and can be controlled only by the dielectric constant of the surrounding environment. As the LL index increases, the energy gap between them decreases, and hence the LL mixing increases. Nevertheless, surprisingly, it has been found that the LL mixing does not change the phase diagram of electron solid phases in higher LLs~\cite{Zhang08}. Also, for $\kappa {\lesssim} 2$ in the $\mathcal{N}{=}0$ LL and $\kappa {\lesssim} 1$ in the $\mathcal{N}{=}1$ LL of MLG, the LL mixing has negligible effect on the $\Bar{\nu}{=}1/3$ Laughlin liquid in the respective LLs~\cite{Peterson14}. In BLG, the energy gap between higher LLs remains approximately the same and is given by $\Delta^{(J{=}2)} {\sim} 4.1 B$ meV. The corresponding LL mixing parameter in BLG is $\kappa{=}2.2/\sqrt{B}$, which decreases with increasing magnetic field. On the other hand, in TLG, the energy separation between the LLs increases with increasing LL index and can be approximated as $\Delta^{(J{=}3)}_{\mathcal{N}}{\sim}0.43 B^{3/2}\sqrt{\mathcal{N}{-}1}$ meV. In the $\mathcal{N}$th LL of TLG, $\kappa{=}21.3/(B\sqrt{\mathcal{N}{-}1})$ for $\mathcal{N}{\geq}3$. In BLG and TLG, the value of $\kappa$ for $B{<}28$ T and $B{<}13$ T, respectively, is larger than that of the MLG. A complete treatment of the effects of LL mixing is beyond the scope of the present work. However, it is possible that as in the higher LLs of MLG, the phase diagram of electron solid and liquid phases is not significantly altered by LL mixing. The energy gap between LLs belonging to the ZLL of multilayer graphene is controlled by the applied electric bias~\cite{Apalkov11} and remains primarily independent of the magnetic field. When the energy of these LLs is close to each other, LL mixing is important and can significantly alter the phase diagram. As we noted earlier, recently, it has been shown that in the ZLL of BLG, a WC gets stabilized at $\Bar{\nu}{=}1/3$ instead of the Laughlin liquid when LL mixing is taken into account~\cite{Le23}. Our results apply only when the LLs in the ZLL of multilayer graphene are sufficiently far from each other.

It would be worth generalizing our results to fractions beyond the Laughlin ones considered here. The FQH state at many of these fractions can be well represented by trial wave functions constructed using the CF~\cite{Jain89} or parton theory~\cite{Jain89b, Wu17, Faugno20a, Dora22, Sharma22, Sharma23}, which can be used to get an accurate estimate of the energy of the liquid state. We note that the inclusion of a trigonal warping term~\cite{Khanna23} (which has been ignored in our calculations) in the four-band Hamiltonian of BLG can potentially stabilize non-Abelian parton states~\cite{Wen91, Wu17, Timmel23} in its ZLL. Additionally, consideration of other phases, such as stripes, nematics, CF Fermi liquid, etc., would provide a more comprehensive phase diagram than the one we have obtained. To make a more thorough comparison with experiments, it would also be important to take into account the effects of LL mixing and screening by gates that we have ignored here. We leave these interesting directions for future exploration.

\section{Acknowledgements}
We acknowledge useful discussions with Arkadiusz W\'ojs. The numerical computations reported in this work were carried out on the Nandadevi supercomputer, which is maintained and supported by the Institute of Mathematical Science's High-Performance Computing Center. Some of the numerical calculations were performed using the DiagHam libraries~\cite{DiagHam}, for which we are grateful to its authors.

\appendix

\section{Dirac equation in planar $J{-}$layer graphene}
\label{sec: Dirac_eqn_J_layer}
The single-particle Hamiltonian for two-dimensional chiral fermions with a chirality index $J$, around the valley $\mathbf{K}$ is given by~\citep{Barlas12}:
\begin{align}
 H_{\mathbf{K}}^{ (J)} = \lambda_{J}
\begin{pmatrix}
0 & (p^{\dagger})^{J} \\
 (p)^J & 0 \\
\end{pmatrix},
\label{eq: chiral_fermion_Hamiltonian}
\end{align}
where $\lambda_{J}$ is a proportionality constant which has dimensions of [energy/(momentum)$^{J}$], $p{=}p_{x}{+}ip_{y}$ is the momentum measured from the Fermi point $\mathbf{K}$ in the Brillouin zone. In the zeroth-order approximation, i.e., neglecting interlayer hopping beyond the adjacent layers, sublattice symmetry breaking, absence of strain, and electric fields, etc., this effective Hamiltonian describes the low-energy physics of \underline{\emph{$ABC{\cdots}$}} stacked $J{-}$layer graphene~\cite{McCann06, Shibata09, McCann12, McCann13, Barlas12, Wu17}. The validity of this approximation and the energy range over which the Hamiltonian of Eq.~\eqref{eq: chiral_fermion_Hamiltonian} applies to $J{-}$LG goes down with the increasing value of $J$. For $J{-}$LG, the proportionality constant $\lambda_{J}{=} {-}\gamma_{1}\big (v_{0}{/}\gamma_{1}\big)^{J}$ as given in the main text [see Eq.~\eqref{eq: two_band_model_of_$J{-}$LG}]. 

In the presence of a uniform magnetic field $(B\hat{z})$ perpendicular to the plane of the 2DES, the canonical momentum $p$ in Eq.~\eqref{eq: chiral_fermion_Hamiltonian} is replaced by the mechanical momentum $\pi{=}p{-}(e/c)\mathcal{A}$ and the Hamiltonian is given by:
\begin{align}
 H_{\mathbf{K}}^{ (J)}=\lambda_{J}
\begin{pmatrix}
0 & (\pi^{\dagger})^{J} \\
\pi^{J} & 0 \\
\end{pmatrix},
\label{eq: Hamiltonian_JLG}
\end{align}
where $\mathcal{A}{=}A_{x}{+}iA_{y}$ and $\mathbf{A}$ is the vector potential for which we choose the Landau gauge $\mathbf{A}{=} (0, Bx, 0)$. The operators $\pi$ and $\pi^{\dagger}$ satisfy the commutation relation $[\pi,\pi^{\dagger}]{=}2\hbar^{2}/\ell^{2}$~\cite{McCann06}, which shows that they act as lowering and raising operators in the space of LL eigenstates $|\mathcal{N}, X\rangle$ corresponding to a nonrelativistic 2DES [see Eq.~\eqref{eq: Action_of_LL_lowering_raising_operator}]. Using the algebra of these operators, the eigenenergies and normalized eigenstates of $H_{\mathbf{K}}^{ (J)}$ can be worked out and are given by:
\begin{widetext}
\vspace{-\baselineskip}
\begin{align*}
&\left.
\begin{aligned}
\bigl\langle\Phi_{\mathbf{K}, \mathcal{N}{=}0, X}^{ (J)}\bigm|&=\bigl (|0, X\rangle, 0\bigr), \quad\\ 
 \bigl\langle\Phi_{\mathbf{K}, \mathcal{N}{=}1, X}^{ (J)}\bigm|&=\bigl (|1, X\rangle, 0\bigr), \\
&\vdotswithin{=} \\
 \bigl\langle\Phi_{\mathbf{K}, \mathcal{N}{=}J{-}1, X}^{ (J)}\bigm|&=\bigl (|J{-}1, X\rangle, 0\bigr), 
\end{aligned}
\right\}
\begin{aligned}
    &\text{constitute the zero-energy Landau levels:}\\
    &E_{\mathbf{K}, \mathcal{N}=0}=E_{\mathbf{K}, \mathcal{N}=1}= \dots=E_{\mathbf{K}, \mathcal{N}=J-1}=0
\end{aligned}
\\
&\kern\nulldelimiterspace\, 
\begin{aligned}
\bigl\langle\Phi_{\mathbf{K}, \mathcal{N}\geq J, X}^{ (J)}\bigm|= \frac{1}{\sqrt{2}}\bigl (\mp|\mathcal{N}, X\rangle, ~|\mathcal{N}{-}J, X\rangle\bigr),  
\end{aligned}
\\
&\kern\nulldelimiterspace\, 
\begin{aligned}
    E_{\mathbf{K}, \mathcal{N}}= \pm\hbar\omega_{J}\sqrt{ \mathcal{N} (\mathcal{N}{-}1)\dots (\mathcal{N}{-}J+1)},
\end{aligned}
\end{align*}
\end{widetext}
where $\hbar\omega_{J}{=}\gamma_{1}\big (2\hbar v_{0}{/} (\gamma_{1}\ell)\big)^{J}$. The positive eigenenergies in the spectrum correspond to the electron states while the negative eigenenergies correspond to the hole states. The zero-energy manifold has a $4J$-fold degeneracy arising from two spins and two valleys, and the quantum number $J$ denotes the orbital degrees of freedom. The orbital degree of freedom labels the LL index $\mathcal{N}{=}0, 1, 2, {\dots} J{-}1$. For $J{=}1,2,3$, the Hamiltonian in Eq.~\eqref{eq: Hamiltonian_JLG} describes the low energy physics of monolayer, bilayer, and trilayer graphene, respectively, in the presence of a constant perpendicular magnetic field. 


\section{Haldane pseudopotentials in $J{-}$layer graphene}
The spherically symmetric Coulomb interactions between electrons in any given LL indexed by $\mathcal{N}$ are conveniently parameterized using the Haldane pseudopotentials~\cite{Haldane83} $V_{m}$, which is the energy of two electrons in a state of relative angular momentum $m$. These Haldane pseudopotentials in the $\mathcal{N}$th LL in the disk geometry are given by (the magnetic length $\ell$ is set to unity for convenience):
\begin{align}
\begin{aligned}
V^{\mathcal{N}}_{m}=&\int \frac{d^{2}\mathbf{q}}{ (2\pi)^2} \frac{2\pi}{q} \left[F_{\mathcal{N}} (\mathbf{q})\right]^{2} e^{-q^2{/2}} ~L_{m} (q^2)\\
=& \int_{0}^{\infty} dq~\left[F_{\mathcal{N}} (q)\right]^{2} e^{-q^2{/2}} ~L_{m} (q^2),
\end{aligned}
\label{eq: Haldane_pps_JLG}
\end{align}
where $L_{m} (x)$ is the Laguerre polynomial, $F_{\mathcal{N}}$ is the so-called form factor, and in going to the last step, we have made use of the spherical symmetry of the interaction whereby the form factor depends only on the magnitude of the planar wave vector $q{=}|\mathbf{q}|$. The form-factor in the $\mathcal{N}$th LL of $J{-}$LG is given by
\begin{equation}
F^{\left (J\right)}_{\mathcal{N}} (q)=e^{-q^2/4}\begin{cases}
	    1 & \mathcal{N}=0 \\
	    L_{1}\Big (\frac{q^2}{2}\Big) & \mathcal{N}=1 \\
	    \vdots \\
	    L_{J-1}\Big (\frac{q^2}{2}\Big) & \mathcal{N}=J-1 \\
	    \frac{1}{2}\Bigg[L_{\mathcal{N}-J}\Big (\frac{q^2}{2}\Big)+L_{\mathcal{N}}\Big (\frac{q^2}{2}\Big) \Bigg] & \mathcal{N} \geq J .
         \end{cases}
\label{eq: Jlayer_graphene_form_factor}
\end{equation}


\subsection{Planar (Disk) pseudopotentials in $J{-}$layer graphene}
\label{sec: disk_pps_JLG}
In the following, we provide analytic expressions of the Coulomb pseudopotentials $V_{m}^{\left (J, ~\mathcal{N}\right)}$ on the disk geometry for different LLs of bilayer graphene $(J{=}2)$ and trilayer graphene $(J{=}3)$. These pseudopotentials are computed using Eq.~\eqref{eq: Haldane_pps_JLG} in conjunction with Eq.~\eqref{eq: Jlayer_graphene_form_factor} for LLs where $\mathcal{N} {\geq} J$, and Eq.~\eqref{eq: ZLL form factor} for the $\mathcal{N}{=}1$ zero-energy LL of bilayer graphene. The computed disk pseudopotentials starting from the $\mathcal{N}{=}1$ zero-energy LL to the $\mathcal{N}{=}4$ LL of bilayer graphene are as follows:
\begin{widetext}
\vspace{-\baselineskip}
\begin{align}
\label{eq: V_ZLL_Coulomb}
V^{\left (J{=}2, ~\mathcal{N}{=}1\right)}_{m} (\theta)&=\frac{\sqrt{\pi}}{32} \Bigl[16~_2F_1\left (\frac{1}{2}, -m;1;1\right)-8~_2F_1\left (\frac{3}{2}, -m;1;1\right)\sin^{2} (\theta)+3~_2F_1\left (\frac{5}{2}, -m;1;1\right)\sin^{4} (\theta) \Bigr],\\[1ex]
\label{eq: V_2_Coulomb_BLG}
 V^{\left (J{=}2, ~\mathcal{N}{=}2\right)}_{m}&=\frac{\Gamma (m-\frac{7}{2})}{131072~\Gamma (m+1)}\Bigl[65536m^4-507904m^3+1299968m^2-1219520m+296625\Bigr],\\[1ex]
 \label{eq: V_3_Coulomb_BLG}
\begin{split}
V^{\left (J{=}2, ~\mathcal{N}{=}3\right)}_{m}&=\frac{\Gamma \left (m-\frac{11}{2}\right)}{8388608~m!}\Bigl[4194304 m^6-73400320 m^5+496074752 m^4\\
&\qquad  -1624174592 m^3+2627746368 m^2-1866183792 m+380654505\Bigr],
\end{split}
\\[1ex]
\label{eq: V_4_Coulomb_BLG} 
\begin{split}
V^{\left (J{=}2, ~\mathcal{N}{=}4\right)}_{m}&=\frac{\sqrt{\pi}}{33554432} \Bigl[ 16777216 \, ~_2F_1\left (\frac{1}{2}, -m;1;1\right)-25165824 \, ~_2F_1\left (\frac{3}{2}, -m;1;1\right)+39321600 \, ~_2F_1\left (\frac{5}{2}, -m;1;1\right)\, \\
&\quad -43909120 ~_2F_1\left (\frac{7}{2}, -m;1;1\right)+35123200 \, ~_2F_1\left (\frac{9}{2}, -m;1;1\right)-19998720 \, ~_2F_1\left (\frac{11}{2}, -m;1;1\right)  \\
&\quad +7835520 \, ~_2F_1\left (\frac{13}{2}, -m;1;1\right)-1921920 \, ~_2F_1\left (\frac{15}{2}, -m;1;1\right)+225225 \, ~_2F_1\left (\frac{17}{2}, -m;1;1\right) \Bigr],
\end{split}
\intertext{where $_2F_1$ is the Gauss hypergeometric function and $\Gamma\left (x\right)$ is the gamma function. The calculated Coulomb pseudopotentials in the disk geometry for trilayer graphene LLs ranging from  $\mathcal{N}{=}3$ to $\mathcal{N}{=}5$ are as follows:}
\label{eq: V_3_Coulomb_TLG}
\begin{split}
    V^{\left (J{=}3, ~\mathcal{N}{=}3\right)}_{m}&=\frac{\sqrt{\pi}}{131072} \Bigl[8 \Bigl\{ 8192 \, ~_2F_1\left (\frac{1}{2}, -m;1;1\right)-6144 \, ~_2F_1\left (\frac{3}{2}, -m;1;1\right)+5760 \, ~_2F_1\left (\frac{5}{2}, -m;1;1\right)\\ 
&\qquad-4640 \, ~_2F_1\left (\frac{7}{2}, -m;1;1\right)+2730 \, ~_2F_1\left (\frac{9}{2}, -m;1;1\right)-945 \, ~_2F_1\left (\frac{11}{2}, -m;1;1\right) \Bigr\}\\
&\qquad+1155 \, ~_2F_1\left (\frac{13}{2}, -m;1;1\right) \Bigr],
\end{split}
\\
\label{eq: V4_Coulomb_TLG}
\begin{split}
    V^{\left (J{=}3, ~\mathcal{N}{=}4\right)}_{m}&=
\frac{\sqrt{\pi}}{33554432}
\Bigl[
128 \Bigl\{131072 \, ~_2F_1\left (\frac{1}{2}, -m;1;1\right)-163840 \, ~_2F_1\left (\frac{3}{2}, -m;1;1\right)+227328 \, ~_2F_1\left (\frac{5}{2}, -m;1;1\right)
  \\
&\qquad -35 \Bigl\{7168 \, ~_2F_1\left (\frac{7}{2}, -m;1;1\right)-6080 \, ~_2F_1\left (\frac{9}{2}, -m;1;1\right)+3816 \, ~_2F_1\left (\frac{11}{2}, -m;1;1\right)
  \\ 
 &\quad -1650 \, ~_2F_1\left (\frac{13}{2}, -m;1;1\right)+429 \, ~_2F_1\left (\frac{15}{2}, -m;1;1\right)\Bigr\}\Bigr\}+225225 \, ~_2F_1\left (\frac{17}{2}, -m;1;1\right) \Bigr],
\end{split}
\\
\label{eq: V5_Coulomb_TLG}
\begin{split}
    V^{\left (J{=}3, ~\mathcal{N}{=}5\right)}_{m}&=
\frac{\sqrt{\pi}}{536870912}
\Bigl[
8 \Bigl\{64 \Bigl\{32 \Bigg (16384 \, ~_2F_1\left (\frac{1}{2}, -m;1;1\right)-28672 \, ~_2F_1\left (\frac{3}{2}, -m;1;1\right)+54528 \, ~_2F_1\left (\frac{5}{2}, -m;1;1\right)  \\ 
&\qquad-80320 \, ~_2F_1\left (\frac{7}{2}, -m;1;1\right)+91420 \, ~_2F_1\left (\frac{9}{2}, -m;1;1\right)-80451 \, ~_2F_1\left (\frac{11}{2}, -m;1;1\right)\Bigg) \\ 
&\qquad+1725108 \, ~_2F_1\left (\frac{13}{2}, -m;1;1\right)-849849 \, ~_2F_1\left (\frac{15}{2}, -m;1;1\right)\Bigr\}+18468450 \, ~_2F_1\left (\frac{17}{2}, -m;1;1\right)  \\ 
&\qquad-3828825 \, ~_2F_1\left (\frac{19}{2}, -m;1;1\right)\Bigr\}+2909907 \, ~_2F_1\left (\frac{21}{2}, -m;1;1\right)
\Bigr].
\end{split}
\end{align}
\end{widetext}


\subsection{Spherical pseudopotentials in $J{-}$layer graphene}
\label{sec: spherical_pps_JLG}
The spherical pseudopotentials, which are obtained by directly solving the Schrodinger equation on the sphere, can be computed following the procedure outlined in Refs.~\cite{Hsiao20, Balram21b}. In particular, for $|\mathcal{N}|{\geq} J$, the pseudopotentials in the LL indexed by $\mathcal{N}$ are given by:
\begin{widetext}
\begin{align}
 V_{L}&=\sum_{m_{1}=-l}^{l}\sum_{m_{2}=-l}^{l}\sum_{m'_{1}=-l}^{l}\sum_{m'_{2}=-l}^{l} \langle L, m|l, m'_{1};l, m'_{2}\rangle \langle l, m_{1};l, m_{2}|L, m\rangle ~\left (1', 2'|V (r)|1, 2 \right)~\delta_{L, m_{1}+m_{2}}\delta_{m_{1}+m_{2}, m'_{1}+m'_{2}},
 \label{eq: Spherical_pp}
\intertext{where $l{=}|Q|{+}\mathcal{N}$ is the shell-angular momentum and $\langle j_{1}, m_{1};j_{2}, m_{2}|j_{3}, m_{3}\rangle$ is the Clebsch-Gordan coefficient. If we take the Coulomb interaction, $V (r)=1/r$, then the quantity $\left (1', 2'|1/r|1, 2 \right)$ is given by:}
\begin{split}
\left (1', 2'|\frac{1}{r}|1, 2 \right) &=\frac{1}{4} \Bigg (
V_{C} (Q+J, Q+J, m_{1}, m_{2}, m'_{1}, m'_{2}, Q, \mathcal{N})+
V_{C} (Q+J, Q, m_{1}, m_{2}, m'_{1}, m'_{2}, Q, \mathcal{N}) \\ 
&\qquad+V_{C} (Q, Q+J, m_{1}, m_{2}, m'_{1}, m'_{2}, Q, \mathcal{N})+
V_{C} (Q, Q, m_{1}, m_{2}, m'_{1}, m'_{2}, Q, \mathcal{N})
\Bigg). \nonumber
\end{split}
\end{align}
\begin{align*}
\intertext{Here, $V_{C} (Q_{1}, Q_{2}, m_{1}, m_{2}, m'_{1}, m'_{2}, Q, \mathcal{N})$ is the two-body Coulomb matrix element for a pair of nonrelativistic fermions, which is given by:}
\begin{split}
V_{C} (Q_{1}, Q_{2}, m_{1}, m_{2}, m'_{1}, m'_{2}, Q, \mathcal{N})&=\frac{e^{2}}{\epsilon R} (2l+1)^{2} (-1)^{Q_1+Q_2-m_1'-m_2'} \times \\
&\qquad\sum_{l'=0}^{2l}\sum_{m'=-l'}^{l'} (-1)^{m'}\begin{pmatrix} l & l' & l \\ m_1' & m' & -m_1\end{pmatrix}\begin{pmatrix} l & l' & l \\ -Q_1 & 0 & Q_1\end{pmatrix}\begin{pmatrix} l & l' & l \\ m_2' & -m' & -m_2\end{pmatrix}\begin{pmatrix} l & l' & l \\ -Q_2 & 0 & Q_2\end{pmatrix} ,
\end{split}
\end{align*}
\end{widetext}
where $\left ([j_{1}, j_{2}, j_{3}];[m_{1}, m_{2}, m_{3}]\right)$ is the Wigner $3j$ symbol, $l{=}|Q|{+}\mathcal{N}$ is the shell-angular momentum and $R{=}\sqrt{l}\ell$ is our choice for the radius of the sphere. Usually, the radius is defined as $R{=}\sqrt{Q}\ell$, and therefore our results, in particular for finite systems, might slightly differ from those given in the literature. 

\section{Overlaps of exact Coulomb ground state with Laughlin state in $J{-}$layer graphene}
\label{sec: overlaps}
Using the above-computed disk pseudopotentials, we have carried out numerical exact diagonalization on the spherical geometry~\cite{Haldane83}. We have calculated overlaps between the exact Coulomb ground state and Laughlin state at $\Bar{\nu}{=}1/3$ and $\Bar{\nu}{=}1/5$, in the various LLs of BLG and TLG. Additionally, in the $\mathcal{N}{=}1$ ZLL of BLG at $\theta{=}\pi/4$ (which is equivalent to the $\mathcal{N}{=}1$ LL of MLG), $\mathcal{N}{=}2$ LL of BLG, and $\mathcal{N}{=}3$ LL of TLG, we have calculated the aforementioned overlaps with the exact Coulomb ground state obtained from the spherical pseudopotentials given in Eq.~\eqref{eq: Spherical_pp}. These overlaps are presented in Tables~\ref{tab: overlaps_Laughlin_exact_N_1_LL_MLG} to~\ref{overlaps_n_5_trilayer_GLL_Laughlin_5}. We note that the overlaps in the $\mathcal{N}{=}1$ LL of MLG are in agreement with those given in Ref.~\cite{Kusmierz18}. The overlaps at filling $\Bar{\nu}{=}1/3$ in the $\mathcal{N}{=}1$ ZLL of BLG for $\theta{=}0$ (LLL) and $\theta{=}\pi/2$ (SLL) can be found in the supplemental material of Ref.~\cite{Balram20b}. Overlaps at $\Bar{\nu}{=}1/5$ can be found in the Table~\Romannum{1} of Ref.~\cite{Balram21}. The low overlaps of the Laughlin state with the exact Coulomb ground state for $N{=}6$, particularly at $\Bar{\nu}{=}1/5$ (see Tables~\ref{tab: overlaps_Laughlin_exact_N_2_LL_BLG} and~\ref{tab: overlaps_Laughlin_exact_N_3_LL_TLG}), can potentially be attributed to the electrons' tendency to form a hexagonal Wigner crystal for this particle number~\cite{Balram21d}. 

\begin{table*}[h]
  \caption{Squared overlaps of the exact Coulomb ground state  $|\Psi^{{\mathcal{N}{=}1~}\rm LL, ~MLG}_{\Bar{\nu}} \rangle$ with the Laughlin state $|\Psi^{\rm Laughlin}_{\Bar{\nu}} \rangle$ at $\Bar{\nu}{=}1/3$ and $1/5$ in the $\mathcal{N}{=}1$ LL of monolayer graphene, which is equivalent to the $\mathcal{N}{=}1$ zero-energy LL of bilayer graphene for $\theta{=}\pi/4$. These overlaps are obtained in the spherical geometry for $N$ electrons at shell angular momentum $2l{=}\Bar{\nu}^{-1} (N{-}1)$ using the disk (D) and spherical (S) pseudopotentials[see Eqs.~\eqref{eq: V_ZLL_Coulomb} and~\eqref{eq: Spherical_pp}, respectively]. A dash ``$-$" indicates numbers that are currently unavailable.}
  \label{tab: overlaps_Laughlin_exact_N_1_LL_MLG}
  \begin{tabular}{c*{5}{c}}
\toprule
$N$ & $\big|\langle \Psi^{{\mathcal{N}{=}1~}\rm LL, ~MLG (S)}_{1/3} | \Psi^{\rm Laughlin}_{1/3} \rangle\big|^{2}$ & $~\big|\langle \Psi^{{\mathcal{N}{=}1~}\rm LL, ~MLG (S)}_{1/5} |\Psi^{\rm Laughlin}_{1/5} \rangle\big|^{2}$ & $~\big|\langle \Psi^{{\mathcal{N}{=}1~}\rm LL, ~MLG (D)}_{1/3} | \Psi^{\rm Laughlin}_{1/3} \rangle\big|^{2}$ & $~\big|\langle \Psi^{{\mathcal{N}{=}1~}\rm LL, ~MLG (D)}_{1/5} |\Psi^{\rm Laughlin}_{1/5} \rangle\big|^{2}$ \\ \midrule
4	& 0.9945	&	0.9736	&	0.9993	&	0.9916	\\
5	& 0.9976    &	0.9956	&	0.9975	&	0.9949	\\ 
6	& 0.9905	&	0.9092	&	0.9834	&	0.8254	\\ 
7	& 0.9907	&	0.9587	&	0.989	&	0.9478	\\ 
8	& 0.9883	&	0.9277	&	0.989	&	0.935	\\ 
9	& 0.9848	&	0.8814	&	0.9871	&	0.9127	\\ 
10	& 0.9821	&	0.8654	&	0.984	&	0.8845	\\ 
11	& 0.9804	&	$-$	    &	0.9807	&	$-$	\\ 
12	& 0.9772    &	$-$	    &	0.9761	&	$-$	\\ 
13	& 0.9744	&	$-$	    &	0.9733	&	$-$	\\ 

\bottomrule
\end{tabular}
   \end{table*}

\begin{table*}[h]
  \caption{Same as Table~\ref{tab: overlaps_Laughlin_exact_N_1_LL_MLG}, but in the $\mathcal{N}{=}2$ LL of bilayer graphene. The exact Coulomb ground state is computed using the disk and spherical pseudopotentials given in Eqs.~\eqref{eq: V_2_Coulomb_BLG} and~\eqref{eq: Spherical_pp}, respectively. }
  \label{tab: overlaps_Laughlin_exact_N_2_LL_BLG}
  \begin{tabular}{c*{5}{c}}
\toprule
$N$ & $\big|\langle \Psi^{{\mathcal{N}{=}2~}\rm LL, ~BLG (S)}_{1/3} | \Psi^{\rm Laughlin}_{1/3} \rangle\big|^{2}$ & $~\big|\langle \Psi^{{\mathcal{N}{=}2~}\rm LL, ~BLG (S)}_{1/5} |\Psi^{\rm Laughlin}_{1/5} \rangle\big|^{2}$ & $~\big|\langle \Psi^{{\mathcal{N}{=}2~}\rm LL, ~BLG (D)}_{1/3} | \Psi^{\rm Laughlin}_{1/3} \rangle\big|^{2}$ & $~\big|\langle \Psi^{{\mathcal{N}{=}2~}\rm LL, ~BLG (D)}_{1/5} |\Psi^{\rm Laughlin}_{1/5} \rangle\big|^{2}$ \\ \midrule
5	&	0.9983	&	0.9909	&	0.9976	&	0.9907	\\ 
6	&	0.9954	&	0.8306	&	0.9857	&	0.7604	\\ 
7	&	0.9937	&	0.9266	&	0.9905	&	0.9192	\\ 
8	&	0.9917	&	0.8670	&	0.9902	&	0.8817	\\ 
9	&	0.9896	&	0.7863	&	0.9898	&	0.8326	\\ 
10	&	0.9868	&	0.7669	&	0.9872	&	0.7961	\\ 
11	&	0.9849	&	$-$	&	0.9843	&	$-$	\\ 
12	&	0.9825	&	$-$	&	0.9806	&	$-$	\\ 
13	&	0.9800	&	$-$	&	0.9777	&	$-$	\\ 
14	&	0.9776	&	$-$	&	0.9755	&	$-$	\\
\bottomrule
\end{tabular}
   \end{table*}

   \begin{table*}
\caption{Squared overlaps of the exact Coulomb ground state $|\Psi^{{\mathcal{N}{=}3~}\rm LL, ~BLG (D)}_{1/3}\rangle$ and the Laughlin state $| \Psi^{\rm Laughlin}_{1/3} \rangle$ at $\Bar{\nu}{=}1/3$ in the $\mathcal{N}{=}3$ LL of bilayer graphene, calculated using the disk (D) pseudopotentials [as given in Eq.~\eqref{eq: V_3_Coulomb_BLG}] in the spherical geometry for $N$ electrons. The table also includes the dimensions of the total orbital angular momentum $L{=}0$ and its $z$ component $L_{z}{=}0$ sectors. $^{\dagger}$ indicates that the ground state does not have $L=0$.}
\label{tab: overlaps_n_3_bilayer_GLL_Laughlin}
\begin{tabular}{c*{5}{c}}
\toprule
$N$ & $2Q$ & $~~$Dimension of $L_{z}=0$ subspace & $~~$Dimension of $L=0$ subspace & $~~$ $|\langle \Psi^{{\mathcal{N}{=}3~}\rm LL, ~BLG (D)}_{1/3} | \Psi^{\rm Laughlin}_{1/3} \rangle|^{2}$\\ \midrule
13  & 36 &  44, 585, 180 	&  21, 660 	&  $^{\dagger}$    \\ 
14  & 39 & 259, 140, 928 	& 100, 123 	&  $^{\dagger}$    \\ 
\bottomrule
\end{tabular} 
\end{table*}

\begin{table*}
\caption{Same as Table~\ref{tab: overlaps_n_3_bilayer_GLL_Laughlin}}, but at $\Bar{\nu}{=}1/5$ in the $\mathcal{N}{=}3$ LL of bilayer graphene.
\label{tab: overlaps_n_3_bilayer_GLL_Laughlin_5}
\begin{tabular}{c*{5}{c}}
\toprule
$N$ & $2Q$ & $~~$Dimension of $L_{z}=0$ of subspace & $~~$Dimension of $L=0$ subspace & $~~$$|\langle \Psi^{{\mathcal{N}{=}3~}\rm LL, ~BLG (D)}_{1/5} | \Psi^{\rm Laughlin}_{1/5} \rangle|^{2}$ \\ \midrule
9  & 40 &   4, 323, 349 	&   2, 082 	&  0.8391 \\ 
10 & 45 &  42, 611, 589 	&  14, 664 	&  0.7898 \\ 
\bottomrule
\end{tabular}
\end{table*}

\begin{table*}
\caption{Same as Table~\ref{tab: overlaps_n_3_bilayer_GLL_Laughlin}, but at $\Bar{\nu}{=}1/3$ in the $\mathcal{N}{=}4$ LL of bilayer graphene. The exact Coulomb ground state is obtained using the disk pseudopotentials given in Eq.~\eqref{eq: V_4_Coulomb_BLG}.}
\label{overlaps_n_4_bilayer_GLL_Laughlin}
\begin{tabular}{c*{5}{c}}
\toprule
$N$ & $2Q$ & $~~$Dimension of $L_{z}=0$ subspace & $~~$Dimension of $L=0$ subspace & $~~$ $|\langle \Psi^{{\mathcal{N}{=}4~}\rm LL, ~BLG (D)}_{1/3} | \Psi^{\rm Laughlin}_{1/3} \rangle|^{2}$ \\ \midrule
13  & 36 &  44, 585, 180 	&  21, 660 	&  $^{\dagger}$ \\ 
14  & 39 & 259, 140, 928 	& 100, 123 	&  $^{\dagger}$ \\ 
\bottomrule
\end{tabular} 
\end{table*}

\begin{table*}
\caption{Same as Table~\ref{tab: overlaps_n_3_bilayer_GLL_Laughlin}, but at $\Bar{\nu}{=}1/5$ in the $\mathcal{N}{=}4$ LL of bilayer graphene. The disk pseudopotentials appropriate for this LL are given in Eq.~\eqref{eq: V_4_Coulomb_BLG}.}
\label{overlaps_n_4_bilayer_GLL_Laughlin_5}
\begin{tabular}{c*{5}{c}}
\toprule
$N$ & $2Q$ & $~~$Dimension of $L_{z}=0$ subspace & $~~$Dimension of $L=0$ subspace & $~~$ $|\langle \Psi^{{\mathcal{N}{=}4~}\rm LL, ~BLG (D)}_{1/5} | \Psi^{\rm Laughlin}_{1/5} \rangle|^{2}$ \\ 
\midrule
9  & 40 &   4, 323, 349 	&   2, 082 	&  $0.49\times 10^{-6}$ \\ 
10 & 45 &  42, 611, 589 	&  14, 664 	&  $0.16\times 10^{-6}$ \\ 
\bottomrule
\end{tabular} 
\end{table*}

\begin{table*}
\caption{Same as Table~\ref{tab: overlaps_Laughlin_exact_N_1_LL_MLG}, but in the $\mathcal{N}{=}3$ LL of trilayer graphene. The exact Coulomb ground state is computed using the spherical and disk pseudopotentials given in Eqs.~\eqref{eq: V_3_Coulomb_TLG} and~\eqref{eq: Spherical_pp}, respectively.}
\label{tab: overlaps_Laughlin_exact_N_3_LL_TLG}
\begin{tabular}{c*{5}{c}}
\toprule
$N$ & $\big|\langle \Psi^{{\mathcal{N}{=}3~}\rm LL, ~TLG (S)}_{1/3} | \Psi^{\rm Laughlin}_{1/3} \rangle\big|^{2}$ & $~\big|\langle \Psi^{{\mathcal{N}{=}3~}\rm LL, ~TLG (S)}_{1/5} |\Psi^{\rm Laughlin}_{1/5} \rangle\big|^{2}$ & $~\big|\langle \Psi^{{\mathcal{N}{=}3~}\rm LL, ~TLG (D)}_{1/3} | \Psi^{\rm Laughlin}_{1/3} \rangle\big|^{2}$ & $~\big|\langle \Psi^{{\mathcal{N}{=}3~}\rm LL, ~TLG (D)}_{1/5} |\Psi^{\rm Laughlin}_{1/5} \rangle\big|^{2}$ \\ \midrule
5	&	0.9631	&	0.9960	&	0.9859	&	0.9946	\\ 
6	&	0.8398	&	0.9312	&	0.9134	&	0.8378	\\ 
7	&	0.9223	&	0.9625	&	0.9536	&	0.9493	\\ 
8	&	0.9071	&	0.9358	&	0.9491	&	0.9357	\\ 
9	&	0.9010	&	0.8996	&	0.9479	&	0.9215	\\ 
10	&	0.8959	&	0.8682	&	0.9365	&	0.8828	\\ 
11	&	0.9031	&	$-$	&	0.9277	&	$-$	\\ 
12	&	0.8868	&	$-$	&	0.9077	&	$-$	\\ 
13	&	0.8820	&	$-$	&	0.9011	&	$-$	\\ 
14	&	0.8768	&	$-$	&	0.8968	&	$-$	\\
\bottomrule
\end{tabular}
\end{table*}

\begin{table*}
\caption{Same as Table~\ref{tab: overlaps_n_3_bilayer_GLL_Laughlin}, but at $\Bar{\nu}{=}1/3$ in the $\mathcal{N}{=}4$ LL of trilayer graphene. The  disk pseudopotentials pertinent to this LL are given in Eq.~\eqref{eq: V4_Coulomb_TLG}.}
\label{overlaps_n_4_trilayer_GLL_Laughlin}
\begin{tabular}{c*{5}{c}}
\toprule
$N$ & $2Q$ & $~~$ Dimension of $L_{z}=0$ subspace & $~~$ Dimension of $L=0$ subspace & $~~$ $|\langle \Psi^{{\mathcal{N}{=}4~}\rm LL, ~TLG (D)}_{1/3}|\Psi^{\rm Laughlin}_{1/3}\rangle|^{2}$   \\ \midrule
11  & 30 &   1, 371, 535 	&   1, 160 	&  0.7252    \\ 
12  & 33 &   7, 764, 392 	&   4, 998 	&  0.6472   \\ 
\bottomrule
\end{tabular} 
\end{table*}

\begin{table*}
\caption{Same as Table~\ref{tab: overlaps_n_3_bilayer_GLL_Laughlin}, but at $\Bar{\nu}{=}1/5$ in the $\mathcal{N}{=}4$ LL of trilayer graphene. The disk pseudopotentials relevant for this LL are given in Eq.~\eqref{eq: V4_Coulomb_TLG}.}
\label{overlaps_n_4_trilayer_GLL_Laughlin_5}
\begin{tabular}{c*{5}{c}}
\toprule
$N$ & $2Q$ & $~~$ Dimension of $L_{z}=0$ subspace & $~~$ Dimension of $L=0$ subspace & $~~$ $|\langle \Psi^{{\mathcal{N}{=}4~}\rm LL, ~TLG (D)}_{1/5}|\Psi^{\rm Laughlin}_{1/5}\rangle|^{2}$     \\ \midrule
9  & 40 &   4, 323, 349 	&   2, 082 	&  0.143    \\ 
10 & 45 &  42, 611, 589 	&  14, 664 	&  0.1633    \\ 
\bottomrule
\end{tabular} 
\end{table*}

\begin{table*}
\caption{Same as Table~\ref{tab: overlaps_n_3_bilayer_GLL_Laughlin}, but at $\Bar{\nu}{=}1/3$ in the $\mathcal{N}{=}4$ LL of trilayer graphene. The disk pseudopotentials for this LL are given in Eq.~\eqref{eq: V5_Coulomb_TLG}.}
\label{overlaps_n_5_trilayer_GLL_Laughlin}
\begin{tabular}{c*{5}{c}}
\toprule
$N$ & $2Q$ & $~~$ Dimension of $L_{z}=0$ subspace & $~~$ Dimension of $L=0$ subspace & $~~$ $|\langle \Psi^{{\mathcal{N}{=}5~}\rm LL, ~TLG (D)}_{1/3}|\Psi^{\rm Laughlin}_{1/3}\rangle|^{2}$ \\ \midrule
11  & 30 &   1, 371, 535 	&   1, 160 	&  0.2958    \\ 
12  & 33 &   7, 764, 392 	&   4, 998 	&  0.1734    \\ 
\bottomrule
\end{tabular} 

\end{table*}
\begin{table*}
\caption{Same as Table~\ref{tab: overlaps_n_3_bilayer_GLL_Laughlin}, but at $\Bar{\nu}{=}1/5$ in the $\mathcal{N}{=}4$ LL of trilayer graphene. The disk pseudopotentials relevant for this LL are given in Eq.~\eqref{eq: V5_Coulomb_TLG}.}
\label{overlaps_n_5_trilayer_GLL_Laughlin_5}
\begin{tabular}{c*{5}{c}}
\toprule
$N$ & $2Q$ & $~~$ Dimension of $L_{z}=0$ subspace & $~~$ Dimension of $L=0$ subspace & $~~$ $|\langle \Psi^{{\mathcal{N}{=}5~}\rm LL, ~TLG (D)}_{1/5}|\Psi^{\rm Laughlin}_{1/5}\rangle|^{2}$  \\ \midrule
9  & 40 &   4, 323, 349 	&   2, 082 	&  $^{\dagger}$    \\ 
10 & 45 &  42, 611, 589 	&  14, 664 	&  $0.144\times 10^{-5}$   \\ 
\bottomrule
\end{tabular} 
\end{table*}


\section{Energies of quasiparticles and quasiholes}
\label{sec: qp_qh_energies}
The quasiparticle (QP) and quasihole (QH) energies above the Laughlin state at $\Bar{\nu}{=}1/ (2s{+}1)$ are computed following the Hamiltonian theory proposed by Murthy and Shankar~\cite{Murthy03, Shankar01}. The energies of the QPs and QHs are calculated using Eqs.~\eqref{eq: quasiparticle energy} and~\eqref{quasihole energy} for $s{=}1{-}4$, in the various LLs of BLG from $\mathcal{N}{=}2{-}5$ and TLG from $\mathcal{N}{=}3{-}5$, and are tabulated in Tables~\ref{table: qp_energies_blg} to \ref{table: qh_energies_tlg}. As described in the main text, the effective interaction $v_{n} (q)$ in Eqs.~\eqref{eq: quasiparticle energy} and~\eqref{quasihole energy} should be replaced by $v_{\mathcal{N}}^{ (J{=}2)} (q)$ for BLG $(J{=}2)$ and by $v_{\mathcal{N}}^{ (J{=}3)} (q)$ for TLG $(J{=}3)$.

\begin{table*}
  \caption{Quasiparticle energies $\Delta_{+}^{\left (J{=}2\right)}\left (\mathcal{N}, s\right)$ above the Laughlin states at fillings $1/ (2s{+}1)$ (where $s{=}1$ to $s{=}$4) in the various LLs of bilayer graphene $(J{=}2)$. The energies are given in units of $e^{2}/ (\epsilon\ell)$.}
  \label{table: qp_energies_blg}
  \begin{tabular}{c*{5}{c}}
\toprule
LLs & $~~\Delta_{+}^{\left (J{=}2\right)}\left (\mathcal{N}, s{=}1\right)$ & $~~\Delta_{+}^{\left (J{=}2\right)}\left (\mathcal{N}, s{=}2\right)$ & $~~\Delta_{+}^{ (J{=}2)}\left (\mathcal{N}, s{=}3\right)$ & $~~\Delta_{+}^{ (J{=}2)} (\mathcal{N}, s{=}4)$\\ \midrule
$\mathcal{N}{=}2$ & $0.105$ & $0.071$ & $0.058$ & $0.05$\\
$\mathcal{N}{=}3$ & $0.099$ & $0.075$ & $0.065$ & $0.057$\\
$\mathcal{N}{=}4$ & $0.09$  & $0.07$  &  $0.062$ & $0.057$\\
$\mathcal{N}{=}5$ & $0.085$ & $0.066$ & $0.059$ & $0.054$ \\
 \bottomrule
\end{tabular}
   \end{table*}
\begin{table*}
  \caption{Quasihole energies $\Delta_{-}^{\left (J{=}2\right)}\left (\mathcal{N}, s\right)$ above the Laughlin states at fillings $1/ (2s{+}1)$ (where $s{=}1$ to $s{=}$4) in the various LLs of bilayer graphene $(J{=}2)$. The energies are given in units of $e^{2}/ (\epsilon\ell)$.}
  \label{table: qh_energies_blg}
  \begin{tabular}{c*{5}{c}}
\toprule
LLs & $~~\Delta_{-}^{\left (J{=}2\right)}\left (\mathcal{N}, s{=}1\right)$ & $~~\Delta_{-}^{\left (J{=}2\right)}\left (\mathcal{N}, s{=}2\right)$ & $~~\Delta_{-}^{ (J{=}2)}\left (\mathcal{N}, s{=}3\right)$ & $~~\Delta_{-}^{ (J{=}2)} (\mathcal{N}, s{=}4)$\\ \midrule
$\mathcal{N}{=}2$ & $0.013$ & $-0.012$ & $-0.016$ & $-0.016$\\
$\mathcal{N}{=}3$ & $-0.009$ & $-0.023$ & $-0.023$ & $-0.021$\\
$\mathcal{N}{=}4$ & $-0.014$  & $-0.026$  &  $-0.026$ & $-0.024$\\
$\mathcal{N}{=}5$ & $-0.016$ & $-0.027$ & $-0.027$ & $-0.025$ \\
 \bottomrule
\end{tabular}
   \end{table*}

   \begin{table*}
  \caption{Same as Table~\ref{table: qp_energies_blg}, but for trilayer graphene $(J{=}3)$.}
  \label{table: qp_energies_tlg}
  \begin{tabular}{c*{5}{c}}
\toprule
LLs & $~~\Delta_{+}^{\left (J{=}3\right)}\left (\mathcal{N}, s{=}1\right)$ & $~~\Delta_{+}^{\left (J{=}3\right)}\left (\mathcal{N}, s{=}2\right)$ & $~~\Delta_{+}^{ (J{=}3)}\left (\mathcal{N}, s{=}3\right)$ & $~~\Delta_{+}^{ (J{=}3)} (\mathcal{N}, s{=}4)$\\ \midrule
$\mathcal{N}{=}3$ & $0.106$ & $0.075$ & $0.062$ & $0.054$\\
$\mathcal{N}{=}4$ & $0.095$ & $0.077$ & $0.067$ & $0.061$\\
$\mathcal{N}{=}5$ & $0.083$  & $0.069$  &  $0.064$ & $0.059$\\
 \bottomrule
\end{tabular}
   \end{table*}

\begin{table*}
  \caption{Same as Table~\ref{table: qh_energies_blg}, but for trilayer graphene $(J{=}3)$.}
  \label{table: qh_energies_tlg}
  \begin{tabular}{c*{5}{c}}
\toprule
LLs & $~~\Delta_{-}^{\left (J{=}3\right)}\left (\mathcal{N}, s{=}1\right)$ & $~~\Delta_{-}^{\left (J{=}3\right)}\left (\mathcal{N}, s{=}2\right)$ & $~~\Delta_{-}^{ (J{=}3)}\left (\mathcal{N}, s{=}3\right)$ & $~~\Delta_{-}^{ (J{=}3)} (\mathcal{N}, s{=}4)$\\ \midrule
$\mathcal{N}{=}3$ & $0.001$ & $-0.018$ & $-0.02$ & $-0.019$\\
$\mathcal{N}{=}4$ & $-0.014$  & $-0.027$  &  $-0.026$ & $-0.024$\\
$\mathcal{N}{=}5$ & $-0.016$ & $-0.029$ & $-0.028$ & $-0.026$ \\
 \bottomrule
\end{tabular}
   \end{table*}
    
\pagebreak

\bibliography{biblio_fqhe}

\end{document}